\documentclass{elsarticle}
\usepackage{amsmath,bm,epsfig,subfigure}
\usepackage{epstopdf}
\usepackage[margin=0.6in]{geometry}
\usepackage{indentfirst}
\usepackage{amssymb}
\usepackage{graphicx}
\usepackage{float}
\usepackage{color}
\usepackage[normalem]{ulem}
\usepackage{lineno,hyperref}
\modulolinenumbers[5]
\usepackage[toc,page]{appendix}
\usepackage{natbib}
\usepackage{subfigure}

\DeclareMathAlphabet{\mathpzc}{OT1}{pzc}{m}{it}

\journal{Journal of Sound and Vibration}
\usepackage{etoolbox}
\makeatletter
\patchcmd{\ps@pprintTitle}
  {Preprint submitted to}
  {(Published in}
  {}{}
\makeatother

\begin{document}

\begin{frontmatter}

\title{Nonlinear dynamics of hidden modes in a system with internal symmetry}

\author{Nathan Perchikov}

\author[]{O.V. Gendelman\corref{mycorrespondingauthor}}
\cortext[mycorrespondingauthor]{Corresponding author}
\ead{ovgend@tx.technion.ac.il}

\address{Faculty of Mechanical Engineering, Technion, Haifa 32000, Israel}

\begin{abstract}
We consider a discrete dynamical system with internal degrees of freedom (DOF). Due to the symmetry between the internal DOFs, certain internal modes cannot be excited by external forcing (in a case of linear interactions) and thus are considered "hidden". If such a system is weakly asymmetric, the internal modes remain approximately "hidden" from the external excitation, given that small damping is taken into account. However, already in the case of weak cubic nonlinearity, these hidden modes can be excited, even as the exact symmetry is preserved. This excitation occurs through parametric resonance. Floquet analysis reveals instability patterns for the explored modes. To perform this analysis with the required accuracy, we suggest a special method for obtaining the Fourier series of the unperturbed solution for the nonlinear normal mode. This method does not require explicit integration of the arising quadratures. Instead, it employs expansion of the solution at the stage of the implicit quadrature in terms of Chebyshev polynomials. The emerging implicit equations are solved by using a fixed-point iteration scheme. Poincar\'{e} sections help to clarify the correspondence between the loss of stability of the modes and the global structure of the dynamical flow. In particular, the conditions for intensive energy exchange in the system are characterized.
\end{abstract}


\end{frontmatter}


\section{Introduction}
\label{sect1}

This work addresses the problem of nonlinear dynamics of internal modes in a discrete mechanical system. Linear dynamical systems with internal symmetries can have "hidden" modes (HM), i.e. modes that cannot be excited by any external excitation. In the same time, it is shown that these can be excited by internal resonance in the presence of interaction nonlinearity -- even if this nonlinearity preserves the symmetry. This mechanism of nonlinear breaking of linear symmetry is interesting both theoretically and from the point of view of possible applications.

The main motivation of the present work is exploration of the relationship between internal symmetry in a simple discrete dynamical system and the localization of energy in the associated phase space. In terms of possible practical applications, one natural realization would be in micro-sensor engineering, where a parametric resonance may be used for enhancement of the device sensitivity. In the case of a hidden mode excited by resonance, the increased sensitivity effect gains new qualitative nature, as elaborated on in Section \ref{sect12}. In a more general context, peculiar dynamics caused by the internal symmetry may be used for design of acoustic metamaterials.

The addressed problem lies within the broader realm of nonlinear normal modes (NNM), their interactions and bifurcations, stability induced mode-switching and interaction of several instability carriers. Although the amount of work done on the problem of nonlinear symmetry breaking in discrete mechanical systems is not vast,  there are several important results in adjacent areas. Mostly these results address the theoretical perspective of nonlinear normal modes (NNM) and various applications, such as oscillations in carbon nano-tubes. Several results are of particular relevance to the investigation.

On the mathematical side, in the general context of NNMs, the issues of stability, integrability and bifurcations are addressed rather thoroughly in \cite{Vakakis1996}, \cite{Kerschen} and \cite{Vakakis2008}. Multiple modes interaction with parametric resonance leading to energy exchange between different modes is presented in \cite{Nayfeh1992}. In that work, a nonlinear discrete mechanical system with quadratic nonlinearities is harmonically excited exhibiting auto-parametric resonance and loss of single-mode stability through period-doubling and Hopf (primary and secondary) bifurcations. The system, originating from flight mechanics, is analyzed by the method of multiple scales. Different dynamic phenomena, such as quasi-periodicity, phase-locking and chaos, are exhibited. In \cite{Rand1992} bifurcations leading to switching between different NNMs are analyzed by asymptotic approximations of integrals of motion, for benchmark two-DOF discrete nonlinear mechanical systems. In \cite{Sofroniou2014}, the dynamics of  a parametrically excited system with two forcing terms is examined. Interactions between instability tongues are studied on the basis of an extended Mathieu equation, and transition to chaos upon inclusion of nonlinearity and damping is established. In \cite{Recktenwald2006} conditions for the collapse of instability tongues in discrete dynamical systems with parametric resonance are investigated.

In \cite{Shi2009} carbon nano-tubes modeled with Lennard-Jones interaction potentials are explored both with MD simulations and structural elastic FE analysis. Parametric-resonance-induced mode switching (energy exchange between longitudinal and transversal vibrational modes) is revealed. In \cite{Shi2010}, parametric-resonance-induced mode transformation in single-walled nano-tubes, modeled as continuous elastic shells, is analyzed through a Mathieu equation and multiple instability tongues are obtained. In \cite{ShiA2009} that analysis is improved by the use of a non-local elasticity model. In \cite{Goncalves2004} parametric-resonance-induced mode-switching is analyzed on the basis of a continuum shallow shell model. Modal interaction is examined through calculation of the instability tongues and exploration of bifurcation diagrams. In \cite{Silva2014} the influence of internal resonances on the dynamics of fluid-filled cylindrical shells is analyzed. In this work the interrelations between parametric resonance and the interaction between the symmetric and the antisymmetric modes were examined by perturbation techniques.

In \cite{Dinklage2000} the issue of hidden modes is addressed, mainly experimentally, in the context of wave propagation in DC glow discharge plasma. It is found there that hidden states with rich spatio-temporal dynamics exist in the system even long after the driving force is gone, producing nonlinear wave--wave interactions. In \cite{Li2009}, hidden dynamics in folding of RNA molecules is characterized experimentally, and through the modeling of quasi-periodic behavior using a Langevin equation with external potential. The authors demonstrated that linearly hidden intra-molecular modes affect the dynamics by yielding damped quasi-periodic responses when becoming nonlinearly excited.

The main objective of the present paper is to describe and explore a relatively simple discrete mechanical system with internal symmetry and "hidden" modes, which can exhibit symmetry breaking due to nonlinearity. This symmetry breaking reveals itself trough excitation of the hidden modes by the mechanism of parametric resonance.

The structure of the paper is as follows. Sections \ref{sect2}-\ref{sect5} address the dynamics of a linear system with perfect symmetry, as well as modifications caused by weak perturbations of this symmetry. Sections \ref{sect7}-\ref{sect9} explore the nonlinear symmetry breaking of a symmetric mode, and Section \ref{sect10} addresses the nonlinear symmetry breaking of an antisymmetric mode. Section \ref{sect11} illustrates the nonlinear instability of single-mode dynamics with the help of Poincar\'e sections. Section \ref{sect12} concludes.

\section{Description of the model and basic setting -- the linear symmetric undamped case}
\label{sect2}

\begin{figure}[H]
\begin{center}
\includegraphics[scale=0.45]{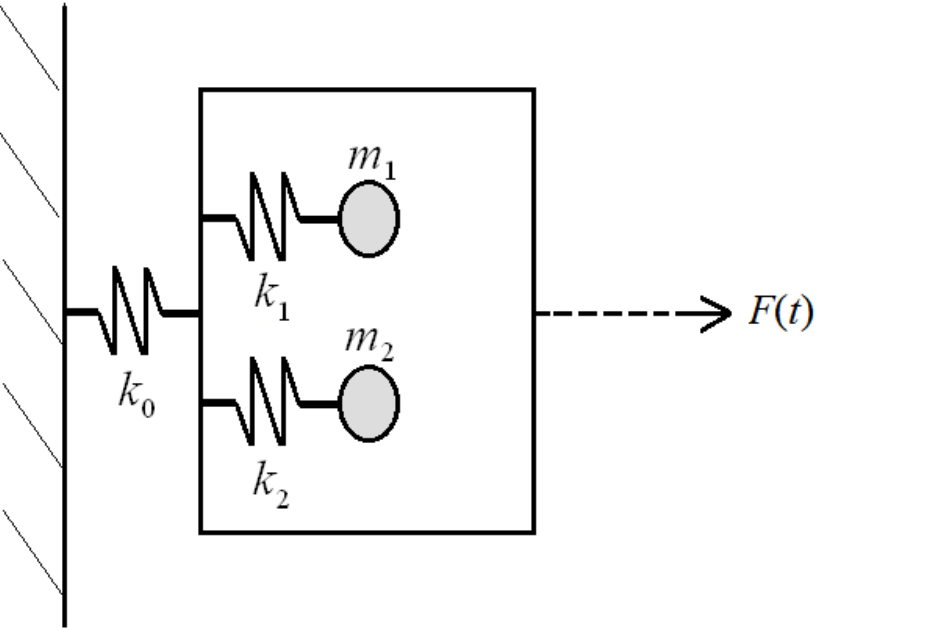}
\end{center}
\caption{\small Schematic plot of the system under investigation.}
\label{Figure1}
\end{figure}

The basic system under consideration includes a symmetric pair of oscillators in a massless box with an external spring (see Fig. \ref{Figure1}).

This system is described by the following equations of motion:
\begin{equation}
\begin{split}
m_1\ddot{x}_1+k_1(x_1-x_0)=0 \\
m_2\ddot{x}_2+k_2(x_2-x_0)=0 \\
k_0 x_0-k_1(x_1-x_0)-k_2(x_2-x_0)=F(t)
\label{eq2.1}
\end{split}
\end{equation}
where $x_1$ and $x_2$ are the displacements of the masses in the horizontal direction, and $x_0$ is the displacement of the rigid massless box containing the masses $m_1$ and $m_2$. The spring constants are denoted by $k_1,k_2$ and $k_0$. In the symmetric case, namely for $m_2=m_1=m, \ k_2=k_1=k$, the system has two normal modes, namely the perfectly symmetric and antisymmetric oscillators. In the case of the antisymmetric mode, the external box remains at rest. Moreover, it is obvious that any external forcing applied to the box is unable to excite this antisymmetric mode. Therefore, it is impossible to reveal its existence if one is only able to excite the external box and to observe its displacement. In other words, there is no possibility to decipher the internal structure of the box. Thus, the antisymmetric mode in this case exemplifies the hidden mode (HM) -- the mode that cannot be excited by any external forcing.

To be specific, external harmonic forcing is introduced and the following notations are used:
\begin{equation}
\begin{split}
F(t)=a \cos{(\Omega t)}, a>0 \\
\omega \triangleq \sqrt{\frac{k}{m}},\xi \triangleq \frac{k_0+k}{k_0+2k},\xi_0 \triangleq \frac{k_0}{k_0+2k},\eta \triangleq \frac{k_0}{k}
\label{eq2.2}
\end{split}
\end{equation}

The solution of Eq. (\ref{eq2.1}) (for the observable quantity, $x_0$) can be written as:
\begin{equation}
\begin{split}
x_0(t)=\frac{a}{m}\left\lbrace\left [ \frac{1-\xi}{\omega^2}-\frac{2(1-\xi)^2}{\Omega^2-\xi_0\omega^2}\cos(\Omega t) \right ]+\right.
\left. \frac{2(1-\xi)^2}{\Omega^2-\xi_0\omega^2}\cos\left(\sqrt{\xi_0}\omega t\right) \right \rbrace
\label{eq2.3}
\end{split}
\end{equation}

This system has evident single resonance:
\begin{equation}
\begin{split}
\Omega \rightarrow \sqrt{\xi_0}\omega \Rightarrow
x_0(t) \to \frac{a}{m} \frac{1-\xi}{\omega^2}\cos\left(\sqrt{\xi_0}\omega t\right)
+\frac{a}{m}(1-\xi)^2\frac{\sin{\left(\sqrt{\xi_0}\omega t\right)}}{\sqrt{\xi_0}\omega} t \underset{t \to \infty}\to \pm\infty
\label{eq2.4}
\end{split}
\end{equation}

The perfect mode hiding described above is realized only in the case of perfect symmetry, which cannot be achieved in realistic systems. The effect of small asymmetry is described in the next section.

\section{Internal mode excitation under small asymmetry $-$ the linear undamped case}
\label{sect3}

The basic setting can be generalized by the introduction of parameter asymmetry. For simplicity and with no real loss of generality, the asymmetry is assumed only in the mass distribution, as follows:
\begin{equation}
m_1=m,m_2=(1+\epsilon) m ; k_2=k_1=k
\label{eq3.1}
\end{equation}

For small asymmetry, the solution for the internal displacements, up to non-resonant $\mathcal{O}(\epsilon)$ terms, can be expressed as follows:
\begin{equation}
\begin{split}
x_1(t) \underset{\epsilon \ll 1} \simeq \frac{a}{m}(1-\xi) \frac{\cos(\omega_{r,2}t)-\cos{(\Omega t)}}{\Omega^2-\omega_{r,2}^2} +
\frac{a}{m}\frac{1-\xi}{2}\frac{\omega_{r,1}^2}{\Omega^2-\omega_{r,2}^2} \frac{\cos(\omega_{r,1}t)-\cos{(\Omega t)}}{\Omega^2-\omega_{r,1}^2}\epsilon
\label{eq3.2}
\end{split}
\end{equation}
where
\begin{equation}
\begin{split}
\omega_{r,1} \triangleq   \frac{\sqrt{2}}{2}\omega\sqrt{\frac{2+\epsilon}{1+\epsilon}\xi+\sqrt{\left(\frac{2+\epsilon}{1+\epsilon}\right)^2\xi^2-\frac{4\xi_0}{1+\epsilon}}} \ , \
\omega_{r,2}\triangleq  \frac{\sqrt{2}}{2}\omega\sqrt{\frac{2+\epsilon}{1+\epsilon}\xi-\sqrt{\left(\frac{2+\epsilon}{1+\epsilon}\right)^2\xi^2-\frac{4\xi_0}{1+\epsilon}}}
\label{eq3.3}
\end{split}
\end{equation}

For small mass asymmetry, one has: $x_2(t,\epsilon)=x_1(t,-\epsilon)+O(\epsilon^2)$. Using this result and the fact that the third equation in Eq. (\ref{eq2.1}) yields the relation: $x_0(t)=a(\xi_0/k_0)\cos{(\omega t)}+(1-\xi)[x_1(t)+x_2(t)]$, two associated resonances are revealed for the observable variable $x_0$. The first resonance is a "strong" one, and is similar to the single resonance in the perfectly symmetric case:
\begin{equation}
\begin{split}
\Omega \rightarrow \omega_{r,2} \underset{\epsilon \ll 1} \approx \sqrt{\xi_0}\omega \Rightarrow  x_0(t)
\to \frac{a}{k_0}\xi_0\cos{(\omega t)}+\frac{a}{k_0}\eta\frac{(1-\xi)^2}{\sqrt{\xi_0}} \sin{\left(\sqrt{\xi_0}\omega t\right)} t \underset{t \to \infty}\to \pm\infty
\label{eq3.4}
\end{split}
\end{equation}

In addition, a second resonance exists which is "weak", or "narrow":
\begin{equation}
\begin{split}
\Omega \rightarrow \omega_{r,1}  \underset{\epsilon \ll 1} \approx (1-\epsilon/4)\omega \Rightarrow x_0(t)
\to \frac{a}{k_0}\xi_0\cos{(\omega t)}-\frac{1}{2}\frac{a}{k_0}\eta\frac{(1-\xi)^2}{1-\xi_0}\left[\cos{(\omega t)}\sin{\left(\frac{\epsilon\omega t}{4}\right)}\right.\\ \left.+\frac{\epsilon}{2}\sin{(\omega t)}\cos{\left(\frac{\epsilon\omega t}{4}\right)}\right] \epsilon \omega t \underset{t \to \infty}\to -\frac{a}{k_0}\frac{\xi_0}{4}\cos{(\omega t)}\sin{\left(\frac{\epsilon\omega t}{4}\right)}\epsilon \omega t\underset{t \to \infty}\to  \pm\infty
\label{eq3.5}
\end{split}
\end{equation}

This additional resonance indicates the presence of internal structure, which is no longer "hidden".  The "weakness" of the second resonance can be understood from Eq. (\ref{eq3.2}). For the first resonance frequency, when $\Omega-\omega_{r,2} \to \epsilon$, the amplitude of $x_0(t)$ is $\mathcal{O}(\epsilon^{-1})$. For the second frequency, when $\Omega-\omega_{r,2} \to \epsilon$, the amplitude is $\mathcal{O}(1)$, due to the multiplication by $\epsilon$ in the end of Eq. (\ref{eq3.2}). Only when  $\Omega-\omega_{r,2} \to \epsilon^2$, the amplitude scales as $\mathcal{O}(\epsilon^{-1})$.  Consequently, the resonance spike is much narrower for the second resonance frequency (amplitude of magnitude, say, 1000 and above, holds for the first resonance in a range of $10^{-3}$ around the singularity, while for the second resonance, only in a range of $10^{-6}$ around the singularity).

\begin{figure}
\begin{center}
\includegraphics[scale=0.8]{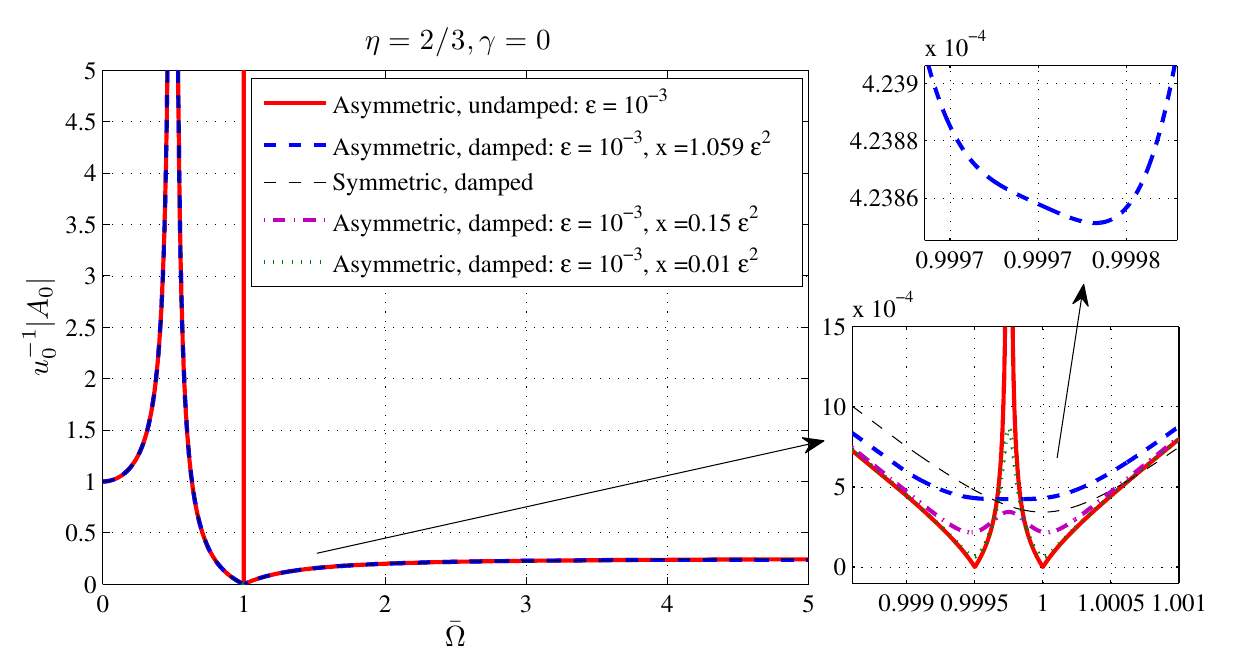}
\end{center}
\caption{\small A typical amplitude-frequency curve with the two emerging resonances in the case of small asymmetry. The first one is "strong", at $\bar{\Omega}=1/2$, and the other is weaker (narrower) at $\bar{\Omega}=1$ ($\bar{\Omega}$ defined in Section \ref{sect4}). The response's nonzero large-frequency limit is an artifact due to zero box mass (not significant for the assumed purposes).}
\label{Figure2}
\end{figure}

A typical two-resonances response curve is shown (in solid red online) in Fig. \ref{Figure2} (the dashed curves are related to the following sections).

If one allows small asymmetry in the internal structure in a realistic scenario, it is also possible to accept that the system under consideration has small viscous damping. This damping will not eliminate the strong main resonance (it will only make its amplitude bounded), but, as it is going to be demonstrated, it can strongly suppress the resonance related to the internal asymmetry. As a result, the amplitude--frequency curve will exhibit only a small kink at the second-resonance frequency (see Section \ref{sect4}), which can even be entirely hidden for somewhat larger damping (see Section \ref{sect5}).

\section{The case of linear interaction with weak asymmetry and small (linear) damping}
\label{sect4}

The equations of motion for the linearly damped slightly asymmetric system take the following form:
\begin{equation}
\begin{split}
m_1\ddot{x}_1+\zeta_1(\dot{x}_1-\dot{x}_0)+k_1(x_1-x_0)=0 \\
m_2\ddot{x}_2+\zeta_2(\dot{x}_2-\dot{x}_0)+k_2(x_2-x_0)=0 \\
k_0 x_0-k_1(x_1-x_0)-\zeta_1(\dot{x}_1-\dot{x}_0)  -k_2(x_2-x_0)-\zeta_2(\dot{x}_2-\dot{x}_0)=F(t)
\label{eq4.1}
\end{split}
\end{equation}

The steady-state equations for the amplitudes are:
\begin{equation}
\begin{split}
(\Omega^2 m_1-i\Omega\zeta_1-k_1)A_1+(i\Omega\zeta_1+k_1)A_0=0 \\
(\Omega^2 m_2-i\Omega\zeta_2-k_2)A_2+(i\Omega\zeta_2+k_2)A_0=0  \\
-(k_1+i\Omega\zeta_1)A_1-(k_2+i\Omega\zeta_2)A_2 +[(k_0+k_1+k_2)+i\Omega(\zeta_1+\zeta_2)]A_0=a
\label{eq4.2}
\end{split}
\end{equation}

In this section, unlike in the previous and following sections, asymmetry in all three parameters is assumed, as follows:
\begin{equation}
\begin{split}
m_1=m, \ m_2=(1+\epsilon)m; \ k_1=k, \ k_2=(1+\gamma)k; \ \zeta_1=\zeta, \ \zeta_2=(1+\delta)\zeta
\label{eq4.3}
\end{split}
\end{equation}

The solution of Eq. (\ref{eq4.2}) for the amplitude of $x_0$, namely, $|A_0|$, takes the following form:
\begin{equation}
\begin{split}
|A_0|=f(\bar{\Omega},x)|A_1|; \ f(\bar{\Omega},x) \triangleq \frac{\sqrt{[1-(1-x)\bar{\Omega}^2]^2+x\bar{\Omega}^6}}{1+x\bar{\Omega}^2} \underset{x,|1-\bar{\Omega}| \ll1} \to \sqrt{(1-\bar{\Omega}^2)^2+x\bar{\Omega}^6} \ ; \
\left |A_1 \right|=u_0 \eta \left |\hat{A}_1 \right|
\label{eq4.4}
\end{split}
\end{equation}
where the expression for an internal mass response amplitude, $|\hat{A}_1|$, is given in \ref{AppendixA1} (due to the fact that its derivation is relatively straightforward, as expected from a solution of a linear problem, and in the same time rather cumbersome), and where the following notation is used:
\begin{equation}
x \triangleq \bar{\zeta}^2, \ u_0 \triangleq \frac{a}{k_0}, \ \bar{\zeta} \triangleq \frac{\zeta}{m\omega}, \ \bar{\Omega} \triangleq \frac{\Omega}{\omega}
\label{eq4.5}
\end{equation}

In the symmetric undamped case, one has: $|A_1|=|f_c^{(0)}|/|f_d^{(0)}|$ (see \ref{AppendixA1}). This gives: $|A_0|\underset{\bar{\Omega}\to \sqrt{\xi_0}}\to (1-\xi_0)^2/0$, which cannot be regularized by an asymptotically small parameter (for a finite $\eta$), and $|A_0|\underset{\bar{\Omega}\to 1}\to 0/0$, which yields $0$ (by l'H\^opital's rule) for the symmetric undamped case, diverges in the undamped case with small asymmetry, and converges to a finite value for the typical slightly-asymmetric weakly-damped case. Such finite-valued "kinks", or local maxima in the $|A_0|$ response curve at the second resonance frequency, are shown for two typical damping values in Fig. \ref{Figure2} in dashed (dark green and purple online) lines.

One important remarks should be made at this point. As may be seen from Eq. (\ref{eq2.1}), the examined system is described in terms of two dynamic degrees of freedom only, the massless box playing the role of nonlinear, nontrivial coupling between the two internal masses. Thus the mode of independent mass vibration is degenerated here. This assumption simplifies the investigation, say, by making even a single Poincar\'e surface section fully representative of global dynamics (for given parameters). An artifact of the assumption is a box-response curve non-decaying at infinite frequencies. This feature is however does not influence the studied effect. The asymmetry-induced second resonance peak is a feature of the existence of two masses, rather than one inside the box. Similarly, as shown in the following sections, the nontrivial dynamic picture emerging is related to the special coupling between the two internal masses created by the box, whether the box has mass or not. Introduction of box mass should, naturally, make the dynamic picture more complex, but it clearly cannot affect the existence of a hidden mode (since for antisymmetric mode dynamics the box mass cancels out from the equations of motion). Box mass can only affect the stability picture for interaction nonlinearities, however, its introduction, rendering the analysis more complex, is expected to have only a quantitative effect in the case of small mass, due to the small energy associated with it (even in the zero mass limit the amplitude is bounded, implying zero energy limit).

\section{Mode hiding}
\label{sect5}

Fig. \ref{Figure2} demonstrates that the second resonance frequency, $\bar{\Omega}=1$, manifests itself even in the symmetric case, which in the undamped limit corresponds to exact anti-resonance in $|A_0|$, corresponding to a zero amplitude of the box.  However, the presence of two slightly asymmetric degrees of freedom converts this anti-resonance into resonance. In the damped case, instead of resonance and anti-resonance (or infinite versus zero amplitude), one obtains a local $minimum$ close to $\bar{\Omega}=1$, related to the mere existence of internal degrees of freedom, and a local $maximum$, due to the existence of $two$ (rather than one) internal masses in asynchronous motion. Consequently, in the slightly damped slightly asymmetric case, the complex internal structure of the box can only reveal itself to the external observer through the local $maximum$ in the $|A_0|$ response curve, in the vicinity of the second resonance frequency -- see the dashed (dark green and purple online) curves in the lower inset in Fig. \ref{Figure2}.

It appears that the linear "hiding" of the $antisymmetric$ internal mode is guaranteed for every practical purpose, if the aforementioned local maximum is eliminated. Then, since in the symmetric slightly damped case there exists a minimum in the response curve near $\bar{\Omega}=1$, and no local $maximum$ near it, (i.e. one observes local $convexity$), it would be reasonable to say that the second mode may be considered hidden if the frequency response curve becomes $convex$ in the neighborhood of $\bar{\Omega}=1$. Local concavity can still be reminiscent of the second internal mode, even without an actual local maximum, but as long as the response curve shows only a local minimum in a strictly convex neighborhood, no conclusion can be drawn regarding the existence of a second, internal mode, at least while one analyzes the response curve itself, and not its higher-order derivatives, which requires much better experimental accuracy.

\subsection{Local convexification of the frequency response of  $|A_0|_{(\bar{\Omega})}$ around $\bar{\Omega}=1$}

In order to assure the local convexity of the response curve around $\bar{\Omega}=1$, one should choose the damping coefficient properly. The higher the asymmetry, the higher should be the damping required for hiding. Thus an increasing function is sought. For no asymmetry, no damping is required for mode hiding. Thus a function increasing from zero is needed. Since the resonance singularities are not of the essential type, it appears that the required function should be expressible as a power series. In the asymptotic limit of small asymmetry, a single, leading power should suffice. Thus the form $x \sim \epsilon^p$ is evident.

Regularization by damping is embodied in removing the resonance singularity, which arises from the denominator of $|\hat{A}_1|$ as given in Eq. (\ref{eqabsA1exp}) in \ref{AppendixA1}, in its undamped limit. From Eqs. (\ref{eq4.4}) and (\ref{eqabsA1exp})-(\ref{eq4.8}), it appears that, at the second resonance frequency, the undamped limit of the denominator of $|\hat{A}_1|$ in Eq. (\ref{eqabsA1exp}), is of order $\mathcal{O}(\epsilon^2)$ (see Eqs. (\ref{eq3.5}), (\ref{eq4.4}) and (\ref{eqabsA1exp})-(\ref{eq4.8})). What is added to this limit case denominator, $f_d^{(0)}$, is to leading-order a term linear with $x$. Thus, for $p>2$, the effect of damping would be insufficient. It would eliminate resonance, but would not affect the shape immediately around it, with the local maximum and the concavity. Thus, it is reasonable to assume $p \le 2$. Next, for $p<2$, one gets $f \sim \sqrt{x}$ around the second resonance frequency, and since $|A_1|$ is regularized (has a finite value) at that frequency for any $p>0$, one would have $|A_0| \underset{\bar{\Omega}\to 1} \sim \sqrt{x}$. The value itself is acceptable, however, if $f$ around $\bar{\Omega}=1$ is simply $\sqrt{x}$, the original local-minimum structure produced by the $(1-\bar{\Omega}^2)^2$ part of $f$ is lost. Although this may be acceptable for hiding, it is an overshoot if one wishes to find the $verge$ of hiding, or in other words, the convexification $bifurcation$. Consequently, it is reasonable to assume that from this perspective, one should look for $p\ge 2$. One thus gets the combined condition: $2\le p \le 2$, which condenses to $p=2$. Thus the following ansatz for the damping coefficient is taken:
\begin{equation}
\label{eq5.1}
x_c=\alpha\epsilon^2
\end{equation}

Next, the hiding requirement can be written as:
\begin{equation}
\label{eq5.2}
|A_0|''_{(\bar{\Omega}_c,x_c)}=0 \  \ , \ |A_0|'''_{(\bar{\Omega}_c,x_c)}=0
\end{equation}

These are two equations with two unknowns, the first -- for the critical damping, $x_c$, and the second determines the bifurcation point, where two adjacent inflection point merge, $\bar{\Omega}_c$. The first equation requires the vanishing of the second derivative, which means that the area around the point stops being concave and is on the verge of becoming convex. The second equation, for the vanishing of the third derivative, guarantees that the curvature-changing point is the one with the extremal curvature in the neighborhood. In other words, by solving Eqs. (\ref{eq5.2}), one guarantees that the most concave point in the neighborhood (of the second undamped-limit resonance frequency) is on the verge of becoming convex. That is that the minimum of the second derivative of the response curve becomes positive. This guarantees the positivity of the second derivative in the entire neighborhood, or quasi-local convexity, which is considered to be sufficient for the qualitative hiding of the internal mode. The details of the solution of Eqs. (\ref{eq5.2}) for $\alpha$ are given in \ref{AppendixA}. Here only the final result is provided, which has an exact representation as a radical:
\begin{equation}
\alpha_c=(2+\sqrt{5})/4\approx 1.059
\label{eq5a}
\end{equation}
(where the subscript stands for ``convexification").

The fact that the result: $\alpha_c\approx 1.059=\mathcal{O}(1)$ was obtained, confirms the correctness of the chosen ansatz.  Hiding by linear damping for small asymmetry is thus established. In terms of the linear damping coefficient itself one therefore gets second (antisymmetric) internal mode hiding if the parametric (mass) asymmetry satisfies:
\begin{equation}
\label{eq5.14}
\zeta_H>\frac{\sqrt{2+\sqrt{5}}}{2}\sqrt{k/m}|\Delta{m}|
\end{equation}
where $\zeta_H$ is the hiding damping and $\Delta{m}$ is the mass asymmetry for the two oscillators inside the box.

Several notes should be made at this point. First, the hiding analysis was performed for only inertial asymmetry, however it is clear that in the small asymmetry limit, stiffness asymmetry plays an equivalent role in the equations. Second, the $linear$ relation obtained for hiding (of the observable quantity) between the problem parameters in both of which the equations of motion are $linear$, is reasonable but not trivial. The emergent convexity is illustrated in the upper inset in Fig. \ref{Figure2}.

\section{Nonlinear normal modes -- the case of cubic nonlinearity: symmetric mode integration}
\label{sect7}

Following the conclusions of the previous section, where it was established that sufficient, yet small damping can almost entirely conceal the effect of the asymmetry on the system's response, it now seems justified to consider the effect of nonlinearity. For this goal we explore the symmetric, undamped system with nonlinear internal potentials. The perspective is to establish the existence of parametric resonance. Therefore, it is reasonable to assume that if parametric resonance (or mode switching/interaction) exists in an undamped symmetric nonlinear system, then asymmetry can only enhance this phenomenon, but would hardly diminish it. Thus for simplicity and better analytic maneuverability, the undamped symmetric case is addressed here. Therefore, the system is still represented by Fig. \ref{Figure1}, however now cubic nonlinearity is added, producing the following equations of motion:
\begin{equation}
\begin{split}
m\ddot{x}_1+k(x_1-x_0)+p(x_1-x_0)^3=0 \\
m\ddot{x}_2+k(x_2-x_0)+p(x_2-x_0)^3=0 \\
k_0 x_0-k(x_1-x_0)-p(x_1-x_0)^3-k(x_2-x_0)-p(x_2-x_0)^3=F(t)
\label{eq6.2}
\end{split}
\end{equation}

In the identification of the modes of the system, we follow the definition of nonlinear normal modes (NNM) as investigated in \cite{Vakakis2008} and \cite{Kerschen}. However, contrary to the damped case, in the conservative case the notion of NNM is more straightforward and in essence consists in a solution of a nonlinear ODE with zero external forcing and initial conditions that lead to complete energy localization in the considered mode.

First, appropriate notation is introduced:
\begin{equation}
\bar{x} \triangleq \frac{x_1+x_2}{2}, \ v \triangleq x_1-x_2, \ y \triangleq \bar{x}-x_0, \  \tilde{p} \triangleq \frac{p}{2k}
\label{eq6.3}
\end{equation}

With these definitions and Eq. (\ref{eq2.2}),  Eqs. (\ref{eq6.2}) can be used to derive the equation for the symmetric mode. Expressing $x_1$ and $x_2$ through $y,v$ and $x_0$ using Eq. (\ref{eq6.3}), one obtains a system of two differential equations and one algebraic equation. Using the algebraic equation and its first and second time derivatives and substituting them into the first two differential equations, and consequently replacing these equations by their sum and difference, two differential equations are obtained, in terms of $v$ and $y$ alone. Substituting $v=0$ in the equation obtained from the sum of the first two original equations in Eq. (\ref{eq6.2}), and setting the force term to zero, a nonlinear differential equation in $y$ alone is obtained, as follows:
\begin{equation}
\begin{split}
\ddot{y}+\left(\sqrt{\xi_0\omega}\right)^2y+(1-\xi_0)\tilde{p}(6y^2\ddot{y}+12y\dot{y}^2+\eta\omega^2y^3)=0
\label{eq6.4}
\end{split}
\end{equation}
(if $v=0$ is substituted in the equation arising when taking the difference of the two first equations in Eqs. (\ref{eq6.2}), a trivial $0=0$ equation is obtained).
Now, setting the force term to zero in the third, algebraic equation from Eqs. (\ref{eq6.2}) expressed in $y,v$ and $x_0$, an algebraic expression is obtained for $x_0$, which when using Definition (\ref{eq6.3}) again, becomes an equation for $\bar{x}$ in terms of $y$ (and $y$ alone, when substituting $v=0$), as follows:
\begin{equation}
\label{eq6.5}
\bar{x}= \xi_0^{-1}y+2(\xi_0^{-1}-1)\tilde{p}y^3+\frac{3}{2}(\xi_0^{-1}-1)\tilde{p}y v^2 \underset{v=0} \to \xi_0^{-1}y+2(\xi_0^{-1}-1)\tilde{p}y^3
\end{equation}

Also, one always has: $x_1=\bar{x}+v/2,x_2=\bar{x}-v/2,x_0=\bar{x}-y$.

Eq. (\ref{eq6.5}) shows that for $v=0$, $y$ is a nonlinear function of $\bar{x}$, through the solution of a cubic equation. Therefore $y$ is the same mode as $\bar{x}$, only in different scaling (the mode emerging for $v=0$, which corresponds to synchronous internal motion). Now, $\bar{x}$ is the displacement of the center of mass of the system, whether the symmetry is broken or not. In that sense, $\bar{x}$ is the symmetric mode. On the other hand, $v$ uniquely defines the motion in a purely antisymmetric regime, where the center of mass remains static. This is why $v$ represents the antisymmetric mode. Thus the logical line is this: $\bar{x}$ is the symmetric mode; for $v=0$ it is equivalent to $y$ (from Eq. (\ref{eq6.5})) -- hence for $v=0$ the symmetric mode is $y$. Since working with $y$, rather than with $\bar{x}$, makes the equations simpler, and apart from this the two are equivalent, the two NNMs of the system will be represented by the symmetric mode $y$ and the antisymmetric mode $v$.

Now, in the absence of forcing or damping, the general expression for the energy of the model system can be expressed as:
\begin{equation}
\label{eq7.1}
\begin{split}
E=E_0=\frac{1}{2}m\dot{x}_1^2+\frac{1}{2}m\dot{x}_2^2+\frac{1}{2}k_0x_0^2
+\frac{1}{2}k(x_1-x_0)^2+\frac{1}{4}p(x_1-x_0)^4+\frac{1}{2}k(x_2-x_0)^2+\frac{1}{4}p(x_2-x_0)^4
\end{split}
\end{equation}

Of obvious interest is the explicit expression for the symmetric mode, $y(t)$. According to the definition of nonlinear normal modes, all the energy in the system is stored in the degrees of freedom associated with the mode (when the system is excited in this mode). Therefore, assuming vibration in the purely symmetric mode and recalling the relation $x_0=\bar{x}-y$ (from Definition (\ref{eq6.3})), the aforementioned expression for the energy simplifies, and, when using Eq. (\ref{eq6.5}) and its first time derivative, the first integral of Eq. (\ref{eq6.4}) is obtained, with $E_0$ as the integration constant, as follows:
\begin{equation}
\label{eq7.3}
\begin{split}
\frac{E_0}{k}=(1+\hat{\eta}+6\hat{\eta}\tilde{p}y^2)^2\omega^{-2}\dot{y}^2+(1+\hat{\eta})y^2+(1+4\hat{\eta})\tilde{p}y^4+4\hat{\eta}\tilde{p}^2y^6
\end{split}
\end{equation}
(with $\hat{\eta}$ defined below).

Also, assuming a static initial condition, the energy can be expressed in terms of the initial amplitude, $Y_0$, as:
\begin{equation}
\label{eq7.5}
\begin{split}
\frac{E_0}{k}=(1+\hat{\eta})Y_0^2+(1+4\hat{\eta})\tilde{p}Y_0^4+4\hat{\eta}\tilde{p}^2Y_0^6
\end{split}
\end{equation}

Consequently, the second integral of motion is obtained in the form of inverse quadrature:
\begin{equation}
\label{eq7.14}
t=\frac{3T}{4}+\frac{1}{2\omega}\int\limits_{0}^{\hat{x}} \frac{\left[1+\hat{\eta}+6\hat{\eta}\hat{\epsilon}x\right]dx}{\sqrt{x(1-x)(Ax^2+Bx+C)}} \ \ , \ \ \forall \ \  3T/4\le t \le T
\end{equation}
where
\begin{equation}
\label{eq7.9}
\begin{split}
A \triangleq 4\hat{\eta}\hat{\epsilon}^2 , \
B \triangleq 4\hat{\eta}\hat{\epsilon}^2+(1+4\hat{\eta})\hat{\epsilon}, \
C \triangleq 4\hat{\eta}\hat{\epsilon}^2+(1+4\hat{\eta})\hat{\epsilon}+1+\hat{\eta}
\end{split}
\end{equation}
and the auxiliary quantities are given by:
\begin{equation}
\label{eq7.4}
\hat{\eta} \triangleq 2k/k_0 \ , \ \hat{\epsilon} \triangleq \tilde{p}Y_0^2 \ , \ \hat{x} \triangleq (y/Y_0)^2 \ , \ T \triangleq t(\hat{x}=1)
\end{equation}

Since the displacement is $T$-periodic, it is enough to describe its dynamics on one period, $T$, and since the Hamiltonian depends smoothly on the square of $\dot{y}$, it is enough to describe its dynamics on a half-period, and the other half can be obtained by smooth reflection. Finally, since the Hamiltonian depends smoothly also on the square of $y$ itself, only $quarter$ of a period is in fact sufficient, and the other quarter can be obtained, yet again, by reflection. Now, the integral in Eq. (\ref{eq7.14}) is expressible as a linear combination of the incomplete elliptic integral of the first kind, $\mathcal{F}(z|m)$, and the incomplete elliptic integral of the third kind, ${\Pi}(n;z|m)$, $z$ being some nonlinear function of $\hat{x}$. Thus, formally Eq. (\ref{eq7.14}) yields an exact implicit solution for the symmetric mode. However, this form of solution is somewhat inconvenient. For instance, for stability analysis by the common approach of Hill's Infinite Determinants method one needs the Fourier expansion of the solution. The fact that the relation $t(\hat{x})$ is a combination of two incomplete elliptic integrals, of the first and third kind, makes it implausible to invert it into $\hat{x}(t)$, which is required in order to obtain the aforementioned Fourier series.

Formally, this obstacle can be easily circumvented. Even a straightforward approach allows the formal derivation of the Fourier series coefficients of $\hat{x}(t)$ without actually having $\hat{x}(t)$ in explicit form. First, it is clear that one can always use a reflection of $t(\hat{x})$ as given in Eq. (\ref{eq7.14}), around $t=T$, to construct an even function for which a cosine series would be sufficient. Next, the coefficients can, in principle, be obtained by changing the integration variable from $t$ to $\hat{x}$ using the relation $dt=t'(\hat{x})d\hat{x}$ and adjusting the integration bounds accordingly, to give:
\begin{equation}
\label{eq7.16}
\begin{split}
c_k =\frac{2}{T/2}\int\limits_{3T/4}^{5T/4}\hat{x}\cos{[2k\pi t/(T/2)]}dt
=\frac{4}{\omega T}\int\limits_{0}^{1}\frac{\cos{\left[k\frac{4\pi}{T}t(\hat{x})\right]}}{\sqrt{1-\hat{x}}}\Psi(\hat{x})\sqrt{\hat{x}}d\hat{x}
\end{split}
\end{equation}
where
\begin{equation}
\label{eq7.17}
\Psi(\hat{x}) \triangleq \frac{1+\hat{\eta}+6\hat{\eta}\hat{\epsilon}\hat{x}}{\sqrt{A\hat{x}^2+B\hat{x}+C}}
\end{equation}

The problem with this formally sufficient formula is that $t(\hat{x})$ has the form of a combination of $\mathcal{F}(z|m)$ and $\Pi(n;z|m)$, which depend on $\hat{x}$ through a sequence of additional nonlinear transformations. As a result, this solution can be handled only numerically. However, the ratio in the integrand in Eq. (\ref{eq7.16}) is problematic for numerical integration due to the combined effect of the (formally integrable) square root singularity near 1 and the cosine that becomes highly oscillatory for large values of $k$. This leads to the lack of convergence of the resulting Fourier series expansion for increasing resolution of the numerical integration grid. The problem therefore requires different solution (which is presented below, in Section \ref{sect8}).

\subsection{Partial explicit integration for the general case}
\label{sect7B}

As aforementioned, the expressions for the time as a function of the squared displacement, as given in Eq. (\ref{eq7.14}), can be represented by analytic, well-studied functions of the variables/parameters, namely, non-invertible combinations of elliptic integrals of the first and third kinds. What is in fact required for stability analysis, at least in its Hill method implementation, is Fourier series expansion, which cannot be calculated directly, for reasons discussed above.

The following section overcomes this problem by employing a specific, numerical-iterative procedure enabling the calculation of Fourier series coefficients of the squared displacement to any degree of accuracy. The algorithm, which will be discussed in the next section, requires only three initial "input" quantities in order to work. These three quantities are directly related to the second integral of motion. These quantities can be derived explicitly for arbitrary values of the parameters.

The quantities in question are the quarter-period, $T/4$, the quarter-period-averaged squared amplitude, $c_0$, and the quarter-period-averaged squared squared-amplitude, which will be denoted by $\psi$. These three quantities, unlike high-frequency Fourier series coefficients, in fact $can$ be calculated directly and later employed in an implicit algorithm constructed for the calculation of higher-frequency Fourier series coefficients.

The first of these three is the quarter-period, $T/4$, which is given by Eqs. (\ref{eq7.14})-(\ref{eq7.4}), and which in principle can be expressed in closed form through complete elliptic functions. However, since the algorithm is numeric and implicit, only the $numeric$ value of $T/4$ would be required, and this can be obtained either by direct evaluation of closed-form elliptic functions or by numerical integration. The latter, in the case of the quarter-period, can be obtained directly and explicitly to a sufficient degree of accuracy (since integrand singularity is non enhanced by oscillatory nature here). Owing to this and to the cumbersome form of the analytic expression in question, the result of the analytic integration of Eq. (\ref{eq7.14}) is not presented here.

The formal expressions for the quarter-period-averaged values of $\hat{x}$ and $\hat{x}^2$ are as follows:
\begin{equation}
\label{eq7B.1}
c_0=\frac{\int\limits_{0}^{1}\frac{x[1+\hat{\eta}+6\hat{\eta}\hat{\epsilon}x]dx}{\sqrt{x(1-x)(Ax^2+Bx+C)}}}{\int\limits_{0}^{1}\frac{[1+\hat{\eta}+6\hat{\eta}\hat{\epsilon}x]dx}{\sqrt{x(1-x)(Ax^2+Bx+C)}}}, \ \psi=\frac{\int\limits_{0}^{1}\frac{x^2[1+\hat{\eta}+6\hat{\eta}\hat{\epsilon}x]dx}{\sqrt{x(1-x)(Ax^2+Bx+C)}}}{\int\limits_{0}^{1}\frac{[1+\hat{\eta}+6\hat{\eta}\hat{\epsilon}x]dx}{\sqrt{x(1-x)(Ax^2+Bx+C)}}}
\end{equation}
where Definisions (\ref{eq7.9}) and (\ref{eq7.4}) are employed.

As in the case with $T/4$, the integrals  given above can be expressed analytically using rather cumbersome formulas, but they will be omitted here, since only the numeric values are required for the aforementioned algorithm. For special values of the parameters, explicit analytic results regarding integration, and in one example even stability analysis, can be obtained. Several such results are summarized in  \ref{AppendixB}.

\section{Implicit computation of Fourier expansion coefficients using Chebyshev polynomials}
\label{sect8}

Before stability analysis of the symmetric mode can be performed, the aforementioned problem of the computation of Fourier expansion coefficients has to be solved. The idea would be to exploit the polynomial form that appears in the first integral of motion and to apply the basis of Chebyshev polynomials. Consequently, Fourier expansion coefficients can be computed using the approach of sequential approximations, employing a fixed-point iteration scheme.

The following definitions of parameter/variable transformation are used:
\begin{equation}
\label{eq8.1}
\hat{T} \triangleq \frac{\omega T}{2\pi}, \ q \triangleq \cos{(4\pi t^{*}/T)}, \ t^{*} \triangleq t-\frac{T}{4}
\end{equation}
where a shift by a quarter period in time is made to obtain symmetric cosine Fourier series in order to compute fewer coefficients. Eqs. (\ref{eq8.1}) yield:
\begin{equation}
\label{eq8.2}
\hat{T}^2[\hat{x}'(t)]^2=4\omega^2(1-q^2)[\hat{x}'(q)]^2
\end{equation}

Expressing $[\hat{x}'(t)]^2$ from Eq. (\ref{eq7.14}), substituting it into Eq. (\ref{eq8.2}), and taking a (negative) square root, one obtains the following form of the quadrature:
\begin{equation}
\label{eq8.3}
\hat{x}'(q)=-\hat{T}\frac{\sqrt{\hat{x}(1-\hat{x})(A\hat{x}^2+B\hat{x}+C)}}{\sqrt{1-q^2}[1+\hat{\eta}(1+6\hat{\epsilon}\hat{x})]}
\end{equation}
(the negative root was taken since during the second quarter-period, where the interpolation is to be made, $\hat{x}(t)$ increases while $q(t)$ decreases -- recall Eq. (\ref{eq7.14})).

The function $\hat{x}(t)$ is even in the interval $t \in [T/4,T/2]$ and can be written in a form of Fourier cosine series expansion, namely:
\begin{equation}
\label{eq8.4}
\hat{x}(t)=c_0+\sum\limits_{j=1}^{\infty}c_j\cos(4j\pi t^*/T)
\end{equation}
where the equality sign is used due to the smoothness of $\hat{x}(t)$ in the considered interval, $c_0$ is the period-averaged value of $\hat{x}(t)$, as given by Eq. (\ref{eq7B.1}) (and thus considered known), and the coefficients $c_{j \geq 1}$ are the sought unknowns.

At this point it is beneficial to apply Chebyshev polynomials, which satisfy the following (defining) relation: $T_n(z)=\cos{(n \arccos{z})}$ (which is true for any $|z|<1$, satisfied in the case in question due to Definition (\ref{eq8.1})), and explicitly to the second of Definitions (\ref{eq8.1}), to obtain the following representation of the squared normalized symmetric mode displacement and the corresponding required derivative:
\begin{equation}
\label{eq8.5}
\hat{x}(q)=c_0+\sum\limits_{j=1}^{\infty}c_j T_j(q) \ , \ \hat{x}'(q)=\sum\limits_{j=1}^{\infty}c_j T'_j(q)
\end{equation}

Substituting Eq. (\ref{eq8.5}) into Eq. (\ref{eq8.3}), one obtains the basic relation in the algorithm, containing infinitely many Fourier series coefficients, and  theoretically holding at infinitely many points $q$ at the interval $q\in (-1,1)$.

Since $\hat{x}(t)$ is smooth, high-frequency Fourier coefficients should decrease exponentially. Therefore the cosine series representation of $\hat{x}(t)$ can be truncated, the number of necessary terms being some finite, possibly large number, $N_c$. If indeed the truncation is justified (that is, $N_c$ is high enough), then the truncated-series version of Eq. (\ref{eq8.3}) with nested expansions should still be valid for every $q$ in the relevant interval (which happens, for instance, for the special case discussed in \ref{sect7A.2}). However, for $N_c$ not high enough, the aforementioned truncated-series equation would only hold at a finite number of points in the interval, namely, $N_q$. For $N_q>N_c$, there may be no solution to the truncated-series equation. For $N_q<N_c$, there may be several solutions. Therefore, a unique solution for the Fourier cosine series coefficients, $\bf{c}$, is possible only for $N_q=N_c$, where for the best solution for a given $N_c$, the correct choice of the $N_q=N_c$ points in the relevant interval has to be made.

Therefore, a truncated, discretized version of the compound of Eqs. (\ref{eq8.3}) and (\ref{eq8.5}) is called for, in the form of a system of $N_c$ nonlinear equations (one equation for every value of $i$ and one unknown for every value of $j$, both attaining integer values from 1 to $N_c$), as follows:
\begin{equation}
\label{eq8.7}
\begin{split}
\forall \ \ i=1,..,N_c: \ \ \sum\limits_{j=1}^{N_c}c_j T'_j(q_i)= -\frac{\hat{T}\sqrt{\sum\limits_{j=1}^{N_c}c_j T_j(q_i)\left[1-\sum\limits_{j=1}^{N_c}c_j T_j(q_i)\right]}}{\sqrt{1-q_i^2}} \frac{\sqrt{A\left[\sum\limits_{j=1}^{N_c}c_j T_j(q_i)\right]^2+B\sum\limits_{j=1}^{N_c}c_j T_j(q_i)+C}}{1+\hat{\eta}+6\hat{\eta}\hat{\epsilon}c_0+6\hat{\eta}\hat{\epsilon}\sum\limits_{j=1}^{N_c}c_j T_j(q_i)}
\end{split}
\end{equation}

This is a system of $N_c\times N_c$ coupled nonlinear equations for the first $N_c$ sufficient components of $\bf{c}$, specified by the parameters $\hat{\eta}$ and $\hat{\epsilon}$ and the auxiliary parameters $A,B,C,\hat{T}$ and $c_0$, defined in Eqs. (\ref{eq7.9}), (\ref{eq7.4}), (\ref{eq7B.1}) and (\ref{eq8.1}). Full characterization of the equations requires the specification of the $1\times N_c$-dimensional vector of points $\bf{q}$ and, of course, explicit expressions for $T_j(q_i)$. Finally, $N_c$ has to be determined according to a suitable convergence criterion.

The points $\bf{q}$ of the application of Eqs. (\ref{eq8.7}) are, in essence, the interpolation points for Chebyshev polynomial decomposition. The optimal choice of these points is traditionally made according to Gauss quadrature theory, aiming to minimize the effect of the Runge-Gibbs phenomenon \cite{Stewart1996}. This leads to the so-called Chebyshev nodes (one notes that the idea behind the Chebyshev-Gauss quadrature is related to the same square-root singularity that appears in Eq. (\ref {eq7.16})).

As the aforementioned choice is indeed found optimal in the considered case, it appears plausible to apply the fundamental Eq. (\ref{eq8.7}) at equidistant times. Points $\bf{q}$ in Eq. (\ref{eq8.7}) are thus chosen at Chebyshev nodes, according to a formula given below. The next issue is the Chebyshev Polynomials, $T_n(z)$, themselves. For reasons to be explained below, the Chebyshev polynomials are calculated using a recurrence relation, as follows:

\begin{equation}
\label{eq8.9}
\begin{split}
T_0(z)=1 \ , \ T_1(z)=z \\ T_{n+1}(z)=2z T_n(z)-T_{n-1}(z)
\end{split}
\end{equation}
(the way in which the recurrence relation is used to derive explicit expressions is explained in the following subsection).

The last issue required for full specification of Eqs. (\ref{eq8.7}) concerns the value of $N_c$ guaranteeing convergence. To check the convergence, the following condition is used:
\begin{equation}
\label{eq8.10}
\sqrt{\left |1-\frac{1}{2}\frac{\textbf{c}^{\top}\textbf{c}}{\psi-c_0^2}\right |} < \delta \ll 1
\end{equation}
where $\bf{c}$ is the vector of Fourier cosine coefficients as arises from the solution of Eqs. (\ref{eq8.7}) for a given value of $N_c$, starting from $c_1$ (that is, excluding $c_0$), $\delta$ is a small number ($10^{-3}$ is taken here), and $c_0$ and $\psi$ are given in Definitions (\ref{eq7B.1}) (the second of which was introduced precisely for being used in a convergence criterion). The idea behind the determination of $N_c$ is thus simply to increase it, starting  from 1, until the inequality in Eq. (\ref{eq8.10}) is first met.

By this the full specification of Eqs. (\ref{eq8.7}), which at this point can be readily solved to produce the desired Fourier series decomposition of $\hat{x}(t)$, as required by Hill's Infinite Determinants method for linear stability analysis, is concluded. The following subsection is dedicated to the strategy that is employed in order to solve Eqs. (\ref{eq8.7}).

\subsection{Solution of Eqs. (\ref{eq8.7})}
\label{sect8A}

Normally, solution of a system of nonlinear equations is a tedious task, but in the examined case it is possible to find a good starting point. Instead of solving Eqs. (\ref{eq8.7}) directly for sufficiently large $N_c$, it is advantageous to start by solving the problem for $n=1$, then to use the obtained  solution as a starting point for the solution of the equations with $n=2$, etc. (until, at $n=N_c$, Condition (\ref{eq8.10}) is fulfilled). More formally, assuming  the best vector of $n-1$ Fourier coefficients is known, one finds the vector of the best $n$ Fourier coefficients.

Now, before giving the formulas for the general algorithm, a simple illustration can be given. For $n=1$, Eqs. (\ref{eq8.7}) can be solved exactly, since as aforementioned, the points $q_i$ are the roots of $T_n(q_i)$. Then, $T_1$ and thus also $c_1$ vanish from the right hand-side, and $c_1$ remains only on the left-hand side, where it is multiplied by $T_1'=1$. Consequently solving for $c_1$ becomes trivial, and one obtains what results in the exact solution for the special $\hat{\epsilon}=\hat{\epsilon}^*(\hat{\eta})$ and $\hat{\epsilon} \ll 1$ cases given in \ref{AppendixB}, but only a first approximation in the general case, and is given by:

\begin{equation}
\label{eq8A.1}
c_1^{(1)}=-\hat{T}\frac{\sqrt{c_0(1-c_0)(Ac_0^2+B c_0+C)}}{1+\hat{\eta}+6\hat{\eta}\hat{\epsilon}c_0}
\end{equation}

At this point Condition (\ref{eq8.10}) is examined, and if it is satisfied, then the problem is solved. If the condition is not satisfied, then another step is necessary, this time with $N_c=2$. This point is a bit less trivial. In order to solve Eqs. (\ref{eq8.7}), for $N_c$, one would need a starting point. In principle, for $N_c=2$, $\bf{c}^{(2)}$ would be a two-component vector, which would serve as an initial guess for next-step subproblems (in case Condition (\ref{eq8.10}) is still not satisfied). In order to obtain the second-step two-component solution $\bf{c}^{(2)}$, one would need a two-component initial guess for it. However, at this point only the one-component initial guess $c_1^{(1)}$ from the previous step is at hand. This indeed may be a problem for the sequential approximations strategy in the general case, however in the examined case, this obstacle may be circumvented.

The remedy comes from the same idea that enabled the solution for $c_1^{(1)}$ explicitly, and is embodied in the fact that the chosen nodes were Chebyshev nodes. Again, as in the first step, since the Chebyshev nodes are the roots of the highest-order Chebyshev polynomial ($T_n$ in the general case and $T_2$ in the example), it turns out that $c_n$ vanishes from the right-hand side of Eqs. (\ref{eq8.7}), and remains only at the left-hand side, where it is multiplied by the non-vanishing derivative $T_n'$ rather than by the vanishing function $T_n$. Owing to this, there remains only one occurrence of $c_n$ in the equations, and moreover a linear one, and thus $c_n$ ($c_2^{(2)}$ in the example) can be solved for explicitly, in terms of the unknown coefficients up to $c_{n-1}$ (or $c_1^{(1)}$ in the example). In order to isolate $c_n^{(n)}$ (that is the $n$-th component at the $n$-th step), one can use the equation corresponding to $q_n$. Then the obtained expression for $c_n$ is inserted in the other $n-1$ equations, and consequently an implicit nonlinear system of $n-1$ equations (at points $q_1,q_2,...,q_{n-1}$) for $n-1$ variables, namely, $c_1^{(n-1)},c_2^{(n-1)},...,c_{n-1}^{(n-1)}$, is obtained.

This procedure provides the initial guess with the correct dimension for the smaller, rank-$(n-1)$ system: it is the solution from the previous step in the sequence of approximations. Moreover, it is not only that one has a feasible starting point, but this starting point is also very probably a good one. The reason for this is that for smooth enough functions, such as $\hat{x}(t)$ (which has no strong singularities, as can be learned from Eq. (\ref{eq7.14})) Fourier coefficients converge fast enough. Due to the smoothness, from some, perhaps higher, $j$ onward, there is order separation between the coefficients, such that an optimal projection with additional coefficients should not strongly affect the values of the coefficients obtained for an optimal projection with a smaller number of coefficients. Numerical investigations confirm that this indeed holds in the considered case.

To conclude this subsection, the complete set of formulas for the proposed sequential approximations algorithm is presented.

First, one defines:
\begin{equation}
\label{eq8A.2}
\begin{split}
A_{ij}^{(n)}\triangleq T_j\left(q_i^{(n)}\right), \ w_i^{(n)} \triangleq T_i(q_n), \ z_i \triangleq T'_i(q_n), \\
M_{ij}^{(n)} \triangleq z_j T'_n\left(q_i^{(n)}\right)-T'_n(q_n)T'_j\left(q_i^{(n)}\right), \\
d_{(n)}^{(m)} \triangleq c_0+\textbf{w}_{(n)}^\top\textbf{c}_{(n)}^{(m)}, \  \textbf{v}_{(n)}^{(m)} \triangleq c_0+\textbf{A}^{(n)}\textbf{c}_{(n)}^{(m)}
\end{split}
\end{equation}
where $(n)$ denotes the number of the current step in the set of sequential approximations (either as a subscript or as a superscript), and the superscript $(m)$ denotes the iteration number in the inner loop of iterations of the fixed-point scheme (presented below).

As aforementioned, the grid points are chosen from Chebyshev nodes, according to the following formula:
\begin{equation}
\label{eq8A.3}
q_i^{(n)} \triangleq \cos{\left(\frac{2i-1}{2n}\pi\right)}
\end{equation}

Next, the inhomogeneous, nonlinear part of the resulting rank-$(n-1)$ system of equations is defined as:
\begin{equation}
\label{eq8A.4}
\begin{split}
g_i^{(n,m)} \triangleq T'_n  \left(q_i^{(n)}\right) (-\hat{T})\sqrt{d_{(n)}^{(m)}\left(1-d_{(n)}^{(m)}\right)}  \frac{\sqrt{A\left(d_{(n)}^{(m)}\right)^2+B d_{(n)}^{(m)}+C}}{\sqrt{1-q_n^2}\left[1+\hat{\eta}\left(1+6\hat{\epsilon}d_{(n)}^{(m)}\right)\right]} \\
-T'_n(q_n) (-\hat{T})\sqrt{v_i^{(n,m)}\left(1-v_i^{(n,m)}\right)} \frac{\sqrt{A\left(v_i^{(n,m)}\right)^2+B v_i^{(n,m)}+C}}{\sqrt{1-\left(q_i^{(n)}\right)^2}\left[1+\hat{\eta}\left(1+6\hat{\epsilon}v_i^{(n,m)}\right)\right]}
\end{split}
\end{equation}
where  the "running" number of the sequential approximation, $(n)$, and the inner loop fixed-point iteration number, $(m)$, are both given in superscripts, as a couplet, $(n,m)$. For Eqs. (\ref{eq8A.2})-(\ref{eq8A.4}), the indices $i j$ must comply to: $i,j=1,...,n-1$.

The aforementioned fixed-point iteration scheme, used for the optimal Fourier-space projection at a given step of the sequence of approximations, is given by:
\begin{equation}
\label{eq8A.6}
\textbf{c}^{(m+1)}_{(n)}=\textbf{M}_{(n)}^{-1}\textbf{g}_{(n)}^{(m)}
\end{equation}

The idea behind this scheme is as follows. After "static condensation" from $n$ equations to $n-1$ equations, there remains a linear part, arising from $\hat{x}(t)$, and a nonlinear part, arising from the integral from Eq. (\ref{eq7.14}). There is but a single way to extract the coefficients $\bf{c}$ from the linear part. This is done by matrix inversion, as shown in Eq. (\ref{eq8A.6}). In contrary, there are five ways to extract the coefficients from the nonlinear part. Thus the symmetric extraction would be from the linear part. Moreover, the integration weakens the effect of the variation in coefficient values, thus it is called for that the implicit part be on the side of the integral. These are intuitive arguments, which explain the choice of the scheme. In the following subsection it is shown by numerical computation that the scheme indeed appears to be a stable and convergent one. To further strengthen the argument, asymptotic convergence analysis is given in \ref{AppendixC}.

Now, assuming that the scheme is convergent, it is applied iteratively with the following convergence criterion:
\begin{equation}
\label{eq8A.7}
\begin{split}
m^{*} = \min{m} \ \ \ \text{s.t.} \ \ \   \left|1-\frac{\left\|\textbf{c}_{(n)}^{(m+1)}\right\|}{\left\|\textbf{c}_{(n)}^{(m)}\right\|}\right|<\delta_m \ll 1
\end{split}
\end{equation}

The value $\delta_m=10^{-10}$ is used here and the scheme is iterated until $m=m^*$. Consequently, the $(n-1)$-component vector $\textbf{c}_{(n)}^{(m^*)}$ is used in order to obtain the corresponding, $n$-th component, according to the following explicit formula:
\begin{equation}
\label{eq8A.8}
\begin{split}
c_n^{(n)}= \frac{(-\hat{T})}{T'_n  (q_n)}  \sqrt{d_{(n)}^{(m^*)}\left(1-d_{(n)}^{(m^*)}\right)}  \frac{\sqrt{A\left(d_{(n)}^{(m^*)}\right)^2+B d_{(n)}^{(m^*)}+C}}{\sqrt{1-q_n^2}\left[1+\hat{\eta}\left(1+6\hat{\epsilon}d_{(n)}^{(m^*)}\right)\right]} -\frac{\textbf{z}_{(n)}^\top\textbf{c}_{(n)}^{(m^*)}}{T'_n  (q_n)}
\end{split}
\end{equation}

Finally, the next step's coefficient vector is constructed as:
\begin{equation}
\label{eq8A.9}
\textbf{c}_{(n+1)}^{(m=0)}=\left(\textbf{c}_{(n,m^*)}^\top,c_n^{(n)}\right)^\top
\end{equation}
where $\textbf{c}_{(n,m^*)}=\textbf{c}_{(n)}^{(m^*)}$.

The left-hand side in Eq. (\ref{eq8A.9}) is the $n$-component initial guess for the fixed-point iterative scheme of the next, $n$-th step in the sequence of approximations. The initial guess in the first step, that is $\textbf{c}_{(2)}^{(m=0)}$, is just $c_1^{(1)}$, as given by Eq. (\ref{eq8A.1}).

\subsection{Numerical illustration of the quality of the Fourier expansion for the symmetric mode obtained by the implicit algorithm}
\label{sect8C}

\begin{figure}
\begin{center}
{\includegraphics[scale=0.6]{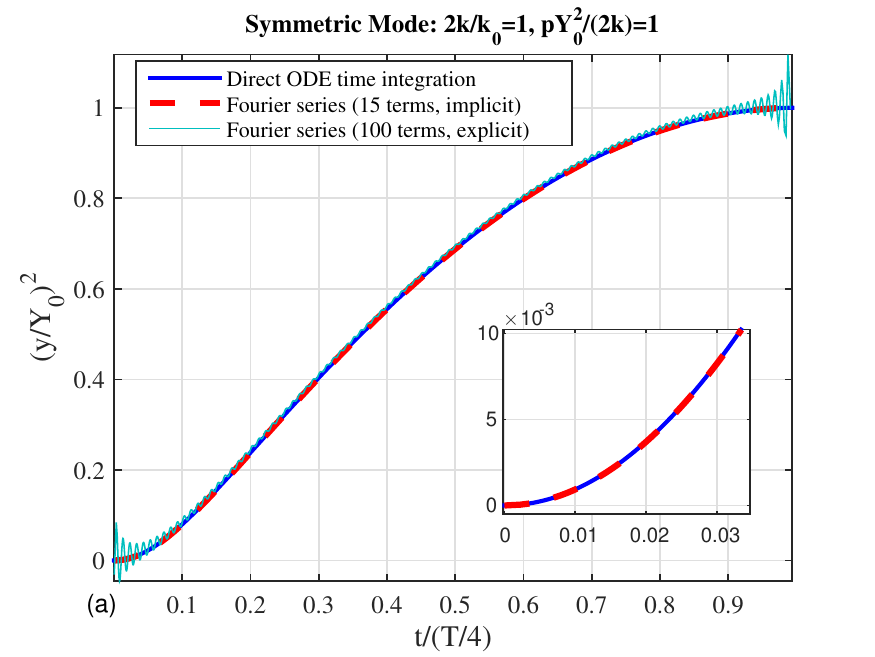}}
{\includegraphics[scale=0.6]{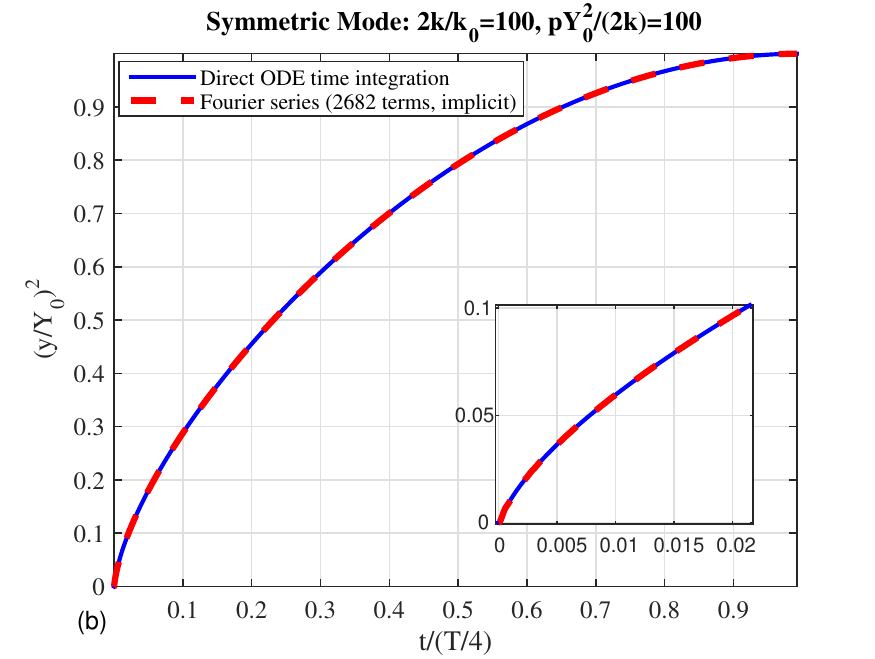}}
\end{center}
\caption{\small Illustration of the quality of Fourier expansion obtained by the implicit algorithm. Comparison of direct numerical time-integration of the ODE (solid, blue online), Fourier expansion by the implicit algorithm (dashed, red online), and Fourier expansion by explicit 'na\"ive' integration (solid, dark cyan online, only in (a)), for (a) $\hat{\eta}=1,\hat{\epsilon}=1$ and (b) $\hat{\eta}=100, \hat{\epsilon}=100$. Close-ups on the initial increase region (hardest to capture by cosine series) are given in insets.}
\label{Figure3}
\end{figure}

The results of the implementation of the algorithm discussed in this section are given in Fig. \ref{Figure3}  for typical values of the parameters. One observes that the application of the Chebyshev-polynomials-based iterative algorithm produces smooth lines evidently coinciding with the curves obtained by direct numerical integration of the equations of motion. These results should be viewed in the perspective of the obvious lack of convergence manifested in the oscillatory non-smooth lines produced by the na\"ive approach to the computation of Fourier coefficients -- the straightforward integration of Eq. (\ref{eq7.16}), shown in the (dark cyan online) solid oscillatory line in Fig. \ref{Figure3}(a).

\section{Linear stability analysis of the symmetric mode by Hill's Infinite Determinants method}
\label{sect9}

Now that the solution for the symmetric mode of the considered system was obtained, it is instructive to explore its stability. The loss of stability of the symmetric mode reveals the "hidden" internal structure of the system. Such mode revealing instability only arises due to nonlinearity.  In a perfectly symmetric linear system it does not take place.

Since the solution for the symmetric mode embodies the assumption $v=0$, it is evident that the obtained solution becomes incorrect whenever the condition $v=0$ is violated. Thus the stability of the symmetric mode, $y$, in the form which was obtained assuming $v=0$, is conditioned by the stability of the (assumed) solution $v=0$ for the antisymmetric mode.

As mentioned in Section \ref{sect7}, one equation from Eqs. (\ref{eq6.2}) is yet unused. This is the equation that one can obtain by taking the difference of the first two equations in Eqs. (\ref{eq6.2}), expressed in terms of $y,v$ and $x_0$, after eliminating $x_0$ using the third, algebraic equation in Eqs. (\ref{eq6.2}), and its first and second time derivatives, expressed in terms of the three aforementioned variables. The emerging equation is as follows:
\begin{equation}
\ddot{v}+\omega^2 (1+6 \tilde{p} y^2) v+\frac{1}{2}\tilde{p} \ \omega^2 v^3=0
\label{eq6B.1}
\end{equation}

One immediately observes that $v=0$ is a valid solution of Eq. (\ref{eq6B.1}), for every bounded $y(t)$. The idea now is to assume a perturbation $\delta{v}(t)$ to $v(t)=0$ and study its possible evolution.

In principle, Eq. (\ref{eq6B.1}) is already the evolution equation for a perturbation to $v(t)=0$. A growing solution of this equation corresponds to instability of the symmetric mode. Taking the first variation of Eq. (\ref{eq6B.1}), denoting the variation in $v$ by: $z \triangleq \delta{v}(t)$, and keeping only linear terms, one obtains the equation of linear stability for the symmetric mode, as follows:
\begin{equation}
\label{eq6B.2}
\ddot{z}(t)+\omega^2[1+6\tilde{p}x(t)]z(t)=0 \ \ ; \ \  x(t) \triangleq \left[\left. y(t,\bar{\epsilon} \ll 1) \right |_{v=0}\right]^2
\end{equation}

Now, since $x(t)$ is a square of a generally periodic function, as evident from Eq. (\ref{eq6.4}) and (\ref{eq7.3}), $x(t)$ is periodic and thus Eq. (\ref{eq6B.2}) is a Hill equation for $z(t)$. Therefore, the problem of the linear stability of the symmetric mode can be investigated by Floquet analysis of Eq. (\ref{eq6B.2}).

Using dimensionless quantities, the linear stability equation for the symmetric mode can be rewritten in the following form:
\begin{equation}
\label{eq9.1}
z''(\lambda)+\Lambda^2[1+6\hat{\epsilon}\hat{x}(\lambda)]z(\lambda)=0 \ \ ; \ \ \Lambda \triangleq \hat{T} \ , \ \lambda \triangleq \omega t^*/\hat{T}
\end{equation}
where $t^*$ is as in Definition (\ref{eq8.1}).

Next, now that the task of Fourier expansion is complete, the squared normalized symmetric-mode displacement, $\hat{x}$, can be cast in the following form (standard for a generalization of the canonic form employed for the Mathieu equation):
\begin{equation}
\label{eq9.3}
\hat{x}(\lambda)=c_0+\sum\limits_{k=1}^{N_c}c_k\cos{(2k\lambda)}
\end{equation}
where the coefficients $c_k$ for $k$  from 1 to $N_c$ are obtained by the method presented in the previous section, and $c_0$ is calculated according to Eq. (\ref{eq7B.1}).

From this point on, stability analysis by Hill's Infinite Determinants method is performed rather standardly and therefore its details are omitted here. However, there are several subtle points that need to be and are addressed below, for which the technical steps of the analysis are required. Those are therefore given in \ref{AppendixD1}.

Several remarks regarding the technical aspects of stability analysis by Hill's method (for the obtained Hill equation with the Fourier expansion coefficients calculated implicitly) should be made at this point. First, the emerging stability-limit (zero determinant) equation (given in \ref{AppendixD1}) appears to have two $\hat{\eta}$ solutions for every finite amplitude, corresponding to the left and right boundaries of the instability tongues. Each of the solutions is found by using a left-or-right perturbed initial $\hat{\eta}$ guess. Second, the emerging stability-limit equation is very complicated even in the Mathieu approximation, let alone in the full Hill case. Therefore it can only be solved numerically, and hence the infinite determinants have to be truncated, corresponding to vanishing infinite frequencies in the periodic solutions for a variation of the symmetric mode from zero. In the case of the largest amplitude examined ($\hat{\epsilon}=100$), several thousands of Fourier coefficients were required for $\hat{x}(t)$ for convergence, and about a hundred $more$ coefficients for convergence in the stability-limit curves. Third, for the large amplitude cases, where about million-entries matrices emerge, the values of the determinants not in the direct vicinity of the root were too large for standard computation tools. Therefore, the zero determinant criterion was substituted by search for the root of the minimum-in-absolute-value eigenvalue of the resulting matrix, rather than the determinant. However, since eigenvalues can be calculated numerically only approximately, unlike determinants, a direct determinant sign-switching condition was employed for verification close enough to the root. Fourth, in order to solve the required nonlinear equation, the Secant Method algorithm (a one-dimensional quasi-Newton method using a finite difference approximation of the derivative) was employed. The Secant Method requires two initial guesses, which can be easily generated by perturbing the known zero-amplitude limit. A convergence criterion of relative root (solution) change norm of $10^{-12}$ was applied.

\begin{figure}[H]
{\includegraphics[scale=0.625]{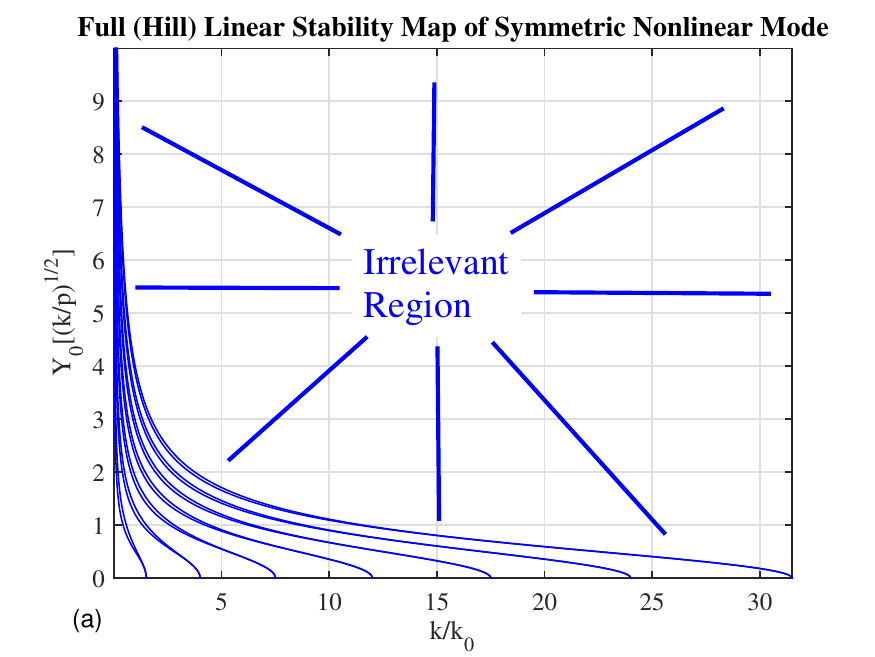}}
{\includegraphics[scale=0.625]{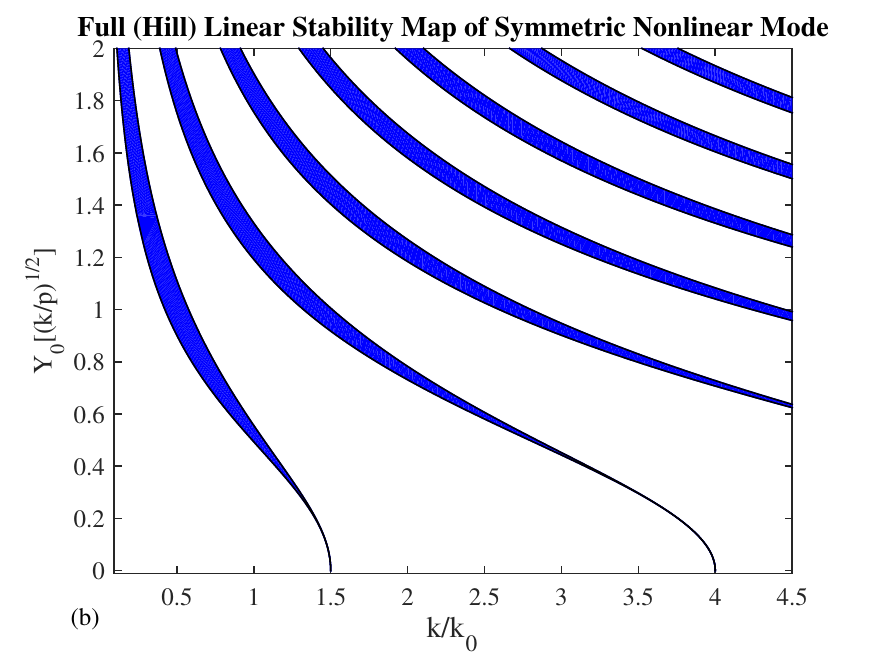}}
\caption{\small Stability map for the symmetric mode for the entire computed range (a) and a representative fully mapped rectangle (b)  -- unstable zones in dark (blue online).}
\label{Figure4}
\end{figure}

The emerging instability tongues are presented in Fig. \ref{Figure4}. The only way to solve the nonlinear equations was to start at the zero amplitude axis. Also, one would want to reach large amplitudes (order of magnitude above the nonlinear elasticity length scale). Due to the special parametrization in the considered case, the tongues "curved-up" strongly, climbing on the $k/k_0\to 0$ axis, leaving a void in a rectangular-domain plot. Hence a close-up on a fully mapped rectangular sub-domain was added in Fig. \ref{Figure4}(b). Fig. \ref{Figure4} shows that with the full Hill calculation, $finitely$ nonlinear symmetry breaking may be considered established. The zero-amplitude limit bifurcation points used as starting points for numerical stability analysis were derived asymptotically. The derivation, along with discussion of the weak-nonlinearity-limit symmetric mode stability, is given in \ref{AppendixD2}.

\section{Dynamics and stability analysis of the antisymmetric mode }
\label{sect10}
\subsection{Integration}
\label{sect10A}

After establishing the existence of a parameter range in which there is nonlinear symmetry breaking of the linearly symmetric mode by instability, the next logical step is to examine the existence and stability of the antisymmetric mode. According to the notation introduced in Section \ref{sect7}, the antisymmetric mode corresponds to zero motion of the center of mass of the system, which leads, following the definitions in Eq. (\ref{eq6.3}), to the following relations:
\begin{equation}
\label{eq10A.1}
\begin{split}
\bar{x}(t) \equiv 0 \underset{Eq. (\ref{eq6.5})}\Rightarrow y(t) \equiv 0 \underset{Eq. (\ref{eq6.3})}\Rightarrow  x_0(t) \equiv 0 \underset{Eq. (\ref{eq6.3})}\Rightarrow x_{1,2} = \pm v/2
\end{split}
\end{equation}

Substituting these relations into the expression for the total energy of the system given in Eq. (\ref{eq7.1}), and using Eqs. (\ref{eq2.2}), (\ref{eq6.3}) and (\ref{eq7.5})-(\ref{eq7.4}), one obtains the first integral of the purely antisymmetric  mode of the system, in the following form:
\begin{equation}
\label{eq10A.2}
[\hat{v}']^2+\hat{v}^2+\frac{1}{4}\hat{\epsilon}\hat{v}^4=4C; \ \hat{v} \triangleq v/Y_0 \ , \ \hat{v}' \triangleq d\hat{v}/d\tau \ , \ \tau \triangleq \omega t
\end{equation}
where $Y_0(E_0,k_0,k,p)$ is perceived merely as convenient parametrization of the total energy in the system.

Computation of the integral results in a closed form expression for the normalized antisymmetric mode displacement as a function of the dimensionless time, using the fifth Jacobi elliptic function 'sd', as follows:
\begin{equation}
\label{eq10A.4}
\hat{v}(\tau)=\frac{\sqrt{4C}\text{sd}\left[(1+4\hat{\epsilon}C)^{1/4}\tau\left| \frac{\sqrt{1+4\hat{\epsilon}C}-1}{2\sqrt{1+4\hat{\epsilon}C}} \right.\right]}{(1+4\hat{\epsilon}C)^{1/4}}
\end{equation}

One notes that for small amplitudes, the simple form $\hat{v}(t)=\sin(\omega t)$ is reproduced, coinciding with the linear undamped infinitesimally asymmetric case second-mode solution (see Section \ref{sect3}).

Eq. (\ref{eq10A.4}) provides the explicit solution for the antisymmetic mode. In order to study its stability, Floquet-Hill strategy is followed, and Fourier-series representation of the solution is called for. The sine series representation of sd$(z|m)$ is presented in \cite{AbramowitzStegun1964}. In the employed parametrization it  produces the following series representation for the antisymmetric mode:
\begin{equation}
\label{eq10A.5}
\hat{v}(\hat{\tau})=\sum\limits_{n=1}^{\infty}V_n\sin(n\hat{\tau})
\end{equation}
where
\begin{equation}
\label{eq10A.6}
\begin{split}
\hat{\tau} \triangleq \Omega_v \tau, \  \hat{m} \triangleq \frac{\sqrt{1+4\hat{\epsilon}C}-1}{2\sqrt{1+4\hat{\epsilon}C}}, \
\Omega_v \triangleq \frac{\pi}{2}\frac{(1+4\hat{\epsilon}C)^{1/4}}{K(\hat{m})}, \
V_n \triangleq \frac{2\pi\sqrt{C}}{(1+4\hat{\epsilon}C)^{1/4}\sqrt{\hat{m}(1-\hat{m})}K(\hat{m})}  \frac{(-1)^{(n-1)/2}\delta_{n,2\mathbb{N}-1}}{\cosh{\left[ \frac{n \pi K(1-\hat{m})}{2K(\hat{m})}\right]}}
\end{split}
\end{equation}
and $K(m)$ is the complete elliptic integral of the first kind (and $\delta_{a,b}$ is Kronecker's delta).

\subsection{Derivation of the Hill equation}
\label{sect10B}

In order to examine the linear stability of the obtained solution, the approach described above in connection with the exploration of the stability of the symmetric mode is followed again here. First, the full equation of motion for the symmetric mode is required (the equation of motion for the antisymmetric mode is already used to represent the dynamics of the antisymmetric mode itself). The term "full" implies that this equation should not include the assumption of single-mode motion, but should rather arise from the equations of motion, Eqs. (\ref{eq6.2}), with the only assumption of zero external excitation.

Finding from Definition (\ref{eq6.3}) that:
\begin{equation}
\label{eq10B.1}
x_{1,2}-x_0=y \pm v/2
\end{equation}
and substituting this, along the way eliminating the external forcing term (for modal dynamics) in the third equation in Eqs. (\ref{eq6.2}), one obtains the equation of motion for the displacement of the box, which reads:
\begin{equation}
\label{eq10B.2}
x_0=y\left(\hat{\eta}+2\hat{\eta}\tilde{p}y^2+\frac{3}{2}\hat{\eta}\tilde{p} v^2\right)
\end{equation}

Since the term in the parentheses is always positive and will remain so even if one adds 1 to the expression multiplying $y$, thus obtaining $\bar{x}$, which should vanish for the antisymmetric mode, it is clear that the antisymmetric mode is only possible for $y=0$, which also leads to $x_0=0$, and thus Eq. (\ref{eq10A.1}) is justified.

Next, taking half of the sum of the first two equations in Eqs. (\ref{eq6.2}) and substituting Eq. (\ref{eq10B.1}), the third of Definitions (\ref{eq6.3}) and the second time derivative of Eq. (\ref{eq10B.2}) into the result, eliminating the second time derivative of $v$ by isolating it from Eq. (\ref{eq6B.1}), one obtains the equation of motion for the symmetric mode for the case of full dynamics, containing only $v$, $y$, their first time derivatives and the second time derivative of $y$, in terms of the dimensionless quantities defined in the first row in Eq. (\ref{eq10A.6}). Next, linearizing with respect to $y(t)$, one obtains the equation for linear perturbations of the antisymmetric mode. Additional mathematical manipulations, described in \ref{AppendixE0}, produce the explicit Hill form of the linear stability equation for the purely antisymmetric mode:
\begin{equation}
\label{eq10B.7}
\tilde{y}''(\hat{\tau})+\frac{\Omega_v^{-2}}{1+\hat{\eta}}\frac{1+(3/2)\hat{\epsilon}[\hat{v}(\hat{\tau})]^2}{1+(3/2)\frac{\hat{\eta}}{1+\hat{\eta}}\hat{\epsilon}[\hat{v}(\hat{\tau})]^2}\tilde{y}(\hat{\tau})=0
\end{equation}
with $\Omega_v$ and $\hat{v}(\hat{\tau})$ given by Eqs. (\ref{eq10A.5})-(\ref{eq10A.6}), and with the Hill function, $h(\hat{\tau})$, being the expression multiplying $\tilde{y}(\hat{\tau})$ in Eq. (\ref{eq10B.7}).

In the next subsection, numerical Floquet analysis of Eq. (\ref{eq10B.7}) is performed using rigorous deconvolution and application of Hill's method, for identification of instability tongues characterizing the antisymmetric mode. Asymptotic analysis of Eq. (\ref{eq10B.7}), resulting in the identification of the number of non-degenerate instability tongues, derivation of starting points at the zero-amplitude limit for numerical instability-tongue boundaries calculation, as well as leading-order asymptotic expansions for the boundaries of the first two instability tongues, using both Mathieu and higher-order approximations, is given in \ref{AppendixE}.

\subsection{Floquet analysis of the Hill equation}
\label{sect10D}

\subsubsection{Fourier decomposition by  numerical deconvolution}
\label{sect10D1}

In order to perform stability analysis by Hill's Determinants Method, the Hill function in Eq. (\ref{eq10B.7}) has to be expressed as a Fourier series.
To this end, first, the numerator and the denominator are decomposed, separately. Both the numerator and the denominator depend linearly on the square of the normalized amplitude of the antisymmetric mode, $\hat{v}(\hat{\tau})$, which was already expressed in a sine series form (Eq. (\ref{eq10A.5})). Thus the first thing one has to do is to perform convolution, since the Fourier series decomposition of the product of two functions (a function and itself, in the present case) equals the (discrete) convolution of the Fourier series decomposition of each of the functions. Multiplying, rearranging and adjusting the indices, one obtains the following expression:
\begin{equation}
\label{eq10D.1}
\begin{split}
[\hat{v}(\hat{\tau})]^2=\frac{V_0^{(2)}}{2}+\sum\limits_{k=1}^{\infty}V_k^{(2)}\cos{\left(2k\hat{\tau}\right)}, \
V_k^{(2)} \triangleq \left [\sum\limits_{m=1,3,5}^{\infty} V_m V_{2k+m}-\frac{1}{2}V_k^2 \delta_{k,2\mathbb{N}-1} \right. \left. -(1-\delta_{k,1}) \sum\limits_{m=1}^{\infty} V_{k-m} V_{k+m}\right ]\delta_{k,\mathbb{N}-1}
\end{split}
\end{equation}

Next the definition of the Hill function, as arising from Eq. (\ref{eq10B.7}), is rearranged to a product form, $f_1(\hat{\tau})h(\hat{\tau})=f_2(\hat{\tau})$, and then $h(\hat{\tau})$  is formally expanded into a Fourier cosine series (as the ratio of two even functions, $h(\hat{\tau})$  is always even):
\begin{equation}
\label{eq10D.4}
h(\hat{\tau})=\frac{H_0}{2}+\sum\limits_{k=1}^{\infty}H_k\cos{\left(2k\hat{\tau}\right)}
\end{equation}

Substituting Eqs. (\ref{eq10D.1}) and (\ref{eq10D.4}) into the aforementioned product-form equation for $h(\hat{\tau})$ that emerges from Eq. (\ref{eq10B.7}), expanding the resulting products of sums, using trigonometric identities to turn cosine products to single index-shifted cosines and performing the convolution, one finally obtains the product-form equation for $h(\hat{\tau})$ as a single Fourier cosine series that should be equal to zero at all times, namely, $g(\hat{\tau})=G_0+\sum\limits_{n=1}^{\infty}G_n\cos{(2n\hat{\tau})}\equiv 0$. Fulfillment of this condition corresponds to the vanishing of the coefficients of this Fourier series. The equations representing the vanishing of these $total$ Fourier cosine series coefficients, formulated as equations for the Fourier cosine coefficients of $h(\hat{\tau})$, take the form of linear algebraic equations.

First, there is the condition $G_0=0$, producing the following explicit equation for $H_0$ in terms of $H_{n \geq 1}$:
\begin{equation}
\label{eq10D.5}
\begin{split}
H_0 = \frac{\Omega_v^{-2}(4+3\hat{\epsilon}V_0^{(2)})}{2(1+\hat{\eta})+(3/2)\hat{\eta}\hat{\epsilon}V_0^{(2)}}
-\frac{3\hat{\eta}\hat{\epsilon}}{2(1+\hat{\eta})+(3/2)\hat{\eta}\hat{\epsilon}V_0^{(2)}}\sum\limits_{n=1}^{\infty}V_n^{(2)}H_n
\end{split}
\end{equation}

Second, taking the set of conditions $G_{n \geq 1}=0$, which contain both $H_0$ and $H_{n \geq 1}$, and eliminating $H_0$ from these equations by isolating it from Eq. (\ref{eq10D.5}), one obtains an infinite set of linear inhomogeneous equations for $H_{n \geq 1}$ alone, which takes the following form in indicial notation:
\begin{equation}
\label{eq10D.6}
\begin{split}
\sum\limits_{m=1}^{\infty}M_{nm} H_m = \mu_n \ \ ; \  \forall \ \  n,m \in \mathbb{N} \ ; \ \mu_n = \frac{3\hat{\epsilon}\Omega_v^{-2}\delta_{n,\mathbb{N}}V_n^{(2)}}{2+2\hat{\eta}+(3/2)\hat{\eta}\hat{\epsilon}V_0^{(2)}} \\
M_{nm} = \left(1+\hat{\eta}+\frac{3}{4}\hat{\eta}\hat{\epsilon}V_0^{(2)} \right)\delta_{n,m}
+ \frac{3}{4}\hat{\eta}\hat{\epsilon}V_{n+m}^{(2)} \delta_{n+m,\mathbb{N}}+ \frac{3}{4}\hat{\eta}\hat{\epsilon}V_{n-m}^{(2)} \delta_{n-2m,\mathbb{N}}\\
+\frac{3}{4}\hat{\eta}\hat{\epsilon}V_{m-n}^{(2)} \delta_{m-n,\mathbb{N}}+\frac{3}{4}\hat{\eta}\hat{\epsilon}V_m^{(2)} \delta_{n,2m}
-\frac{3}{4}\hat{\eta}\hat{\epsilon}\frac{3\hat{\eta}\hat{\epsilon}V_n^{(2)}V_m^{(2)}}{2+2\hat{\eta}+(3/2)\hat{\eta}\hat{\epsilon}V_0^{(2)}}
\end{split}
\end{equation}

Finally, Eqs. (\ref{eq10D.6}) are truncated, exchanging $\mathbb{N}$ by $\mathbb{N}\leq\hat{N}$ everywhere in these equations (this is the reason for the presence of Kronecker's delta in the first term in the third row in Eq. (\ref{eq10D.6})), and deconvolution is performed by solving the truncated system numerically, by a standard algorithm for the solution of linear systems, thus producing $H_{n \geq 1}$. Coefficient $H_0$ is then derived using Eq. (\ref{eq10D.5}).

The issue of convergence is less problematic here than it was in the case of the symmetric mode, since accuracy can be controlled explicitly by choosing a higher value of $\hat{N}$. Moreover, the accuracy in the Fourier series decomposition of $h(\hat{\tau})$ is controlled by the accuracy in the Fourier series decomposition of $\hat{v}(\hat{\tau})$, which does not pose a problem since $V_m$ converges rather rapidly with $m$, due to the hyperbolic cosine in the denominator. Several tens of coefficients are computed  to suffice in the amplitude range already examined in the investigation of the stability of the symmetric mode. Having obtained the Fourier cosine coefficients of the Hill function, one performs the stability analysis by Hill's (truncated) Infinite Determinants method, the same way it was done in the previous section for the symmetric mode.

\subsubsection{Linear stability maps for the antisymmetric mode}
\label{sect10D2}

\begin{figure}[h]
\begin{center}
{\includegraphics[scale=0.6]{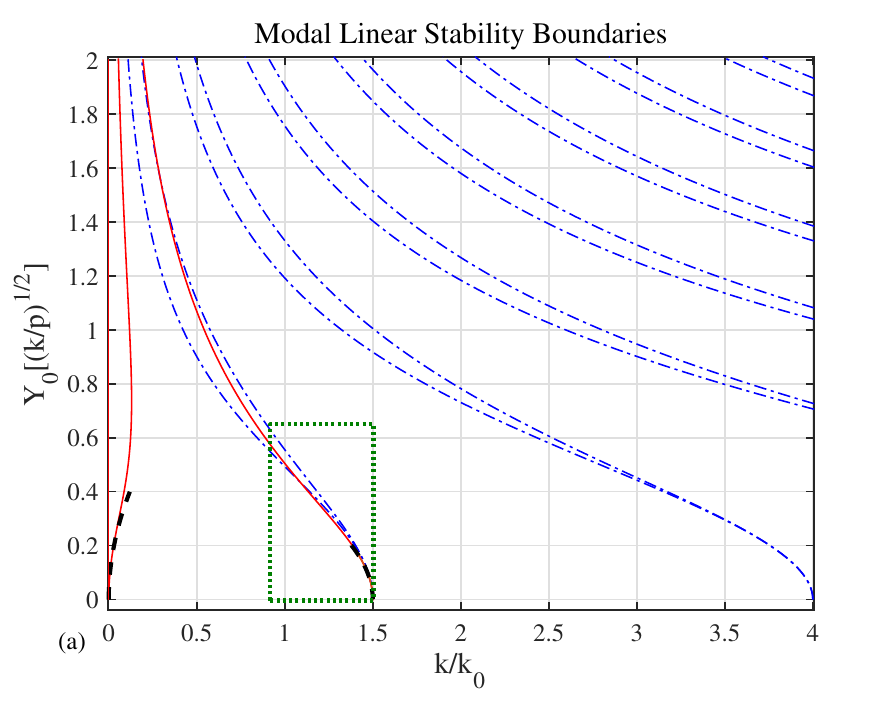}}
{\includegraphics[scale=0.6]{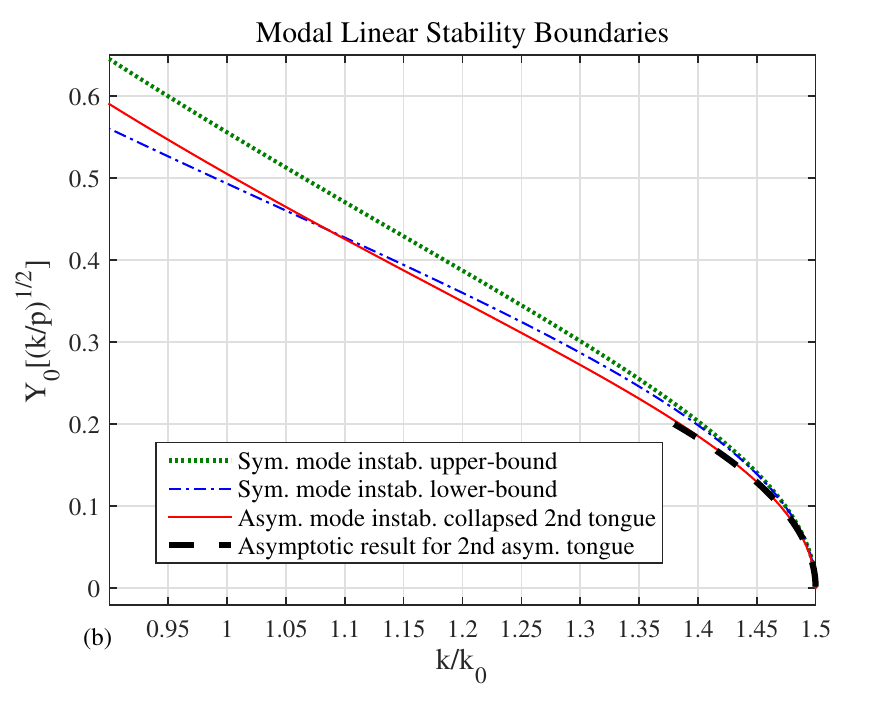}}
\end{center}
\caption{\small $1^{st}$ (finite-width) and $2^{nd}$ (degenerate) instability tongues of the purely antisymmetric mode (in solid red online) along with asymptotic expansions for the tongue boundaries (dashed black) as computed in \ref{AppendixE}, on top of the multiple (finite-width) instability tongues of the symmetric mode (in dash-dot blue online), in an entirely mapped rectangle in the parameter plane (a), and in a close-up (corresponding to the dot-line green online rectangle in (a)) on the second (collapsed) instability tongue of the antisymmetric mode, where it crosses the lower (or left) boundary of the first instability tongue of the symmetric mode (b).}
\label{Figure5}
\end{figure}

Plots of the first (finite-width) instability tongue of the purely antisymmetric mode, and the second tongue, which is (at least nearly) a collapsed, zero-width one, are presented in Fig. \ref{Figure5}, along with asymptotic approximations for their boundaries, as given in \ref{AppendixE}, on top of the instability tongues of the symmetric mode (for better understanding of mode interaction).

\section{Full dynamics and Poincar\'e sections}
\label{sect11}

The dynamical system representing general Hamiltonian motion for the examined mechanism is obtained by assuming no damping, no external forcing and no parametric asymmetry, and is embodied in Eqs. (\ref{eq6.2}) for $F(t)\equiv 0$. In order to gain additional insight into the system's dynamics, it is more convenient to formulate the governing equations in terms of the symmetric mode $y$ and the antisymmetric mode $v$. The second-order equation for the antisymmetric mode is already given in Eq. (\ref{eq6B.1}). It is reformulated here in terms of dimensionless quantities, as defined in Eqs. (\ref{eq6.3}), (\ref{eq7.9})-(\ref{eq7.4}) and (\ref{eq10A.2}), producing the evolution equation for $\hat{v}'$ (where the differentiation, here and onward, is with respect to dimensionless time $\tau \triangleq \omega t$):
\begin{equation}
\label{eq11A.1}
\hat{v}''=-(1+6\hat{\epsilon}\hat{y}^2)\hat{v}-\frac{1}{2}\hat{\epsilon}\hat{v}^3
\end{equation}

Taking half of the sum of the two first equations in Eqs. (\ref{eq6.2}) (with zero forcing), after eliminating $x_0$ using the third equation therein, and after eliminating $v''$ by isolating it from Eq. (\ref{eq11A.1}), and using only dimensionless quantities, the same ones that were used in order to obtain Eq. (\ref{eq11A.1}), leads to the second equation of full dynamics, namely, the (scaled time) evolution equation for $\hat{y}'$:
\begin{equation}
\label{eq11A.2}
\begin{split}
\hat{y}''=-\frac{2\hat{y}+4\hat{\epsilon}\hat{y}^3+24\hat{\eta}\hat{\epsilon}\hat{y}\hat{y}'^2+12\hat{\eta}\hat{\epsilon}\hat{v}\hat{y}'\hat{v}'-3\hat{\epsilon}\hat{y}\hat{v}^2[2\hat{\eta}-1+\hat{\eta}\hat{\epsilon}(12\hat{y}^2+\hat{v}^2)]+6\hat{\eta}\hat{\epsilon}\hat{y}\hat{v}'^2}{2+2\hat{\eta}+12\hat{\eta}\hat{\epsilon}\hat{y}^2+3\hat{\eta}\hat{\epsilon}\hat{v}^2}
\end{split}
\end{equation}

The complementary equations for the box displacement, the displacement of the center of mass and each mass displacement are given in Eqs. (\ref{eq10B.1})-(\ref{eq10B.2}), but they are reproduced here in dimensionless form (in which $x_0,\bar{x},x_1$ and $x_2$ are all scaled by $Y_0$ and thus capped):
\begin{equation}
\label{eq11A.3}
\hat{x}_0=\hat{y}\left( \hat{\eta}+2\hat{\eta}\hat{\epsilon}\hat{y}^2+\frac{3}{2}\hat{\eta}\hat{\epsilon}\hat{v}^2\right), \  \hat{\bar{x}}=\hat{y}+\hat{x}_0, \  \hat{x}_{1,2}=\hat{\bar{x}} \pm \frac{\hat{v}}{2}
\end{equation}

In order to integrate the equations of motion given above, initial conditions have to be specified. A physically natural choice would, perhaps, be setting the initial position of the box, $x_0$ and the initial velocity of the box, $\dot{x}_0$, since these quantities are externally controllable. If in addition, initially the symmetric mode is assumed, then Eq. (\ref{eq11A.3}) and its time derivative can be used to derive initial conditions for $y$ and $y'$ (in the dimensionless version, one of the initial conditions in this case would be $\hat{y}=1)$. If initially the antisymmetric mode is assumed, then the four initial conditions can be chosen in a more trivial fashion. However, if the objective is not to simply integrate the equations but rather to investigate the associated flow topology, additional measures have to be taken, as elaborated on below.

Although the basic phase-space vector is: $\bf{w}=(\hat{y},\hat{v},\hat{y}',\hat{v}')^\top$, for a given value of the energy $E_0$ (or its dimensionless stiffness-ratio-dependent parametrization, $\hat{\epsilon}$), the dynamic flow occupies only a three-dimensional sub-space in the four-dimensional space, $\mathpzc{span}(\bf{w})$. Therefore, one may expect that by observing a projection of the occupied sub-space on a section where one of the variables is fixed, important insights may be gained as to the topology of the three-dimensional manifold of the flow, by observing its trajectories on a certain plane. The flow of Eqs. (\ref{eq11A.1})-(\ref{eq11A.2}) is next examined for fixed energy, $E_0$ (or, really,  $\hat{\epsilon}$) on representative Poincar\'e surface sections.

In order to investigate possible topological phenomena in the flow-occupied phase space, it seems natural to examine trajectories on Poincar\'e Sections (PS), where it is relatively easier to detect things such as a torus changing to a higher genus topological object as the system energy varies. In order to observe that, PS have to be constructed for different energy levels. All the points on PS have thus to be iso-energetic. In order to map different areas of the phase space (and hence on PS), different starting points should be used, corresponding to the same energy. Therefore, initial conditions have to be chosen such that the energy would be the same in a single PS. This means that three initial conditions can be chosen freely, satisfying only an inequality constraint representing the energy in the system, while the fourth initial condition should be calculated from the energy and the first three initial conditions.

Since for the broader problem of understanding localization there is particular interest in the loss of stability of the symmetric mode, and since the problem of the instability of the symmetric mode is more complicated (and even requires  special procedures for the Fourier decomposition) and the resulting instability diagram is richer and more delicate, it is opted to perform PS such that the evolution of the symmetric mode can be observed on a plane. Therefore, $\hat{y}$-$\hat{y}'$ planes are investigated, where $\hat{v}$ and $\hat{v}'$ are not independent. Either $\hat{v}$ or $\hat{v}'$ can be calculated from the known constant energy for a PS, using the first integral of motion, and the second variable in that pair should be fixed on a PS. Since the two options are equivalent, it is chosen to perform PS for the condition $\hat{v}=0$. Thus all the calculated PS depict $(\hat{y},\hat{y}')$ pairs at different times, all corresponding to $\hat{v}=0$, for a large enough number of cycles ($\hat{v}=0$ occurrences), such that the entire flow (on a section) completes a period. For a fixed value of the energy on a particular PS, the energy level is constant and thus $\hat{v}'$ is a dependent variable that technically is integrated using Eq. (\ref{eq11A.1}), but in principle is not an active space dimension of the flow. Therefore, all PS corresponding to $\hat{v}=0$ show the entire picture in $(\hat{y},\hat{y}')$ coordinates, once the (adequately parametrized) energy level $\hat{\epsilon}$ (and of course also the stiffness ratio, $\hat{\eta}$) is specified. Since the energy is not used for the calculation of $\hat{v}'$ for every integration step, but rather use is made of Eq. (\ref{eq11A.1}), the dimensionless general-dynamics version of the first integral of motion is not presented here, it is enough to know that it exists. In order to generate iso-energetic IC on a PS, all corresponding to the chosen section condition, $\hat{v}=0$, the following formula is employed:
\begin{equation}
\label{eq11B.1}
\begin{split}
\hat{v}'_0=2\sqrt{(1+\hat{\eta})(1-\hat{y}_0^2)+(1+4\hat{\eta})\hat{\epsilon}(1-\hat{y}_0^4)
+4\hat{\eta}\hat{\epsilon}^2(1-\hat{y}_0^6)-(1+\hat{\eta}+6\hat{\eta}\hat{\epsilon}\hat{y}_0^2)^2\hat{y}'^2_0 }
\end{split}
\end{equation}

Using this formula, the flow through the PS can be characterized by choosing different $(\hat{y}|_0,\hat{y}'|_0)$ pairs and plotting all the intersections of the corresponding flow with the PS, until the cross-section of the flow on a PS is completed (one full flow period). For every energy level, $\hat{\epsilon}$, the feasible values of $(\hat{y}|_0,\hat{y}'|_0)$ are bounded. The bound is obtained if one assumes that all the energy is initially stored in the symmetric mode. Then the pair $(\hat{y},\hat{y}')$, either initially or at arbitrary time, has to lie on a PS plane within the domain defined by:
\begin{equation}
\label{eq11B.2}
\begin{split}
|\hat{y}'| \leq \frac{\sqrt{(1+\hat{\eta})(1-\hat{y}^2)+(1+4\hat{\eta})\hat{\epsilon}(1-\hat{y}^4)+4\hat{\eta}\hat{\epsilon}^2(1-\hat{y}^6)}}{{1+\hat{\eta}+6\hat{\eta}\hat{\epsilon}\hat{y}^2}}
\end{split}
\end{equation}

\begin{figure}[H]
    \begin{center}
    \subfigure[]
    {
     \includegraphics[scale=0.28]{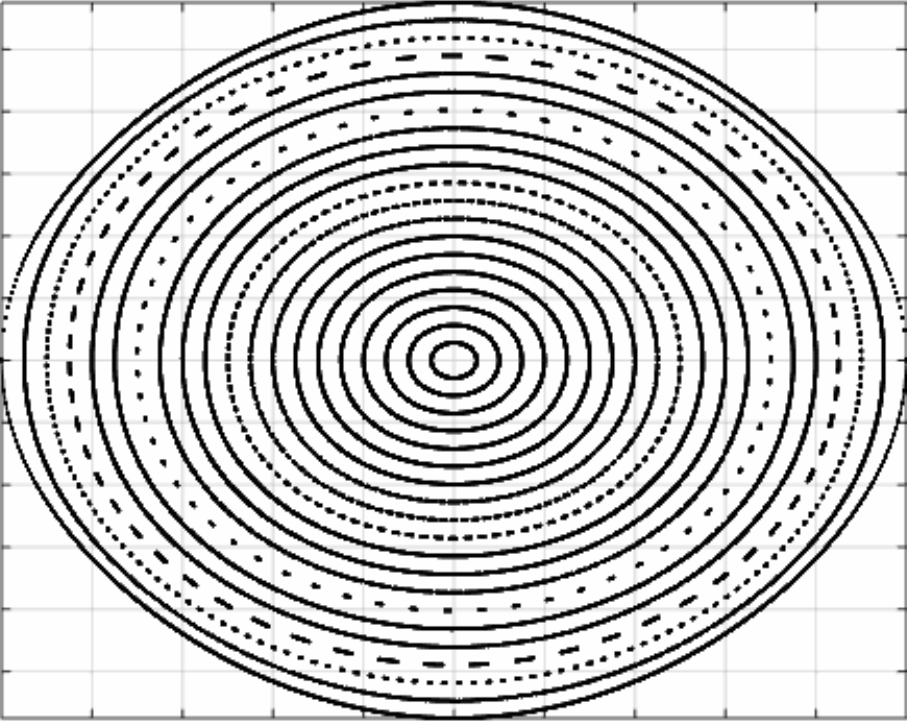}
    }
    \subfigure[]
    {
      \includegraphics[scale=0.28]{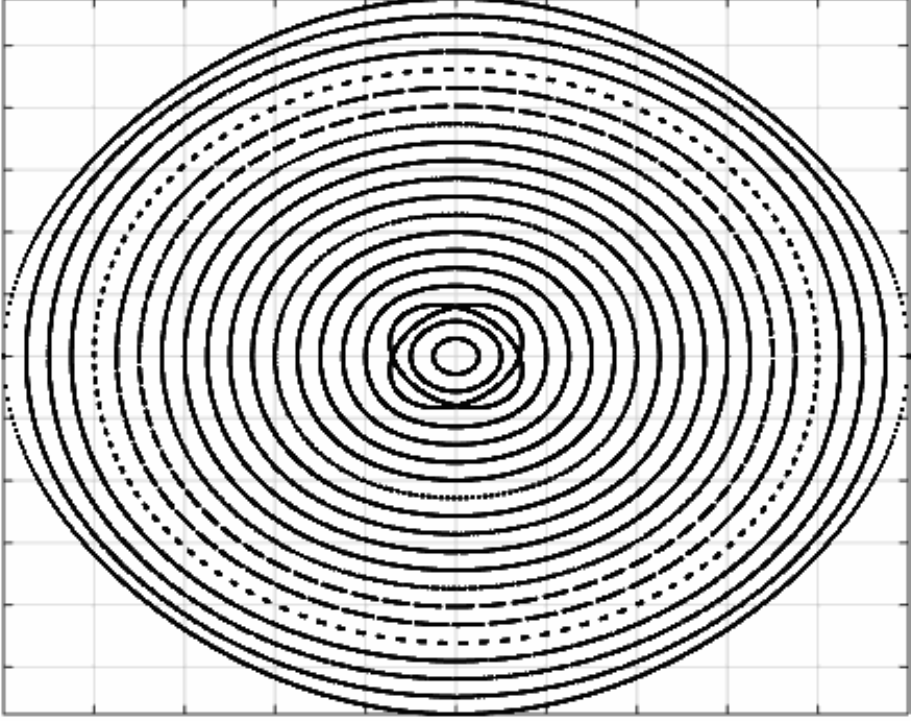}
    }
    \subfigure[]
    {
      \includegraphics[scale=0.28]{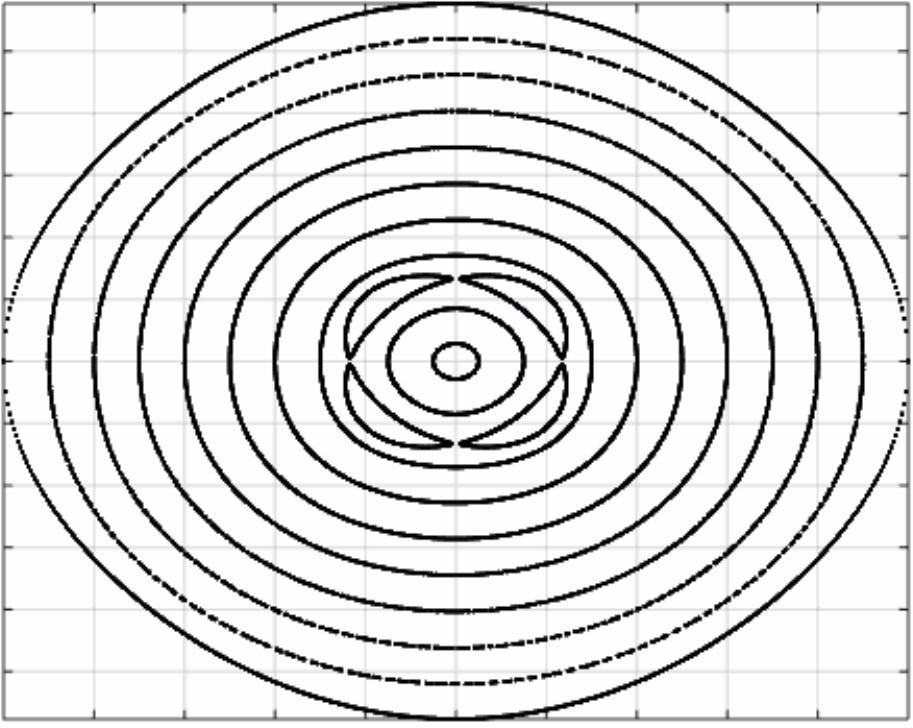}
    }
      \subfigure[]
    {
      \includegraphics[scale=0.28]{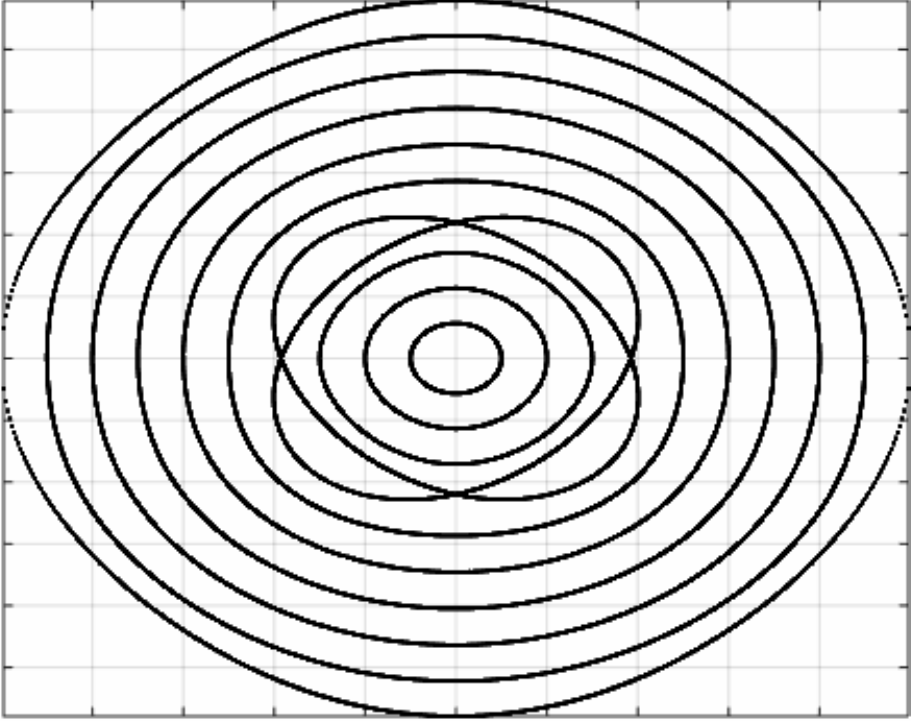}
    }
  \end{center}
  \caption{\small Poincar\'e $\hat{y}$-$\hat{y}'$ planar sections at $\hat{v}=0$ for $\hat{\eta}=5/2$ and $\hat{\epsilon}=0.048$ (a), $\hat{\epsilon}=0.049$ (b), $\hat{\epsilon}=0.05$ (c), and $\hat{\epsilon}=0.0528$ (d).}
\label{Figure6}
\end{figure}

\begin{figure}[H]
\begin{center}
\subfigure[]
    {
     \includegraphics[scale=0.28]{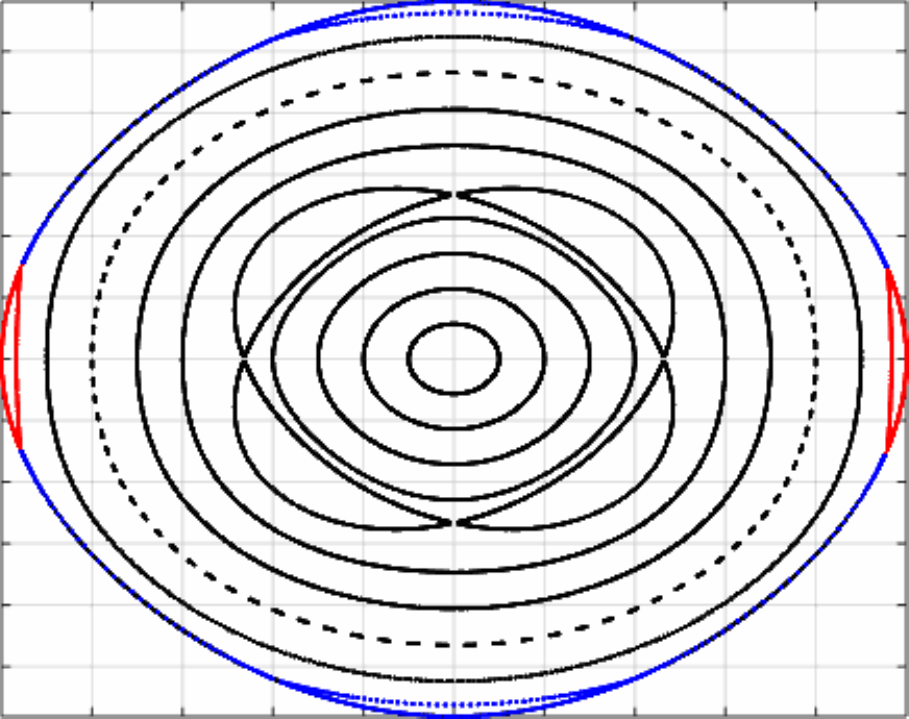}
    }
    \subfigure[]
    {
      \includegraphics[scale=0.28]{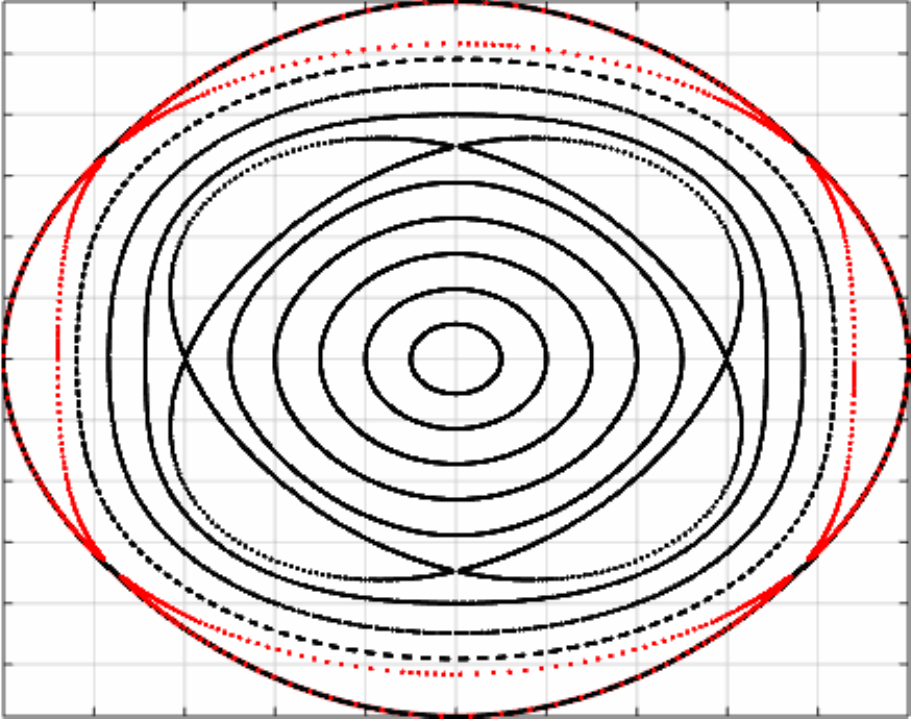}
    }
    \subfigure[]
    {
      \includegraphics[scale=0.28]{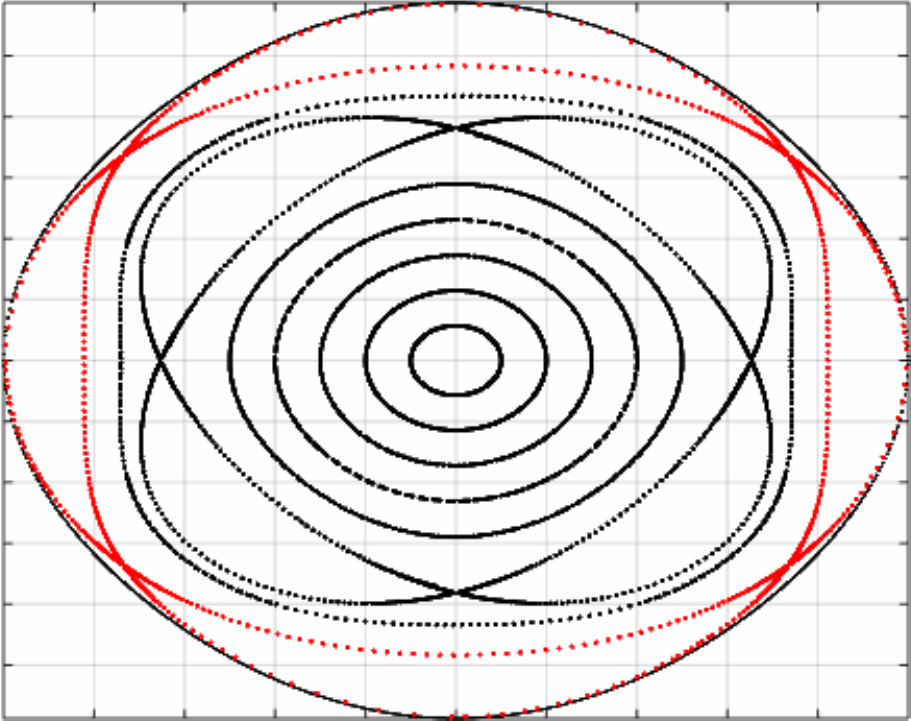}
    }
      \subfigure[]
    {
      \includegraphics[scale=0.28]{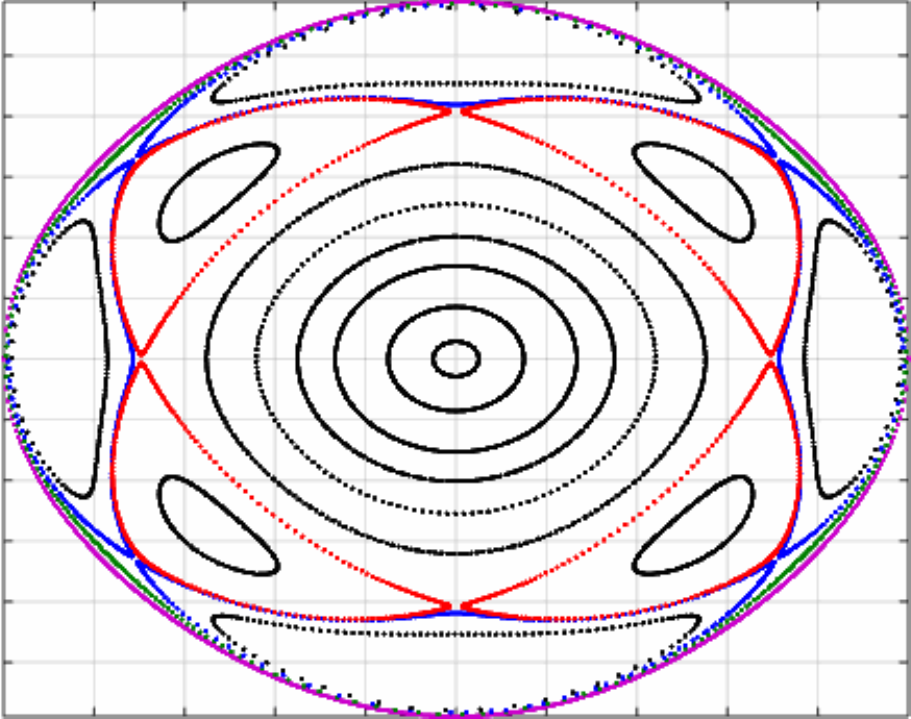}
    }
\end{center}
\caption{\small  Poincar\'e $\hat{y}$-$\hat{y}'$ planar sections at $\hat{v}=0$ for $\hat{\eta}=5/2$ and $\hat{\epsilon}=0.055$ (a), $\hat{\epsilon}=0.06$ (b), $\hat{\epsilon}=0.0625$ (c), and $\hat{\epsilon}=0.0645$ (d).}
\label{Figure7}
\end{figure}

\begin{figure}[H]
\begin{center}
\subfigure[]
    {
     \includegraphics[scale=0.28]{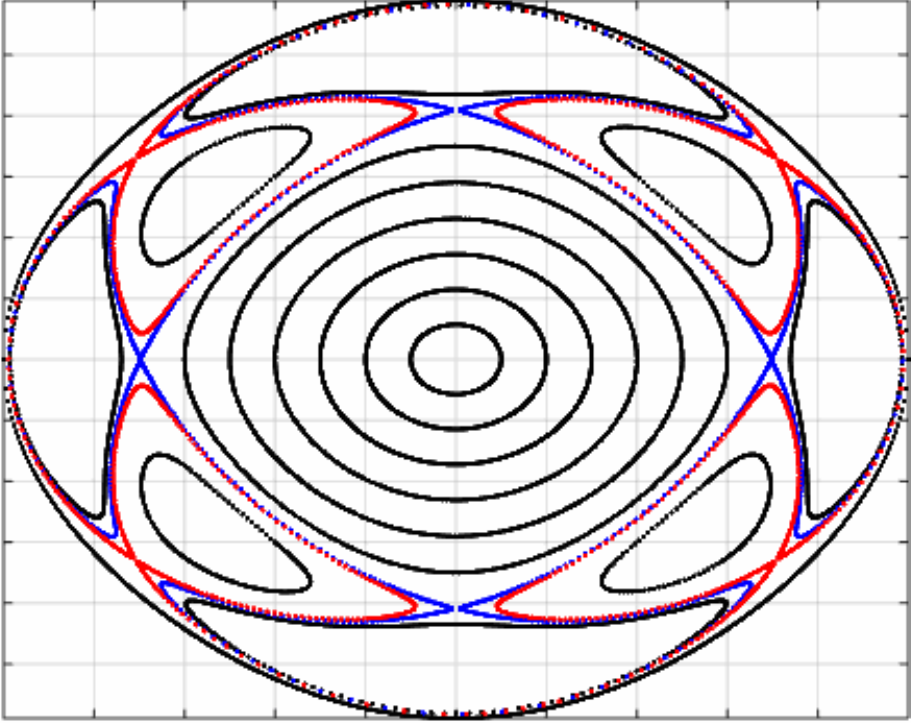}
    }
    \subfigure[]
    {
      \includegraphics[scale=0.28]{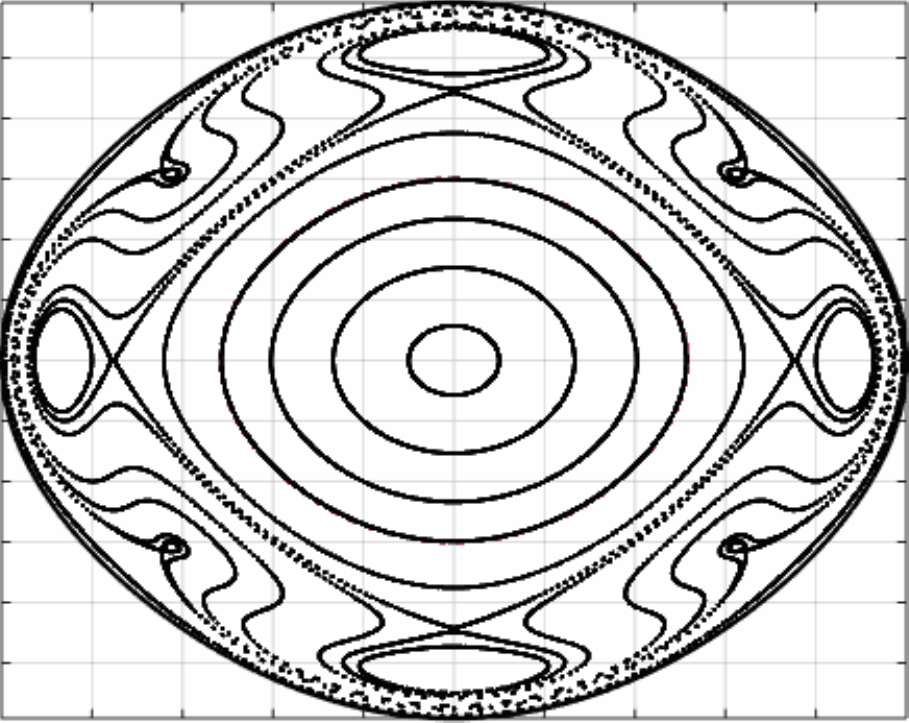}
    }
    \subfigure[]
    {
      \includegraphics[scale=0.28]{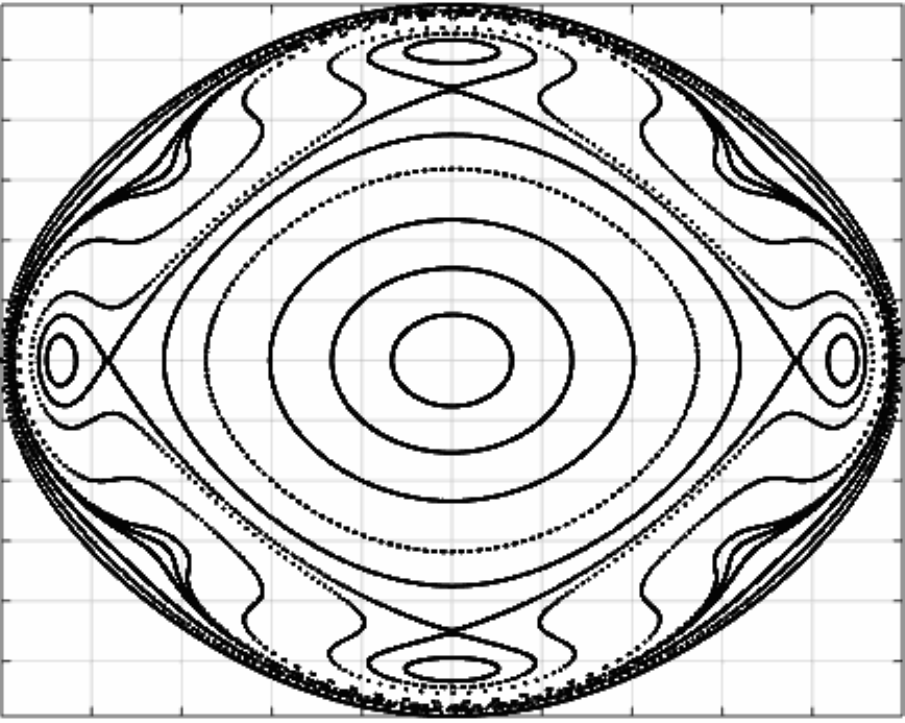}
    }
      \subfigure[]
    {
      \includegraphics[scale=0.28]{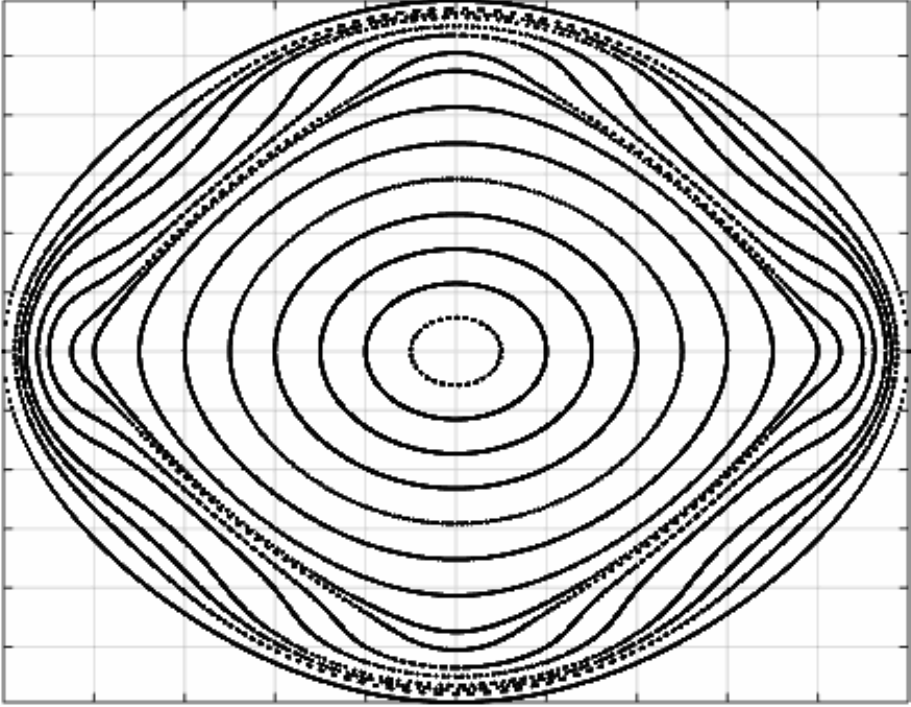}
    }
\end{center}
\caption{\small Poincar\'e $\hat{y}$-$\hat{y}'$ planar sections at $\hat{v}=0$ for $\hat{\eta}=5/2$ and  $\hat{\epsilon}=0.0646$ (a), $\hat{\epsilon}=0.067$ (b), $\hat{\epsilon}=0.0675$ (c), and $\hat{\epsilon}=0.07$ (d).}
\label{Figure8}
\end{figure}

\begin{figure}[H]
\begin{center}
\subfigure[]
    {
     \includegraphics[scale=0.28]{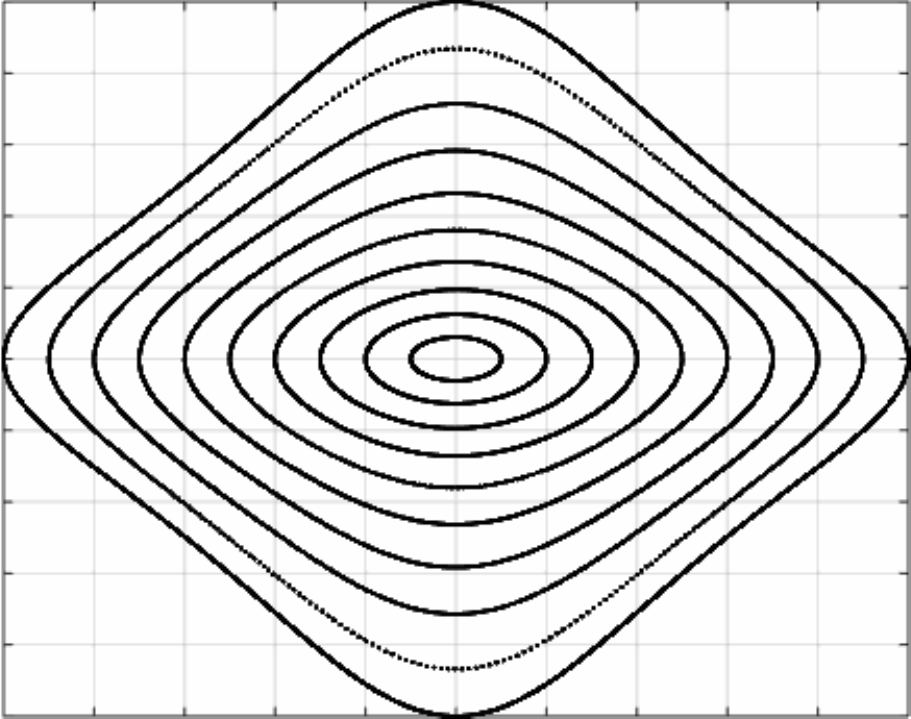}
    }
    \subfigure[]
    {
      \includegraphics[scale=0.28]{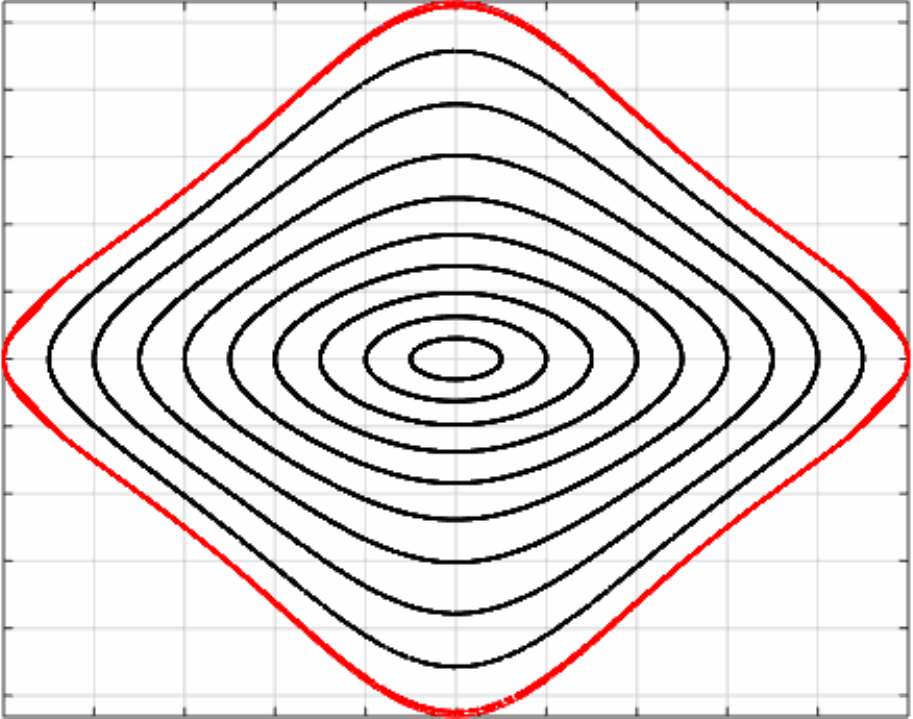}
    }
    \subfigure[]
    {
      \includegraphics[scale=0.28]{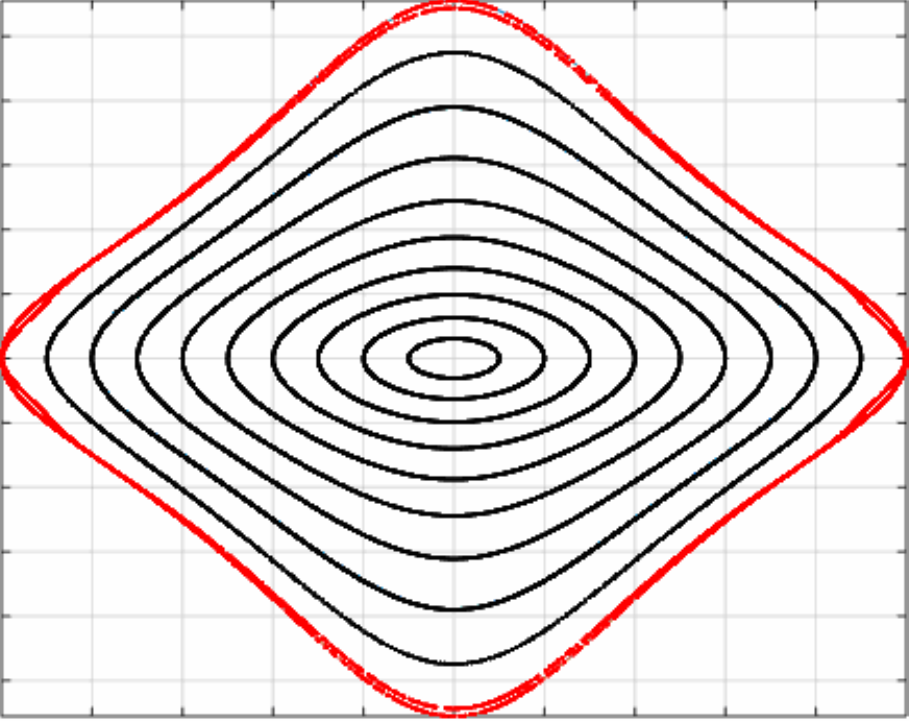}
    }
      \subfigure[]
    {
      \includegraphics[scale=0.28]{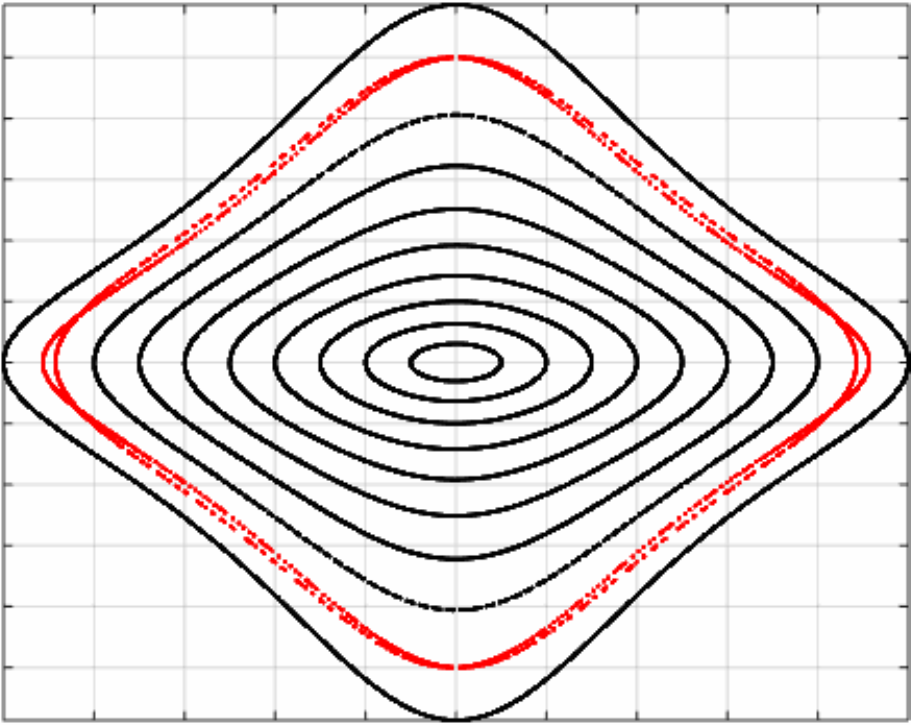}
    }
\end{center}
\caption{\small Poincar\'e $\hat{y}$-$\hat{y}'$ planar sections at $\hat{v}=0$ for $\hat{\eta}=5/2$ and  $\hat{\epsilon}=0.5357$ (a), $\hat{\epsilon}=0.6050$ (b), $\hat{\epsilon}=0.66$ (c), and $\hat{\epsilon}=0.73$ (d).}
\label{Figure9}
\end{figure}

One obvious measure of nonlinear stability analysis using the PS would be to choose initial conditions corresponding to the equality sign in Eq. (\ref{eq11B.2}), and to see whether at any time before a period on the PS is completed, there appears a point at finite distance from that very curve, closer to the center. This would correspond to energy flowing from the symmetric mode to the antisymmetric one (note that on $(\hat{y},\hat{y}')$ planes, the antisymmetric mode is represented by a single point in the center). The results of such analysis on PS are presented in Fig.s  \ref{Figure6}-\ref{Figure9}.

In all the results presented in those figures, the integration of the dynamical system, the nontrivial equations in which are given in Eqs. (\ref{eq11A.1})-(\ref{eq11A.2}), was performed numerically, using the Crank-Nicolson (CN) type updating scheme and trapezoidal approximation of the integral of error \cite{CN}. The symmetry of the CN scheme results in a non-growing, oscillating energy error, a fact which is of major importance in the examined case, where flow topology is to be examined and where modes interaction is of main interest. Last, due to the fact that the $second$ instability tongue of the antisymmetric mode was only reproduced for a bifurcation point at $\eta=3$, and since Fig. \ref{Figure5} shows interesting proximity between the instability tongue boundaries  for values of $\hat{\eta}$ between 2 and 3, a representative value of stiffness ratio was chosen in this range and all the PS calculations were performed for $\hat{\eta}=5/2$.

Figures  \ref{Figure6}-\ref{Figure9} below present the PS for $\hat{\eta}=5/2$, for increasing values of $\hat{\epsilon}$, starting (following Fig. \ref{Figure5}) from below the second antisymmetric mode instability tongue and up to somewhat above the second symmetric mode instability tongue.

Fig. \ref{Figure6}(a) shows the flow at its most simple topology. One sees only concentric ovals, revealing a toroidal structure and implying the absence of parametric resonance. In fact, for all the plots in Fig. \ref{Figure6}, the symmetric mode appears stable, which is in line with Fig. \ref{Figure5}, which shows that all the amplitudes examined in Fig. \ref{Figure6}, are below the first symmetric mode instability tongue. However, only Fig. \ref{Figure6}(a) is below the second instability tongue of the purely antisymmetric mode. It appears from Fig. \ref{Figure6}(a) and (b) that for $\hat{\epsilon}$ between 0.048 and 0.049, the  antisymmetric mode becomes unstable. The evidence for this is the fact that in Fig. \ref{Figure6}(b) one observes KAM islands emanating from the central point, which represents the antisymmetric mode. Such KAM islands are absent from Fig. \ref{Figure6}(a). This transition in PS topology corresponds well with Fig. \ref{Figure5}, which shows that for $\hat{\eta}=5/2$ there is a collapsed linear instability tongue exactly at the normalized amplitude of 0.3113 (which is equivalent to $\hat{\epsilon}=0.048456$).

This result is very interesting since it demonstrates the importance of even a degenerate, or collapsed linear instability tongue, which appears to be responsible for modification of the flow topology by creating KAM islands. The loss of stability of the antisymmetric mode can be inferred from the appearance of the four KAM islands in the center, which change the topological structure of the flow. One can appeal here to a theorem by Kozlov \cite{Kozlov1983} that for Hamiltonian systems with two dynamic degrees of freedom, phase flow with toroidal structure of genus higher than 2, which is what can be seen in Fig.s  \ref{Figure6}(b)--(d), implies the loss of stability through parametric resonance. Another interesting feature in Fig. \ref{Figure6} is that the produced KAM islands are repelled from the center as the amplitude is increased, perhaps implying increasing regained stability of the antisymmetric mode.

Fig. \ref{Figure7} shows continuing expansion and drift from the center of the "antisymmetric" KAM islands upon energy increase from (a) to (d) (note that here, as well as in the previous figure, the "antisymmetric" KAM islands are depicted only in terms of their separatrix). However, here a new phenomenon emerges, already in Fig. \ref{Figure7}(a), corresponding to $\hat{\epsilon}=0.055$, which is equivalent to a normalized amplitude of about 0.33, the value corresponding to the "entrance" into the first symmetric instability tongue, which, as shown in Fig. \ref{Figure5} spans from 0.32 to 0.34, for $\hat{\eta}=5/2$. At this critical energy level, as prognosed by linear stability analysis, four new KAM islands are created, exactly at the perimeter (their separatrix is shown in color online, to distinguish between the four islands). Fig. \ref{Figure7}(a) is plotted for somewhat higher an amplitude than the value critical for linear stability. Upon further increasing the energy, the "symmetric" KAM islands continue to grow. The "antisymmetric'' KAM islands are further repelled from the center and the "symmetric" islands are slightly repelled from the perimeter, as shown in Fig. \ref{Figure7}(c), until the symmetric mode becomes linearly stable again, as demonstrated by the emerging (green online) oval contour conformal to the perimeter from inside. On the other hand, the "symmetric" islands bend in the edges until their separatrices (plotted in blue online) almost coincide with those of the "antisymmetric" islands (in red online). This is the verge of the $third$ topological bifurcation (following the creation of "antisymmetric" KAM islands at $\hat{\epsilon}=0.0485$ and the creation of "symmetric" KAM islands at $\hat{\epsilon}=0.0535$), namely the merge of the two sets of islands, occurring at $\hat{\epsilon}=0.06455$, and shown before merge in Fig. \ref{Figure7}(d) and after merge in Fig. \ref{Figure8}(a). This (global) KAM-islands bifurcation can be associated with a bifurcation of limiting phase trajectory in the averaged flow (see  \cite{GM} and references therein).

Fig. \ref{Figure8}(a) is rather interesting, as it shows the PS of the flow immediately after the aforementioned global KAM-islands bifurcation. This plot shows the maximum exchange of energy,  for a starting point close to the symmetric mode. The bifurcation manifests itself in the fact that as the amplitude increases slightly above $\hat{\epsilon}=0.06455$, the system, initiated at nearly perfectly symmetric conditions, becomes antisymmetric to almost double the extent (in terms of the amount of energy stored in the antisymmetric mode) corresponding to a value slightly lower than $\hat{\epsilon}=0.06455$. Fig. \ref{Figure8}(b) and (c) shows "shrinkage" up to disappearance of "antisymmetric" KAM islands upon further energy increase, as well as "shrinkage" of "symmetric" islands, which later also disappear, as shown in Fig. \ref{Figure8}(d), for yet higher amplitudes. Fig. \ref{Figure8}(c) establishes the existence of a fourth topological change, at about $\hat{\epsilon}=0.0675$, where the total number of KAM islands jumps from eight back to four. Fig. \ref{Figure8}(d) shows a fifth topological change at about $\hat{\epsilon}=0.07$, where the total number of KAM islands jumps back to zero, which is the value corresponding to $\hat{\epsilon}< 0.048$. Fig.s  \ref{Figure6}-\ref{Figure8} replicate the stripes-pattern energy-dependence of linear stability to more general phase flow topology properties. Increase in energy complicates the topology, and further increase simplifies it. Fig. \ref{Figure9} further establishes this phenomenon by exhibiting the topological evolution of PS when crossing the second instability tongue of the symmetric mode. Interestingly enough, new ''antisymmetric'' KAM islands were not encountered here, either because the third degenerate asymmetric linear instability tongue crosses the $\hat{\eta}=5/2$ line for higher amplitudes than were mapped. In fact, there is no clear evidence of the existence of a third or higher-index tongues.

Fig. \ref{Figure9}(a) shows perfectly toroidal flow of genus 2 at amplitudes (from $\hat{\epsilon}=0.07$) up to about $\hat{\epsilon}=0.55$, where (only) symmetric mode KAM islands emerge again at the perimeter of the flow in the PS. This time, the number of islands is not 4 but rather 6. Fig. \ref{Figure9}(b) shows the still narrow emerging "symmetric" KAM islands (in red online) slightly above the critical amplitude (for better appearance). This is another topological bifurcation.  Fig. \ref{Figure9}(c) shows the same islands as in Fig. \ref{Figure9}(b), only thicker and more apparent. It should be noted, though, that the second-symmetric-tongue-KAM-islands are not large enough to demonstrate how infinitesimal asymmetry in the initial conditions grows to a finite one. It does grow to a finite one, but remains relatively small, unlike in the case of the first symmetric instability tongue. Fig. \ref{Figure9}(d) shows that indeed the aforementioned KAM islands do not grow further, but rather emanate from the perimeter to the center, rendering the symmetric mode linearly stable again. The shift of the "symmetric" KAM islands is reminiscent of the shift outward exhibited by the "antisymmetric" islands for lower amplitudes.

\section{Conclusions}
\label{sect12}
The results presented above explore the concept of "hidden" modes in a discrete mechanical system. Such modes appear due to internal symmetry in the system, and can exist both for linear and nonlinear interaction forces. In the simple setting considered, in the linear case, the antisymmetric mode cannot be excited by any external forcing. This property of the system is robust, in the sense that the mode remains almost non-excitable in the case of small asymmetry and small damping. More rigorously speaking, the amplitude-frequency curve reveals the existence of the internal structure only if high-order derivatives are considered. However, in the presence of finite nonlinearity, the internal structure may reveal itself even in the perfectly symmetric case, due to possible loss of stability of the symmetric mode through parametric resonance.  Detailed exploration of such resonance required us to develop a special method for efficient computation of Fourier expansion directly from integral quadratures, without resorting to explicit integration. Poincar\'{e} sections relate the parametric instabilities to the global structure of the dynamical flow. In particular, it was demonstrated that global "merging" of KAM tori, which appeared due to the instability of both the symmetric and the antisymmetric modes, leads to intensive energy exchange inside the system. Possible application of the results obtained in this work could lie in the realm of strictly mechanical systems, for instance, in MEMS sensors, where the advantage of the emergence of internal resonance often exploited for signal enhancement could be further enhanced by internal-to-external (and vice versa) resonant energy flow. A special feature of resonance-increased sensitivity in the examined system is that loss of stability of the symmetric mode results in a growing perturbation in the $only$ observable quantity in the system, which makes the resonance qualitatively more pronounced by (possibly) making the (still hypothetical) device sensitivity spatially non-uniform. Further applications may include explicitly physical natural or synthetic systems, where the "hidden" modes could be directly related to compacton solutions.

\section*{Acknowledgments}

The authors are grateful to the Israel Science Foundation (grant 838/13) for financial support.

\renewcommand{\theequation}{A1.\arabic{equation}}
\setcounter{equation}{0}
\setcounter{figure}{0}
\appendix
\section{Derivation details for the linear asymmetric case}

\subsection{Internal mass response amplitude for general parametric asymmetry}

\label{AppendixA1}

The expression for the (normalized) amplitude of the displacement of the first internal mass in the linear regime with general parametric asymmetry ($\gamma$ for stiffness and $\delta$ for damping) takes the following form (the solution for the second mass can be consequently deduced trivially):

\begin{equation}
\begin{split}
\label{eqabsA1exp}
\left |\hat{A}_1 \right|= \\= \frac{\sqrt{ \left(f_c^{(0)}+\tilde{f}_c^{(2)} x\right)^2 \left(f_d^{(0)}+\tilde{f}_d^{(2)} x\right)^2   +\left[\left(f_c^{(1)}\right)^2\left(f_d^{(0)}+\tilde{f}_d^{(2)} x\right)^2 +\left(f_d^{(1)}\right)^2\left(f_c^{(0)}+\tilde{f}_c^{(2)} x\right)^2 \right ] x
+\left(f_c^{(1)} f_d^{(1)}\right)^2x^2}}{\left(f_d^{(0)}+\tilde{f}_d^{(2)} x\right)^2+\left(f_d^{(1)}\right)^2x}
\end{split}
\end{equation}
where the following auxiliary functions
\begin{equation}
\begin{split}
f_c^{(0)} \triangleq c_0+\bar{c}_2 \bar{\Omega}^2, \ f_c^{(1)} \triangleq c_1 \bar{\Omega}+c_3 \bar{\Omega}^3, \
\tilde{f}_c^{(2)} \triangleq \tilde{c}_2\bar{\Omega}^2, \\ f_{d}^{(0)} \triangleq d_0+\bar{d}_2 \bar{\Omega}^2 + \bar{d}_4 \bar{\Omega}^4, \
f_d^{(1)} \triangleq d_1 \bar{\Omega}+d_3 \bar{\Omega}^3+d_5\bar{\Omega}^5, \ \tilde{f}_d^{(2)} \triangleq \tilde{d}_2\bar{\Omega}^2+\tilde{d}_4\bar{\Omega}^4
\label{eq4.6}
\end{split}
\end{equation}
and the following auxiliary parameters
\begin{equation}
\begin{split}
c_0 \triangleq -(1+\gamma), \ c_1 \triangleq -(2+\gamma+\delta), \ c_3 \triangleq 1+\epsilon \\
d_0 \triangleq -\eta(1+\gamma), \ d_1 \triangleq -\eta(2+\gamma+\delta), \ d_3 \triangleq (2+\eta)(2+\delta+\epsilon)+2\gamma+\gamma\epsilon+\delta\epsilon, \
d_5 \triangleq -(2+\delta)(1+\epsilon)
\label{eq4.7}
\end{split}
\end{equation}
and
\begin{equation}
\begin{split}
\bar{c}_2 \triangleq 1+\epsilon, \ \tilde{c}_2 \triangleq 1+\delta \\
\bar{d}_2 \triangleq 2(1+\eta)+(1+\eta)\epsilon+(2+\eta)\gamma+\gamma\epsilon, \
\tilde{d}_2 \triangleq \eta(1+\delta), \
\bar{d}_4 \triangleq -(2+\eta+\gamma)(1+\epsilon), \
\tilde{d}_4 \triangleq -(2+\epsilon)(1+\delta)
\label{eq4.8}
\end{split}
\end{equation}
are used.

\setcounter{figure}{0}

\subsection{Mode hiding by analytic local convexification}
\label{AppendixA}

Eqs. (\ref{eq5.2}) determine $x_c$, which should be close to zero, and $\bar{\Omega}_c$, which should be close to unity. These equations can, in principle, be solved with no additional assumptions. However, since two coupled nonlinear equations with a small parameter, $\epsilon$, may be hard to solve, especially if one wishes to obtain as much analytical a result as possible, good starting points are beneficiary. A better starting point for $x_c$ than zero is given in the ansatz in Eq. (\ref{eq5.1}). It can now be argued that a better starting point than unity for $\bar{\Omega}_c$ would be the second undamped-limit resonance frequency, as given in Eq. (\ref{eq3.5}), which in dimensionless form becomes:
\begin{equation}
\label{eq5.3}
\bar{\Omega}_c=1-\epsilon/4
\end{equation}

The reason for choosing this value as a starting point is that in the small asymmetry limit this value corresponds to the second resonance frequency, which in the damped case becomes the point of local maximum, and thus maximum concavity (the narrow high spike has to be of negative curvature of large absolute value). Increasing the damping, or $\alpha$, one decreases the value of the maximum, and thus also the curvature, significantly, but hardly shifts the point of the occurrence of that local maximum. Therefore, the maximum concavity remains approximately at the same place. Consequently, the condition that convexification works on the point of maximum convexity is in fact the identification of $\bar{\Omega}_c$. Thus one may say that the first equation in Eqs. (\ref{eq5.2}) is for $x_c$, or $\alpha$, and the second equation, namely, $|A_0|'''_{(\bar{\Omega}_c,x_c)}=0$, is for $\bar{\Omega}_c$. Consequently, one should find $\bar{\Omega}_c(\alpha )$ (which should be a weak function) that would satisfy the equation $|A_0|'''_{(\bar{\Omega}_c,x_c)}=0$ for every $\alpha$. It turns out that in the small asymmetry limit, $\bar{\Omega}_c(\alpha )$ is a constant equal to the expression given in Eq. (\ref{eq5.3}). In order to prove it, one should evaluate $|A_0|'''_{(\bar{\Omega}_c,x_c)}$ for unspecified $\alpha$ and show that it vanishes for $\bar{\Omega}_c=1-\epsilon/4$. In addition, it should be shown that it vanishes $only$ there. The latter is better to start with. In stricter terms, it should be shown that there is but a single maximum concavity point in the examined neighborhood. This is reasonable since there is but a single resonance-originating spike in the smaller-than-necessary-for-hiding damping case, and the functional structure, which shows only one locally maximum concavity point is already manifested there. In addition, if the third derivative vanishes at $\bar{\Omega}_c=1-\epsilon/4$, and there are also other solutions for the assumed ansatz, then they also must be linear in $\epsilon$, due to the functional form of $|A_1|$ discussed above. However, this is problematic. For too small a damping coefficient, there can be only one point of maximum concavity, the one originating from the undamped-limit resonance. For sufficient damping, the response curve becomes rather smooth in the relevant area, which allows one obtaining $|A_0|'''$ at a second solution using a Taylor series approximation extending from the first solution:

\begin{equation}
\label{eq5.4}
\begin{split}
0=|A_0|'''_{\bar{\Omega}_{c,2}}= |A_0|'''_{\bar{\Omega}_{c,1}}+|A_0|^{(4)}_{\bar{\Omega}_{c,1}}(\bar{\Omega}_{c,2}-\bar{\Omega}_{c,1})+(1/2)|A_0|^{(5)}_{\bar{\Omega}_{c,1}}(\bar{\Omega}_{c,2}-\bar{\Omega}_{c,1})^2=\\
=|A_0|^{(4)}_{\bar{\Omega}_{c,1}}(\bar{\Omega}_{c,2}-\bar{\Omega}_{c,1})+(1/2)|A_0|^{(5)}_{\bar{\Omega}_{c,1}}(\bar{\Omega}_{c,2}-\bar{\Omega}_{c,1})^2
\end{split}
\end{equation}

If $\bar{\Omega}_{c,1}=1-\epsilon/4$ is a solution, and it corresponds to the minimum of the second derivative, then $|A_0|^{(4)}_{\bar{\Omega}_{c,1}}>0$. It is also true that $(\bar{\Omega}_{c,2}-\bar{\Omega}_{c,1})=\mathcal{O}(\epsilon)$, since it is a difference of two terms linear in $\epsilon$. If $|A_0|^{(5)}_{\bar{\Omega}_{c,1}}=0$, then $\bar{\Omega}_{c,2}-\bar{\Omega}_{c,1}=0$ (exactly). If $|A_0|^{(5)}_{\bar{\Omega}_{c,1}}\neq 0$, then, due to orders separation, one again has: $\bar{\Omega}_{c,2}-\bar{\Omega}_{c,1}=0$ (an $\mathcal{O}(\epsilon)$ term cannot be compensated by an $\mathcal{O}(\epsilon^2)$ term). In any case, if one assumes: $\bar{\Omega}_{c,2}-\bar{\Omega}_{c,1}\neq 0$, then one gets: $\bar{\Omega}_{c,2}-\bar{\Omega}_{c,1}=0$, which is a contradiction.  Consequently, it is evident that: $\bar{\Omega}_{c,2}-\bar{\Omega}_{c,1}=0$.

It has thus been proven (by contradiction) that if a solution of the equation $|A_0|'''_{\bar{\Omega}_{c}}=0$ exists, then it is unique. There is a candidate solution given in Eq. (\ref{eq5.3}), which may be suggested drawing from the undamped-limit scenario. It is next shown that the assumption in Eq. (\ref{eq5.3}) is indeed $a$ solution of the second equation in Eqs. (\ref{eq5.2}) (and thus is the only solution).

First, recalling Eq. (\ref{eq4.4}), an explicit form of the required third derivative of $|A_0|$ is obtained, using the derivatives of $f(\bar{\Omega})$,  by the chain rule:
\begin{equation}
\label{eq5.5}
|A_0|'''_{\bar{\Omega}_{c}}=u_0\eta \left(f'''|_{\bar{\Omega}_{c}}|\hat{A}_1|_{\bar{\Omega}_{c}}+3f''|_{\bar{\Omega}_{c}}|\hat{A}_1|'_{\bar{\Omega}_{c}}+3f'|_{\bar{\Omega}_{c}}|\hat{A}_1|''_{\bar{\Omega}_{c}}+f|_{\bar{\Omega}_{c}}|\hat{A}_1|'''_{\bar{\Omega}_{c}} \right)
\end{equation}

Next, using the second, asymptotically sufficient form of $f(\bar{\Omega})$ from Eq. (\ref{eq4.4}), and employing Eqs. (\ref{eq5.1}) and (\ref{eq5.3}), one obtains the derivatives of $f$ at the required point, in terms of the unknown $\alpha$ and the small parameter $\epsilon$:
\begin{equation}
\label{eq5.6}
f|_{\bar{\Omega}_{c}}=(\alpha+1/4)^{1/2}\epsilon, \ f'|_{\bar{\Omega}_{c}}=-(\alpha+1/4)^{-1/2}, \ f''|_{\bar{\Omega}_{c}}=4\alpha(\alpha+1/4)^{-3/2}\epsilon^{-1}, \ f'''|_{\bar{\Omega}_{c}}=12\alpha(\alpha+1/4)^{-5/2}\epsilon^{-2}
\end{equation}

Now, in order to obtain the derivatives of $|\hat{A}_1|$, the following is done. Instead of directly differentiating the cumbersome expression in Eq. (\ref{eqabsA1exp}), one can take a different approach. First, realizing that it is the neighborhood of the second resonance frequency, for small asymmetry, that is of interest, the auxiliary variable $\beta=\mathcal{O}(1)$, defined by the relation:
\begin{equation}
\label{eq5.7}
\bar{\Omega}=1-\beta\epsilon
\end{equation}
is introduced.

Substituting Eqs. (\ref{eq5.7}) and (\ref{eq5.1}) into the expression for $|\hat{A}_1|$ given with the auxiliary quantities in Eqs. (\ref{eqabsA1exp})-(\ref{eq4.8}), contracting identical powers of $\epsilon$ appearing in the numerator and denominator, such that only positive powers remain, and taking the $\epsilon\ll 1$ limit, in accordance with the small asymmetry assumption (which eliminates all remaining positive powers of $\epsilon$), one obtains the following expression for $|\hat{A}_1|_{(\alpha,\beta)}$:
\begin{equation}
\label{eq5.8}
|\hat{A}_1|_{(\alpha,\beta)}=\frac{\sqrt{(1-2\beta)^2(1-4\beta)^2+[4(1-2\beta)^2+(1-4\beta)^2]\alpha+4\alpha^2}}{(1-4\beta)^2+4\alpha}
\end{equation}

Using Eq. (\ref{eq5.7}), the derivatives required for Eq. (\ref{eq5.5}) can readily be obtained in terms of the derivatives of the function in the left-hand side of Eq. (\ref{eq5.8}), as follows:
\begin{equation}
\label{eq5.9}
\begin{split}
|\hat{A}_1|_{\bar{\Omega}_{c}}=|\hat{A}_1|_{(\alpha,\beta=1/4)}, \
|\hat{A}_1|'_{\bar{\Omega}_{c}}=-\frac{1}{\epsilon}\left.\left(\frac{\partial}{\partial{\beta}}|\hat{A}_1|_{(\alpha,\beta)}\right)\right|_{\beta=1/4} \\ \ |\hat{A}_1|''_{\bar{\Omega}_{c}}=\frac{1}{\epsilon^2}\left.\left(\frac{\partial^2}{\partial{\beta^2}}|\hat{A}_1|_{(\alpha,\beta)}\right)\right|_{\beta=1/4}, \ \hat{A}_1|'''_{\bar{\Omega}_{c}}=-\frac{1}{\epsilon^3}\left.\left(\frac{\partial^3}{\partial{\beta^3}}|\hat{A}_1|_{(\alpha,\beta)}\right)\right|_{\beta=1/4}
\end{split}
\end{equation}

Next, the four functions in Eq. (\ref{eq5.9}) can be calculated directly using the general expression given in Eq. (\ref{eq5.8}), to give:
\begin{equation}
\label{eq5.10}
\begin{split}
|\hat{A}_1|_{(\alpha,\beta=1/4)}=\frac{1}{2}\alpha^{-1/2}(\alpha+1/4)^{1/2}, \
\left.\left(\frac{\partial}{\partial{\beta}}|\hat{A}_1|_{(\alpha,\beta)}\right)\right|_{\beta=1/4} =-\frac{1}{2}\alpha^{-1/2}\frac{1}{(\alpha+1/4)^{1/2}}\\ \left.\left(\frac{\partial^2}{\partial{\beta^2}}|\hat{A}_1|_{(\alpha,\beta)}\right)\right|_{\beta=1/4}=-\frac{1}{8}\alpha^{-3/2}\frac{1+8\alpha}{(\alpha+1/4)^{3/2}}, \ \left.\left(\frac{\partial^3}{\partial{\beta^3}}|\hat{A}_1|_{(\alpha,\beta)}\right)\right|_{\beta=1/4}=\frac{3}{8}\alpha^{-3/2}\frac{1+8\alpha+32\alpha^2}{(\alpha+1/4)^{5/2}}
\end{split}
\end{equation}

Finally, substituting Eqs. (\ref{eq5.10}), (\ref{eq5.9}) and (\ref{eq5.6}) into Eq. (\ref{eq5.5}), the required result is obtained:
\begin{equation}
\label{eq5.11}
|A_0|'''_{\bar{\Omega}_{c}=1-\frac{\epsilon}{4}}=0 \ \forall \ \alpha
\end{equation}
This result is asymptotically exact (which means -- up to higher-order corrections in $\epsilon$). This means that half of the convexification problem, finding the point where the second derivative obtains its extremum, is solved. Next,  the value of $\alpha$ for which this extremum vanishes should be found. In addition, it should be shown that the extremum is a minimum. There is still one unknown, $\alpha$, and one equation, the first of Eqs. (\ref{eq5.2}). First, recalling Eq. (\ref{eq4.4}) again, one obtains an explicit form of the now $second$ derivative of $|A_0|$, using the derivatives of $f(\bar{\Omega})$, and employing the chain rule, as follows:
\begin{equation}
\label{eq5.12}
|A_0|''_{\bar{\Omega}_{c}}=u_0\eta \left(f''|_{\bar{\Omega}_{c}}|\hat{A}_1|_{\bar{\Omega}_{c}}+2f'|_{\bar{\Omega}_{c}}|\hat{A}_1|'_{\bar{\Omega}_{c}}+f|_{\bar{\Omega}_{c}}|\hat{A}_1|''_{\bar{\Omega}_{c}} \right)
\end{equation}

Next, substituting Eqs. (\ref{eq5.12}), (\ref{eq5.10}), (\ref{eq5.9}) and (\ref{eq5.3}) into the first of Eqs. (\ref{eq5.2}) and rearranging, one obtains the convexification condition for $\alpha$ in the form of a simple quadratic equation, as follows:
\begin{equation}
\label{eq5.13}
\alpha^2-\alpha-1/16\ge 0
\end{equation}
where the weak inequality sign means that $\alpha$ has to guarantee a positive second derivative (at its minimum) for convexification, and the equality sign should produce the critical value.

The right-hand side in Eq. (\ref{eq5.13}) has two roots, $(2\pm\sqrt{5})/4$, only one of which is positive, and thus feasible for the examined case (since real, rather than imaginary damping is assumed). Consequently, one has: $\alpha_c=(2+\sqrt{5})/4\approx 1.059$.

Eq. (\ref{eq5.13}) was obtained without sign changes, and thus the positive branch of the parabola guarantees convexity in the neighborhood of the second resonance frequency. Thus, taking $\alpha>\alpha_c$ guarantees that at its extremum, the second derivative of $|A_0|$ becomes positive. Regarding the fact that the extremum is a minimum, one could examine the fourth derivative, but simpler argumentation can be used. It is apparent that after hiding, the minimum of $|A_0|$ lies in the vicinity of $\bar{\Omega}$ (due to the effect of $f$). This means that for the hiding damping, to the left (enough) of $\bar{\Omega}$, the curve is convex. The same is true to the right (enough) of $\bar{\Omega}$. This means that $|A_0|''$ is positive to the left and to the right of the area where it is negative. This means that its extremum is negative and that it is a local minimum (a negative value is $smaller$ than its positive surroundings). Thus the extremum is a $minimum$, and therefore if it is positive, then the entire second derivative is positive in that area. Consequently, quasi-local convexification under the single condition $\alpha>\alpha_c$ is guaranteed.

\renewcommand{\theequation}{B.\arabic{equation}}
\setcounter{equation}{0}

\section{Explicit analytic results regarding integration and stability -- special parameter values}
\label{AppendixB}

It appears to be instructive to explore several special parameter values cases, for which simplified analytic results regarding symmetric mode integration and stability can be obtained.

\subsection{The $\hat{\epsilon}\ll 1$ case}
\label{sect7A.1}

If one takes the $\hat{\epsilon}\ll 1$ limit in Eq. (\ref{eq7.9}), then Eq. (\ref{eq7.14}) simplifies to:
\begin{equation}
\label{eq7A.1}
t-\frac{3T}{4}=\frac{\sqrt{1+\hat{\eta}}}{2\omega}\int\limits_{0}^{\hat{x}}\frac{dx}{\sqrt{x(1-x)}}
\end{equation}
which can be integrated analytically and inverted, to yield:
\begin{equation}
\label{eq7A.2}
y(t)=Y_0\cos{\left(\frac{\omega t}{\sqrt{1+\hat{\eta}}}\right)}=Y_0\cos{\left(\sqrt{\xi_0}\omega t\right)}
\end{equation}

This result is identical to the $\hat{\epsilon}\to 0$ limit of what one obtains using the Poincar\'{e}-Lindstedt method, for which stability analysis can be performed within the Mathieu equation approximation. The outcome is the emergence of multiple instability tongues. However, since the result is only rigorous in the small amplitude limit, and in this limit the existence of small damping (the same damping that hides parametric asymmetry) may eliminate instability (see \cite{Verhulst2008}, for example), it is a fact that analysis in the $\hat{\epsilon} \ll 1$ limit is insufficient for establishing nonlinear breaking of linear symmetry. Thus the further analysis performed in Sectsions \ref{sect7}-\ref{sect9}.

\subsection{The $\hat{\epsilon}=\hat{\epsilon}^{*}(\hat{\eta})$ case}
\label{sect7A.2}

Interestingly enough, derivation of Fourier series coefficients by direct integration is problematic due to the first ratio term in the integrand in the second row in Eq. (\ref{eq7.16}), the one multiplying $\Psi(\hat{x})$, and not due to $\Psi(\hat{x})$ itself, which is smooth and regular (due to the positivity of $A,B$ and $C$). Nevertheless, it is precisely the form of $\Psi(\hat{x})$ that can render the problem solvable analytically, for a special relation between the parameters.

The parabola in the square root in the denominator of $\Psi(\hat{x})$ can always be presented as the product of two linear terms. Now, there are two special cases. It the first special case, one of the linear terms would be proportional to the numerator of $\Psi(\hat{x})$. In the second special case, the two terms would be identical, i.e. the parabola would have a multiple root. These two cases are special because in each of them, there would formally be a different elliptic integral required for the derivation of say, $T$ or $t$, leading to different functional construction in analytic integration. Interestingly enough, for all feasible values of the parameters (but only for them), these two cases coincide, that is, there appears to be a single functional relation between the parameters, for which, the parabola in the denominator of $\Psi(\hat{x})$ has a multiple root $and$ is in fact proportional to the numerator of $\Psi(\hat{x})$.

The functional relation between the parameters for which this twofold effect occurs is:
\begin{equation}
\label{eq7A.3}
\hat{\epsilon}^{*}(\hat{\eta})=\frac{1-8\hat{\eta}}{12\hat{\eta}} \ , \ 0< \hat{\eta}<1/8
\end{equation}

The corresponding $\Psi(\hat{x})$ for this case is noteworthy because it becomes a constant:
\begin{equation}
\label{eq7A.4}
\Psi(\hat{x},\hat{\epsilon}=\hat{\epsilon}^{*})=3\sqrt{ \hat{\eta}}
\end{equation}

Consequently, the quarter-period and the time at the last quarter-period become:
\begin{equation}
\label{eq7A.5}
\frac{T}{4}=\frac{3\pi\sqrt{\hat{\eta}}}{2\omega} \ , \
t-\frac{3T}{4}=\frac{3\sqrt{\hat{\eta}}}{2\omega}\int\limits_{0}^{\hat{x}}\frac{dx}{\sqrt{x(1-x)}}
\end{equation}
which is similar to the result obtained in the $\hat{\epsilon} \ll 1$ case. The corresponding (also similar) displacement history becomes:
\begin{equation}
\label{eq7A.6}
\left. y(t,\hat{\epsilon}=\hat{\epsilon}^{*})\right|_{\hat{\eta}<1/8}=Y_0\cos{\left(\frac{\omega t}{3\sqrt{\hat{\eta}}}\right)}
\end{equation}

This result is very interesting, since it is an exact expression for a NNM, with no asymptotic approximations, but rather for finite values of the parameters, and it is merely a simple cosine function, much like the solution in the linear case, only with a different frequency, $\omega/(3\sqrt{\eta})$ instead of $\omega/\sqrt{1+\hat{\eta}}$.

One notes that the frequencies become identical at $\hat{\eta}=1/8$ and are similar around that value, which means that the $\hat{\epsilon}=\hat{\epsilon}^{*}$ case is not only functionally similar but exactly coincides with the $\hat{\epsilon} \ll 1$ case in the appropriate limit. Also noteworthy is the fact that in the same time not only a result for finite parameters, but also an asymptotic result was obtained, for the double asymptotic condition: $\hat{\epsilon} \gg 1,\hat{\eta} \ll 1$, as can be observed by taking the appropriate limit in Eq. (\ref{eq7A.3}).

Next, Eq. (\ref{eq7A.6}) can be used to check for instability. This time, due to the fact that $\hat{x}(t)$ would contain only a single cosine (with double the frequency of $y(t)$), linear stability could be examined for the nonlinear solution by Floquet analysis of the corresponding Mathieu equation, and it would be exact for all amplitudes (and not just in the small amplitude limit, as in the case later addressed in \ref{AppendixD2}).

The simplest way to perform that analysis would be as follows. First, one needs to calculate the values of $r$ and $s$ corresponding to the considered case, to substitute for the values given in Eq. (\ref{eq6B.6}).

Substituting Eq. (\ref{eq7A.6}) into Eq. (\ref{eq6B.2}) employing Eq. (\ref{eq7A.3}), and bringing it to the form in Eq. (\ref{eq6B.5}), one gets the following result:
\begin{equation}
\label{eq7A.7}
r =\frac{9}{4}-9\hat{\eta} \ \ , \ \ s = \frac{9}{8}-9\hat{\eta}
\end{equation}
which means that the two Mathieu equation coefficients are parametrized by a single parameter, and thus one should have a line in the $r$-$s$ plane for which the solution in Eq. (\ref{eq7A.6}) is true and thus Mathieu equation-based stability analysis is relevant.

Next, the two simple equations in Eqs. (\ref{eq7A.7}) can be solved for $\hat{\eta}$ in order to be brought to explicit form and produce the equation of the aforementioned line in the $r$-$s$ plane, which would be:
\begin{equation}
\label{eq7A.7}
s = r- \frac{9}{8}  \ \ , \ r \in \left(\frac{9}{8},\frac{9}{4}\right)
\end{equation}
which is a finite-length straight line. A known result \cite{Verhulst2008} gives the small $s$ limit of the right (rather than the left) boundary of the first Mathieu-limit instability tongue as: $s=r-1$, the quadratic correction for which is positive, since the first canonic tongue is convex. Consequently, since: $r-9/8<r-1$, one sees that the emerging line lies $below$ the first Mathieu instability tongue, and thus the corresponding solution is $stable$.

Therefore this special case analysis although providing interesting insights does not lead to the definite establishing of the existence of nonlinear breaking of linear symmetry. More general parametric combinations thus need to be examined.

As a concluding remark, one should note the curious fact that for: $\hat{\eta}=1/9,\hat{\epsilon} \ll 1$, the symmetric mode follows: $\hat{y}(t)=\cos{(\sqrt{\xi_0}\omega t)}$ (where $\sqrt{\xi_0}=3/\sqrt{10}\approx 0.95$), which is the symmetric mode of the first resonance frequency in the linear infinitesimally asymmetric undamped case, whereas for one special (higher) amplitude, for $\hat{\eta}=1/9,\hat{\epsilon} =1/12$, the symmetric $stable$ mode follows: $\hat{y}(t)=\cos{(\omega t)}$, which is also exactly the symmetric mode of the $second$ resonance frequency in the linear infinitesimally asymmetric undamped case (Eqs. (\ref{eq3.4}), (\ref{eq3.5})).

\subsection{The $\hat{\eta} \ll 1,\hat{\epsilon}=\mathcal{O}(1)$ case}
\label{sect7A.3}

The small $\hat{\eta}$ is in fact already covered in the previous subsection, but only for large amplitudes, where no instability is found. The small $\hat{\eta}$-small amplitude case is also covered (in \ref{sect7A.1}) and no sufficient instability is established (if small, asymmetry linearly hiding damping is assumed). Within the realm of asymptotic approximation it is therefore only left to examine the small $\hat{\eta}-$finite amplitude range.

Substituting the assumption $\hat{\eta} \ll 1, \hat{\epsilon} =\mathcal{O}(1)$ into Eqs. (\ref{eq7.9}), one sees that the numerator in the integrand in Eq. (\ref{eq7.14}) tends to unity and the corresponding values in the denominator become: $A \to 0, B \to \hat{\epsilon}, C \to 1+\hat{\epsilon}$. By contracting $\hat{\epsilon}^{-1/2}$, a standard elliptic integral is obtained with $P^{-1/2}_3(x)$ as the integrand ($P_3(x)$ being a cubic polynomial in canonic form). Using numerous properties of elliptic integrals as described in \cite{AbramowitzStegun1964}, one obtains the quarter period as:
\begin{equation}
\label{eq7A.8}
\frac{T}{4}=\frac{1}{\omega}\frac{K\left(\frac{\hat{\epsilon}}{1+2\hat{\epsilon}}\right)}{\sqrt{1+2\hat{\epsilon}}}
\end{equation}
where $K(m)$ is the complete elliptic integral of the first kind.

Since for this case the second integral of motion is given by a single standard elliptic integral function, it can be inverted analytically to give a closed-form expression for the symmetric mode, which becomes:

\begin{equation}
\label{eq7A.9}
\left.y(t,\hat{\epsilon})\right|_{\hat{\eta}\to 0}=Y_0  \ \text{cn} \left(\sqrt{1+2\hat{\epsilon}} \ \omega t\left|\frac{\hat{\epsilon}}{1+2\hat{\epsilon}}\right.\right)
\end{equation}
where $\text{cn}(z|m)$ is the second Jacobi elliptic function.

As expected, this solution reproduces the small-amplitude approximation's result in the double limit $\hat{\eta}\to 0,\hat{\epsilon}\to 0$, owing to the fact that $K(0)=\pi/2$ and $\text{cn}(z|0)=\cos(z)$.

Now, since it would be perhaps most natural to examine the stability of $y(t)$ using Hill's method, there arises a need for a Fourier series expansion of $\hat{x}(t)$. Since the initial condition was chosen such that the solution would be an even function (and 'cn' is indeed even), a cosine series for $\hat{y}^2(t)$ would suffice. Using a cosine series representation given for $\text{cn}(u|m)$ in \cite{AbramowitzStegun1964}, one obtains the following representation for $\hat{y}(t)$:
\begin{equation}
\label{eq7A.10}
\hat{y}(\hat{\mathcal{T}})=\sum\limits_{n=1}^{\infty}C_n\cos{\left(n\hat{\mathcal{T}}\right)}, \ \hat{\mathcal{T}} \triangleq \frac{\pi\sqrt{1+2\hat{\epsilon}}\omega t}{2K\left(\frac{\hat{\epsilon}}{1+2\hat{\epsilon}}\right)}, \ C_n \triangleq \frac{\pi\sqrt{1+2\hat{\epsilon}}\delta_{n,2\mathbb{N}-1}}{\sqrt{\hat{\epsilon}}K\left(\frac{\hat{\epsilon}}{1+2\hat{\epsilon}}\right)\cosh{\left[ \frac{n\pi K\left(\frac{1+\hat{\epsilon}}{1+2\hat{\epsilon}}\right)}{2K\left(\frac{\hat{\epsilon}}{1+2\hat{\epsilon}}\right)}\right]}}
\end{equation}
where $\delta_{p,q}$ is Kronecker's Delta, and $\mathbb{N}$ is any natural number.

Using the cosine series representation for $\hat{y}$, one can now derive a cosine series representation for $\hat{x}$ using direct discrete convolution, yielding the following result:
\begin{equation}
\label{eq7A.12}
\hat{x}(\hat{\mathcal{T}})=\frac{X_0}{2}+\sum\limits_{n=1}^{\infty}X_n\cos{\left(2n\hat{\mathcal{T}}\right)}
\end{equation}
where
\begin{equation}
\label{eq7A.13}
\begin{split}
X_n \triangleq \sum\limits_{m=1,3,5}^{\infty} C_m C_{2n+m}+\frac{1}{2}C_n^2 \delta_{n,2\mathbb{N}-1}+(1-\delta_{n,1}) \sum\limits_{m=1}^{\infty} C_{n-m} C_{n+m}
\end{split}
\end{equation}

Next, rewriting Eq. (\ref{eq6B.2}) to the normalized time, $\hat{\mathcal{T}}$, one obtains the final form of the Hill equation suitable for analysis by Hill's method for arbitrary finite order amplitudes and $\hat{\eta} \ll 1$:
\begin{equation}
\label{eq7A.14}
z''(\hat{\mathcal{T}})+\frac{4}{\pi^2}\frac{\left[K\left(\frac{\hat{\epsilon}}{1+2\hat{\epsilon}}\right)\right]^2}{1+2\hat{\epsilon}}[1+6\hat{\epsilon}\hat{x}(\hat{\mathcal{T}} )]z(\hat{\mathcal{T}})=0
\end{equation}

Eqs. (\ref{eq7A.10})-(\ref{eq7A.14}) are sufficient for the implementation of Hill's Infinite Determinants method for the establishing of the stability of $y(t)$, as given by the second Jacobi elliptic function in the $\hat{\eta} \ll 1,\hat{\epsilon}=\mathcal{O}(1)$ limit.

\renewcommand{\theequation}{C.\arabic{equation}}
\setcounter{equation}{0}

\section{Asymptotic convergence analysis of the fixed-point iterative scheme in Eq. (\ref{eq8A.6})}
\label{AppendixC}

General analytic convergence analysis of a multi-dimensional fixed-point iteration scheme is normally quite tedious a task. Here a different strategy is followed and asymptotic approach is followed instead. Before examining the stability of the scheme in Eq. (\ref{eq8A.6}), second-order expansion of the associated quantities in terms of the squared dimensionless amplitude, $\hat{\epsilon}$, is performed. The fundamental quantities required for the implicit algorithm are $\hat{T}$ and $c_0$. These quantities are given explicitly by Eqs. (\ref{eq7.14})-(\ref{eq7.4}), (\ref{eq7B.1}) and (\ref{eq8.1}). Taking second-order asymptotic expansions in $\hat{\epsilon}$ of the full expressions, one gets:
\begin{equation}
\label{eqAppC.1}
\hat{T} \underset{\hat{\epsilon} \ll 1} \to \sqrt{1+\hat{\eta}}-\frac{3}{4}\frac{\hat{\epsilon}}{\sqrt{1+\hat{\eta}}}+\frac{48\hat{\eta}+57}{64(1+\hat{\eta})^{3/2}}\hat{\epsilon}^2+\mathcal{O}(\hat{\epsilon}^3)
\end{equation}
\begin{equation}
\label{eqAppC.2}
c_0 \underset{\hat{\epsilon} \ll 1} \to \frac{1}{2}\left[1-\frac{1}{8}\frac{1-8\hat{\eta}}{1+\hat{\eta}}\hat{\epsilon}-\frac{40\hat{\eta}^2-8\hat{\eta}-3}{16(1+\hat{\eta})^2}\hat{\epsilon}^2\right]+\mathcal{O}(\hat{\epsilon}^3)
\end{equation}

The next step is to use these expansions and to apply the implicit algorithm in sequential approximations, only instead of understanding the Eq. (\ref{eq8A.6}) as an iterative scheme one should view it as an implicit nonlinear equation and solve it by asymptotic analysis, starting from $n=2$. Substituting Eqs. (\ref{eqAppC.1}) and (\ref{eqAppC.2}) into the 'implicit equation' version of Eq. (\ref{eq8A.6}), along with the associated definitions, taking $n=2$ and performing rigorous asymptotic analysis of the resulting implicit equation, an asymptotic expansion is obtained for $c_1$. Using the three obtained expansions in Eq. (\ref{eq8A.8}) and performing rigorous asymptotic analysis of the resulting explicit expression, the second-order expansion for $c_2$ is obtained. Repeating the process for higher values of $n$, one obtains second-order asymptotic expansions for the coefficients $c_{n \ge3}$, as well as formal (and as it turns out, vanishing) corrections for $c_1$ and $c_2$. The results are given below:
\begin{equation}
\label{eqAppC.3}
\begin{split}
c_1 \underset{\hat{\epsilon} \ll 1} \to -\frac{1}{2}+\mathcal{O}(\hat{\epsilon}^3) \\
c_2 \underset{\hat{\epsilon} \ll 1} \to \frac{1-8\hat{\eta}}{16(1+\hat{\eta})}\hat{\epsilon}+\frac{120\hat{\eta}^2-24\hat{\eta}-9}{32(1+\hat{\eta})^2}\hat{\epsilon}^2+\mathcal{O}(\hat{\epsilon}^3) \\
c_{n \ge3} \underset{\hat{\epsilon} \ll 1} \to \mathcal{O}(\hat{\epsilon}^3)
\end{split}
\end{equation}

These results indicate that for a second-order expansion it is sufficient to take $c_1=-1/2$ and $c_{n \ge 3}=0$, which corresponds to $n=2$ and a one-dimensional iterative scheme $c_1^{(m+1)}=G(c_1^{(m)})$. The scheme can be considered convergent under the condition: $|G'(-1/2)|<1$. Differentiating Eq. (\ref{eq8A.4}), substituting the result into Eq. (\ref{eq8A.6}), substituting the known solution for $c_1=-1/2$ in the result and recalling Eq. (\ref{eq8.3}) which becomes valid at the root, one obtains the following expression for the derivative:
\begin{equation}
\label{eqAppC.4}
G'(c_1)=\frac{q_1q_2}{q_1-q_2}c_1[H(c_0+q_2 c_1)-H(c_0+q_1 c_1)]
\end{equation}
where $q_{1,2}=\pm 1/\sqrt{2}$ and $H(z)$ is given by:
\begin{equation}
\label{eqAppC.5}
H(z) \triangleq \frac{1-2z}{2z(1-z)}-\frac{6\hat{\eta}\hat{\epsilon}}{1+\hat{\eta}(1+6\hat{\epsilon}z)}+\frac{2A z+B}{2z(1-z)(A z^2+B z+C)}
\end{equation}
and $A,B,C$ are given by Eq. (\ref{eq7.9}). The values of $c_1$ and $c_0$ are taken from the second-order expansions given above and use is made of the recursive definitions for Chebyshev polynomials to obtain $T_1,T_2,T'_1,T'_2$.

Rigorous second-order asymptotic analysis of Eq. (\ref{eqAppC.4}) then produces the following result:
\begin{equation}
\label{eqAppC.6}
G'(c_1) \underset{\hat{\epsilon} \ll 1} \to -1+\frac{80\hat{\eta}^2+176\hat{\eta}-39}{32(1+\hat{\eta})^2}\hat{\epsilon}^2+\mathcal{O}(\hat{\epsilon}^3)
\end{equation}

The small amplitude/weak nonlinearity limit convergence condition for the iterative fixed-point scheme then becomes:
\begin{equation}
\label{eqAppC.7}
\hat{\eta} \left | _{\hat{\epsilon} \ll 1}^{stab.} \right. >\frac{\sqrt{679}-22}{20} \approx 0.20288
\end{equation}

Under this condition one has: $-1<G'(c_1)<0$, which formally guarantees stability. Numerical analysis reveals that optimal convergence rate is retained for: $\hat{\eta}>1/4$. This result is rigorously (asymptotically) applicable also for the multidimensional case. The spectral norm of the gradient of the scheme would be small than unity in absolute value under the condition in Eq. (\ref{eqAppC.7}). This is confirmed numerically. Finally, although rigorous as a limit, order comparison shows that the result should hold for dimensionless amplitudes up to about 2/3, for which the error is less than about 1 percent (the error is a sixth-order correction in amplitude). This covers all the stability map region relevant for all the revealed Poincar\'{e} Sections bifurcations. Thus the tongue boundaries obtained with the scheme can be even formally trusted. To conclude, two notes should be made. First is that as one can observe from Eq. (\ref{eqAppC.7}), for, say, $\hat{\eta}>1/4$,  increasing nonlinearity only enhances the convergence condition fulfillment, at least as long as the asymptotics is valid. Second, although in the $\hat{\eta}<1/5,\hat{\epsilon} \ll1$ case the scheme is divergent, as shown asymptotically in \ref{AppendixD2}, there is no instability to search for in that region, and for higher amplitudes the third-order correction presumably renders the scheme convergent again, as this is what numerical investigation suggests.

\renewcommand{\theequation}{D.\arabic{equation}}
\setcounter{equation}{0}

\section{Symmetric mode stability analysis details}
\label{AppendixD}

\renewcommand{\theequation}{D.\arabic{equation}}
\setcounter{equation}{0}

\subsection{Symmetric mode general stability analysis details -- Hill's method}
\label{AppendixD1}

Substituting Eq. (\ref{eq9.3}) into Eq. (\ref{eq9.1}), reorganizing and using a complex representation of the cosine, one obtains a form of Hill's equation as follows:
\begin{equation}
\label{eq9.4}
\begin{split}
z''(\lambda)+\left[\hat{r}+\sum\limits_{k=1}^{N_c}\hat{s}_k \left(e^{2 i k\lambda}+e^{-2i k \lambda}\right)\right]z(\lambda)=0
\end{split}
\end{equation}
where
\begin{equation}
\label{eq9.5}
\begin{split}
\hat{r} \triangleq \Lambda^2(1+6\hat{\epsilon}c_0) \ , \ \hat{s}_k \triangleq 3\hat{\epsilon}\Lambda^2c_k
\end{split}
\end{equation}

Next, $z(\lambda)$ can be rewritten as the general solution of Eq. (\ref{eq9.4}), according to the well-known result of Floquet theory applied to the Hill equation \cite{Ward}, as follows:
\begin{equation}
\label{eq9.6}
z(\lambda)=e^{i\gamma\lambda}\sum\limits_{m=-\infty}^{\infty}\hat{z}_m e^{2i m\lambda}
\end{equation}

Substituting Eq. (\ref{eq9.6}) into Eq. (\ref{eq9.4}), one obtains a set of equations, that can formally be written in the following form:
\begin{equation}
\label{eq9.7}
\hat{s}_k \hat{z}_{m-k}+[\hat{r}-(\gamma+2m)^2]\hat{z}_m+\hat{s}_k\hat{z}_{m+k}=0
\end{equation}
where Einstein's summation convention is assumed for $k$, and $m \in \mathbb{Z}$.

The stability requirement (of the identical zero solution for the antisymmetric mode) then reduces to the condition that guarantees periodic solutions for all initial conditions. In other words, stability is guaranteed if $\gamma$ is real for every $\hat{\textbf{z}}$. Thus the stability condition can be obtained if one assumes $\gamma\in \mathbb{R}$, then writes Eqs. (\ref{eq9.7}) explicitly and requires that the determinant of the resulting system of equations vanishes.

In order to obtain the boundaries of the linearly stable domain in the $\hat{\eta}$-$\hat{\epsilon}$ plane, one needs to compute the critical values of $\gamma$. Then an equation for $\hat{r}$ and $\hat{\textbf{s}}$ will be obtained. It relates $\hat{\eta}$ and $\hat{\epsilon}$ and allows derivation of limit-stability curves.

When the multipliers become identical, a secular solution emerges, and this is the instability. Since the arguments in the exponents have to be imaginary in the stable or critically-stable case, the multipliers are less then unity in absolute value. In the shift to instability, they are equal to unity in absolute value. If they are identical and a secular solution emerges, they can be either both $1$ or both $-1$. When a multiplier equals 1, the exponent argument is $2\pi i$, and when the multiplier equals $-1$, the exponent argument is $\pi i$. A characteristic multiplier is the ratio between the solution in consecutive periods. The period of the periodic function in the fundamental solution is as in the Hill function, that is $\pi$. Consequently, the prefactor before the sum in Eq. (\ref{eq9.6}) becomes the ratio between the values of $z$ at (normalized) times $\lambda=0$ and $\lambda=\pi$, and thus the characteristic multiplier is just the prefactor at $\lambda=\pi$, which in the considered case is: $e^{i\gamma\pi}$. Therefore the two cases of critical stability/instability correspond to:
\begin{equation}
\label{eq9.8}
\gamma^{cr}_1=1,\gamma_2^{cr}=2
\end{equation}
\cite{Arnold}, \cite{Ward}. Consequently, by substituting Eq. (\ref{eq9.8}) into Eq. (\ref{eq9.7}), rearranging and solving for the nontrivial solution (for an arbitrary $\hat{\textbf{z}}$), one obtains the linear stability limits of the symmetric mode.

\subsection{Asymptotic analysis of symmetric mode stability for weak nonlinearity}
\label{AppendixD2}

It is possible to obtain the zero amplitude limit-points of bifurcation of the instability tongues for the symmetric mode by asymptotic expansion of Eqs. (\ref{eq7.14})-(\ref{eq7.4}), Eqs. (\ref{eq7B.1}),(\ref{eq8.1}),(\ref{eq8A.1}),(\ref{eq9.1})-(\ref{eq9.3}),(\ref{eq9.5}) and Eq. (\ref{eq9.7}). However, more direct and somewhat easier asymptotic approach is employed next. To start, small-amplitude approximation of $y(t)$ can be obtained by the Poincar\'{e}-Lindstedt  method (which is the simplest perturbation technique allowing the elimination of artificial secular terms in the solution, much due to the cubic nonlinearity).
First, without loss of generality, one can assume a static initial condition of the form $y(0)=Y_0,\dot{y}(0)=0$. Then the Poincar\'{e}-Lindstedt (version of the multiple time scale) method produces the following leading-order approximation for $y$ (solving Eq. (\ref{eq6.4}) for modal initial conditions):
\begin{equation}
\begin{split}
\left. y(t,\bar{\epsilon} \ll 1) \right |_{v=0}=Y_0\left\lbrace \cos{\Theta} -\frac{9\xi_0-8}{32}\bar{\epsilon}[\cos{\Theta}
 -\cos{(3\Theta)}]\right \rbrace+\mathcal{O}(\bar{\epsilon}^2)
\end{split}
\label{eq6A.1}
\end{equation}
where
\begin{equation}
\begin{split}
\Theta \triangleq \left(1+\frac{3}{8}\xi_0\bar{\epsilon}\right)\sqrt{\xi_0}\omega t, \ \bar{\epsilon} \triangleq \frac{p Y_0^2}{k}
\end{split}
\label{eq6A.2}
\end{equation}

This solution is next checked for stability in order to detect possible symmetry breaking already in the small amplitude regime. Consequently, defining:
\begin{equation}
\begin{split}
\mathcal{T} \triangleq \Theta +\pi/2
\end{split}
\label{eq6B.4}
\end{equation}
and using Eqs. (\ref{eq6A.1}), (\ref{eq6A.2}) and (\ref{eq6B.2}), when only the leading-order term from Eq. (\ref{eq6A.1}) is retained, the Mathieu equation is obtained for $z(\mathcal{T})$, as follows:
\begin{equation}
\label{eq6B.5}
z''(\mathcal{T})+[r-2s\cos(2\mathcal{T})]z(\mathcal{T})=0
\end{equation}
where the parameters $r$ and $s$ are given (to leading order) by:
\begin{equation}
\label{eq6B.6}
r \triangleq \xi_0^{-1}+\left(\frac{3}{2}\xi_0^{-1}-\frac{3}{4}\right)\bar{\epsilon} , \ s \triangleq \frac{3}{4}\xi_0^{-1}\bar{\epsilon}
\end{equation}

Having obtained the parameters $r$ and $s$, defining the Mathieu equation in terms of the natural parameters of the system, one can now characterize the stability range using the known analytic result regarding the bifurcation points from which instability tongues emerge at the zero amplitude limit.

The naturally most interesting is the first tongue, which represents 1:2 resonance, and which corresponds to evident instability even in the infinitesimal nonlinearity limit. The first instability tongue remains close enough to the zero-amplitude axis even in the presence of infinitesimal damping \cite{Verhulst2008}, as what the considered system may have and what accounts for parametric asymmetry hiding, as shown in Section \ref{sect5}.

The first instability tongue is characterized by the following instability conditions in the canonic Mathieu plane \cite{Verhulst2008}: $s>|r-1|$. From Eqs. (\ref{eq2.2}), (\ref{eq7.4}) and (\ref{eq6B.6}) in the $\epsilon \to 0$ limit, we find that: $r>1$ and thus the first-tongue instability condition is: $s>r-1$. Substituting Eq. (\ref{eq6B.6}) into this condition we obtain the following condition for the existence of the first instability tongue:
\begin{equation}
\label{eq6B.7}
\bar{\epsilon}<-4/3
\end{equation}
which is impossible for a convex interaction potential in view of Definitions (\ref{eq6A.2}). Consequently, the first canonical (strongest) instability tongue does not emerge for the considered system.

It is noteworthy that the first instability tongue of the Ince-Strutt diagram, the one in which the two boundaries of instability differ linearly in terms of amplitude, is totally absent in the considered case, due to the specifics of the employed parametrization. Thus in the considered case the first instability tongue is the second tongue from the Ince-Strutt diagram, the one in which the difference between the two boundaries is quadratic in the amplitude.

The edges of the additional instability tongues have the following zero-amplitude limit in the employed parametrization:
\begin{equation}
\label{eq6B.8}
\left. \frac{k}{k_0}\right|^{cr}_{Y_0\to 0}=\frac{N^2-1}{2}, \ \forall \ N \in \mathbb{N}
\end{equation}

These additional tongues, in an amplitude range for which the Mathieu equation limit is asymptotically relevant, are used as starting points for exact stability analysis in Section \ref{sect9}.

The result obtained here is important because according to it the presence of even infinitesimal damping, which can hide parametric asymmetry, renders the symmetric mode stable in the weakest nonlinearity limit, since the higher-order tongues emerge as higher power of the nonlinearity and are barely present even in the undamped case, unlike the first tongue which grows linearly with energy.

Nevertheless, as shown in Section \ref{sect9}, $finite$ (even if small) nonlinearity, does produce symmetry breaking, only by a more delicate parametric resonance mechanism, corresponding to the 1:1 resonance second tongue.

\renewcommand{\theequation}{E.\arabic{equation}}
\setcounter{equation}{0}

\section{Antisymmetric mode stability analysis details}
\label{AppendixE1}

\subsection{Detailed derivation of Hill's equation for the antisymmetric mode}
\label{AppendixE0}

As noted in Section \ref{sect10B}, the dynamic equation for the symmetric mode in terms of the second integral of the antisymmetric mode and the problem parameters, all in dimensionless form, produce, after linearization with respect to an identical zero solution for the symmetric mode, the following linear full-form second-order periodic-coefficient ODE:

\begin{equation}
\label{eq10B.3}
\hat{a}(\hat{\tau})\hat{y}''(\hat{\tau})+\hat{b}(\hat{\tau})\hat{y}'(\hat{\tau})+\hat{c}(\hat{\tau})\hat{y}(\hat{\tau})=0 \ \ ; \ \ \hat{y} \triangleq y/Y_0
\end{equation}
where
\begin{equation}
\label{eq10B.4}
\begin{split}
\hat{a}(\hat{\tau}) \triangleq 2(1+\hat{\eta})\Omega_v^2+3\hat{\eta}\hat{\epsilon}\Omega_v^2[\hat{v}(\hat{\tau})]^2, \
\hat{b}(\hat{\tau}) \triangleq 12\hat{\eta}\hat{\epsilon}\Omega_v^2\hat{v}(\hat{\tau})\hat{v}'(\hat{\tau}) \\
\hat{c}(\hat{\tau}) \triangleq 2+3(1-2\hat{\eta})\hat{\epsilon}[\hat{v}(\hat{\tau})]^2
+6\hat{\eta}\hat{\epsilon}\Omega_v^2[\hat{v}'(\hat{\tau})]^2-3\hat{\eta}\hat{\epsilon}^2[\hat{v}(\hat{\tau})]^4
\end{split}
\end{equation}

Here, one notes that $\hat{b}(\hat{\tau}) =2\hat{a}'(\hat{\tau})$, and, therefore, by defining:
\begin{equation}
\label{eq10B.5}
\tilde{y}(\hat{\tau})\triangleq \hat{a}(\hat{\tau})\hat{y}(\hat{\tau})
\end{equation}
one can eliminate the first ''stretched-time'' derivative of $\hat{y}$, transforming the linear stability equation for $\hat{v}$ into the canonic form of the Hill equation:
\begin{equation}
\label{eq10B.6}
\tilde{y}''(\hat{\tau})+\frac{\hat{c}(\hat{\tau})-\hat{a}''(\hat{\tau})}{\hat{a}(\hat{\tau})}\tilde{y}(\hat{\tau})=0
\end{equation}
(note that $\hat{a}(\hat{\tau})>0$).

Substituting Eq. (\ref{eq10B.4}) into Eq. (\ref{eq10B.6}) and eliminating $\hat{v}''(\hat{\tau})$ from the result by isolating it from the dimensionless, stretched-time, linearized with respect to $\hat{y}(\hat{\tau})$ (or $\tilde{y}(\hat{\tau})$) version of Eq. (\ref{eq6B.1}), one obtains the final, explicit Hill equation for antisymmetric mode linear stability, as given in Eq. (\ref{eq10B.7}).

\subsection{Asymptotic analysis of antisymmetric mode stability for weak nonlinearity}
\label{AppendixE}

Taking the zero amplitude limit in Eq. (\ref{eq10A.6}), recalling that $K(m\to0)\to\pi/2$, one finds that $\Omega_v\to1$ in this case and thus the Mathieu parameter $r$, as defined, say, in Eq. (\ref{eq6B.5}), has the zero amplitude limit of $r \to \xi_0$ (interestingly enough, this is just the opposite of the symmetric mode case, for which the result: $r \to \xi_0^{-1}$ was obtained in the zero amplitude limit).

Since the bifurcation points of the instability tongues are designated by $r\to N^2, N \in \mathbb{N}$, and $\xi_0=1/(1+\hat{\eta})$, where $\hat{\eta}>0$, it appears that the only bifurcation point-producing parameter-plane tongue of antisymmetric mode instability occurs at the limit $\hat{\eta} \to 0$. First-order approximations for the boundaries of this instability tongue in the Mathieu equation limit are next established.

From Eq. (\ref{eq10A.4}) it is known that in the small amplitude limit, the solution becomes a single sine, which directly implies that the Fourier coefficients given in Eq. (\ref{eq10A.6}) decay with $\hat{\epsilon}$. The Hill function in the Mathieu limit should be linear in the amplitude, and thus, since the Fourier coefficients in Eq. (\ref{eq10B.7}) are multiplied by the amplitude, it can be deduced that for a linear expansion in $\hat{\epsilon}$ of the Hill function, the first term in  Eq. (\ref{eq10A.4}), or, in other words, a single sine approximation of $\hat{v}(\hat{\tau})$, should be sufficient for the Mathieu equation limit (this argumentation could be wrong if the linear term vanishes in the final expression, but this does not happen in the considered case). Consequently, performing the asymptotic derivation for the Hill function, one gets:
\begin{equation}
\label{eq10C.1}
\begin{split}
\hat{v}(\hat{\tau}) \approx V_1\sin{(\hat{\tau})} \Rightarrow h(\hat{\tau}) \triangleq
 \frac{\Omega_v^{-2}}{1+\hat{\eta}}\frac{1+(3/2)\hat{\epsilon}[\hat{v}(\hat{\tau})]^2}{1+(3/2)\frac{\hat{\eta}}{1+\hat{\eta}}\hat{\epsilon}[\hat{v}(\hat{\tau})]^2} \approx \\ \frac{\Omega_v^{-2}}{1+\hat{\eta}}\frac{1+(3/4)\hat{\epsilon}V_1^2-(3/4)\hat{\epsilon}V_1^2\cos{(2\hat{\tau})}}{1+(3/4)\frac{\hat{\eta}}{1+\hat{\eta}}\hat{\epsilon}V_1^2[1-\cos{(2\hat{\tau})}]} \underset{\hat{\epsilon}\to 0}\to \frac{\Omega_v^{-2}}{1+\hat{\eta}} \left \lbrace 1+(3/4)\hat{\epsilon}\frac{V_1^2}{1+\hat{\eta}}[1-\cos{(2\hat{\tau})}] \right \rbrace
\end{split}
\end{equation}
where in order to get a linear expansion in $\hat{\epsilon}$, one needs a linear expansion of $\Omega_v^{-2}$, and an $\mathcal{O}(1)$ expansion of $V_1^2$.

Making use of the following asymptotic expansions of the complete elliptic integral of the first kind \cite{AbramowitzStegun1964}:
\begin{equation}
\label{eq10C.2}
\begin{split}
K(\hat{m}) \underset{\hat{m}\to 0} \to \frac{\pi}{2}\left(1+\frac{1}{4}\hat{m}\right), \
K(1-\hat{m}) \underset{\hat{m}\to 0} \to \ln{4}-\frac{1}{2}\ln{\hat{m}}
\end{split}
\end{equation}
and deriving by asymptotic expansion from Eq. (\ref{eq10A.6}) that:
\begin{equation}
\label{eq10C.3}
\hat{m} \underset{\hat{\epsilon}\to0} \to (1+\hat{\eta})\hat{\epsilon} \ , \ \Omega_v\underset{\hat{\epsilon}\to0} \to 1+\frac{3}{4}\hat{m}
\end{equation}
one finds that:
\begin{equation}
\label{eq10C.4}
V_1 \underset{\hat{\epsilon}\to0} \to 2\sqrt{1+\hat{\eta}}
\end{equation}
and finally, that:
\begin{equation}
\label{eq10C.5}
h(\hat{\tau}) \underset{\hat{\epsilon}\to 0} \to \tilde{r}-2\tilde{s}\cos{(2\hat{\tau})} \ , \ \tilde{r} \triangleq \frac{1+(3/2)(1-\hat{\eta})\hat{\epsilon}}{1+\hat{\eta}} \ , \ \
\tilde{s} \triangleq \frac{(3/2)\hat{\epsilon}}{1+\hat{\eta}}
\end{equation}

\subsubsection{First instability tongue}
\label{sect10C1}

According to the classical result regarding Mathieu equations, instability tongues bifurcate from the zero amplitude axis for $\tilde{r}=N^2,N \in \mathbb{N}$ (in the parametrization of Eq. (\ref{eq10C.5})). Due to the form of $\tilde{r}$ in Eq. (\ref{eq10C.5}), Definition (\ref{eq7.4}), and the positivity of linear spring-stiffness coefficients, it appears that the only Mathieu equation limit instability tongue exist for $\hat{\eta} \to 0$, where $\tilde{r} \to 1^2$.

Taking a first-order asymptotic expansions of $\tilde{r}$ and $\tilde{s}$ using Eqs. (\ref{eq10C.5}), retaining only terms linear in either $\hat{\epsilon}$ or $\hat{\eta}$ (but not in their product, which for $\hat{\epsilon}\to 0,\hat{\eta}\to 0$ would be a higher-order correction), one obtains the following expressions for the Mathieu approximation parameters applicable to the first instability tongue:
\begin{equation}
\label{eq10C1.1}
\tilde{r}^{(1)}\underset{\hat{\eta},\hat{\epsilon}\to 0} \to 1-\hat{\eta}+\frac{3}{2}\hat{\epsilon} \ , \ \ \tilde{s}^{(1)}\underset{\hat{\eta},\hat{\epsilon}\to 0} \to \frac{3}{2}\hat{\epsilon}
\end{equation}

Next, one recalls the well-known result regarding the functional description of the boundaries of the first instability tongue for the Mathieu equation in the canonic representation (as in Eq. (\ref{eq10C.5})) \cite{Verhulst2008}, namely:
\begin{equation}
\label{eq10C1.2}
\tilde{r}^{(1)} \underset{\tilde{s}^{(1)} \to 0} \to 1 \pm \tilde{s}^{(1)}
\end{equation}

Combining Eqs. (\ref{eq10C1.1}) and (\ref{eq10C1.2}), one obtains two first-order small-amplitude asymptotic approximations for each of the two  boundaries of the first instability tongue of the purely antisymmetric mode in the natural (rather than canonic) parametrization, which reads:
\begin{equation}
\label{eq10C1.3}
\hat{\eta}^{(1)}_- \underset{\hat{\epsilon}\to 0} \to 0\ \ , \ \ \hat{\eta}^{(1)}_+ \underset{\hat{\epsilon}\to 0} \to 3\hat{\epsilon}
\end{equation}

The first of these boundaries is found to coincide with the exact boundary established by analyzing the full Hill equation. The second boundary is found to be a good approximation for dimensionless amplitude up to about 0.3 (it is plotted along with the exact analysis instability tongue boundaries in Fig. \ref{Figure5} at the end of Section \ref{sect10}).

Several things should be noted at this point. The first is that the already obtained result is noteworthy. Unlike in the case of the analysis of the stability of the symmetric mode, where $all$ the tongues were present, $except$ for the first one, the "widest" one, which vanished, leaving only a $quadratically$ non-degenerate (second) tongue as the dominant one, which in view of possible small (linear asymmetry hiding) damping might have vanished entirely as well at small amplitudes, requiring full Hill analysis for establishing real instability, here, in the case of the purely antisymmetric mode, there is only one instability tongue, but it is the first one and thus the strongest one, and it is linearly non-degenerate, and thus is sufficient to assure the emergence of instability, for small enough stiffness ratios, even for small amplitudes (and probably even in the presence of infinitesimal linear damping).

The second point is that the Mathieu equation approximation which shows the emergence of all the tongues except for the first one for the symmetric mode, shows only the first tongue for the antisymmetric mode and no $additional$ tongues. It is true that only the bifurcation points at zero amplitudes are considered, but in a conservative system, tongues cannot start at finite amplitudes. Knowing that some small damping might be present, full Hill analysis should be (and is indeed) performed (in Section \ref{sect10}). Still however, it is a fact that the immediate result from the Mathieu approximation is that only $one$ instability tongue emerges, finite and instability-establishing as it may be. In the following subsection it is examined by higher-asymptotics linear stability analysis, whether additional, degenerate instability tongues emerge, which may reflect the existence of finite, non-degenerate $nonlinear$ $instability$ tongues (this issue is also elaborated on in Section \ref{sect11}).

\subsubsection{Second instability tongue}
\label{sect10C2}

Appealing to the canonic form of the Mathieu equation shows that in the feasible parameter range there appears to be only one, the first, instability tongue, and $no$ additional $canonic$ tongues. However, one can formally asymptotically examine if there exist additional instability tongues in the natural (rather than canonic) ($\hat{\eta},\hat{\epsilon}$) parametrization.

For example, if the already stretched time $\hat{\tau}$ is further rescaled to $\bar{\tau}=\hat{\tau}/2$, the following, modified Hill function is obtained from Eq. (\ref{eq10C.1}), instead of Eq. (\ref{eq10C.5}):
\begin{equation}
\label{eq10C2.1}
\bar{h}(\bar{\tau}) \underset{\hat{\epsilon}\to 0} \to \frac{4+6(1-\hat{\eta})\hat{\epsilon}}{1+\hat{\eta}}-\frac{12\hat{\epsilon}}{1+\hat{\eta}}\cos{(4\bar{\tau})}
\end{equation}

The period of $\bar{h}(\bar{\tau})$ is $\pi/2$. In order to perform Floquet analysis, it should now be assumed that the period of $\tilde{y}(\tau)$, as emerges from Eq. (\ref{eq10B.7}), is also $\pi/2$, or $\pi$, but then one would reproduce the result of the previous section. Instead, a periodic symmetric mode with a (larger) period of $2\pi$ now is assumed. This is reasonable since in the absence of resonance of multiplicity 1 and 2 for finite values of $\hat{\eta}$, the weaker but still possible resonance would be of multiplicity 4, corresponding to the response being 4 times slower than the (parametric) excitation. Consequently, one would still decompose $\tilde{y}(\bar{\tau})$ in the manner in which it is done in Section \ref{sect9}, and the critical values of the characteristic exponent (divided by $i$) would still be 1 or 2. For $\gamma_{cr}=2$, one would get $\tilde{y}(\bar{\tau})$ with a period of $\pi$, which is twice as large as the period of $\bar{h}(\bar{\tau})$, and is already covered in the Mathieu approximation treatment. Therefore, the only additional tongue could be obtained for $\gamma_{cr}=1$, producing resonance with multiplicity 4.

The first odd-index tongue corresponding to $\gamma_{cr}=1$ can be obtained using Eqs. (\ref{eq10C2.1}) and (\ref{eq9.7}). If one substitutes the parameters of the Hill function recognizable as $r$ and $s$, from Eq. (\ref{eq10C2.1}) into Eq. (\ref{eq9.7}), takes the zero amplitude limit and sets $\gamma=\gamma_{cr}=1$, one obtains an equation which only gives zero for arbitrary coefficients $\hat{z}_k$, if the condition $r=(1+2m)^2$ is satisfied for $m\in\mathbb{Z}$, or explicitly:
\begin{equation}
\label{eq10C2.2}
\frac{4}{1+\hat{\eta}|_{\hat{\epsilon\to 0}}}=(1+2m)^2,m\in\mathbb{Z}
\end{equation}

This equation has only one solution for a positive $\hat{\eta}$, namely $\hat{\eta}|_{\hat{\epsilon\to 0}}=3$ (interestingly enough this is also the bifurcation point of the first non-vanishing instability tongue of the $symmetric$ mode).

Now, in order to obtain asymptotic approximations for the tongue boundaries, in addition to just the bifurcation point, the following reasonable form (a bifurcation point can hardly have a zero slope, or strictly quadratic dependence at its origin) is assumed:
\begin{equation}
\label{eq10C2.3}
\hat{\eta}\underset{\hat{\epsilon}\to 0} \to 3-\beta \hat{\epsilon}
\end{equation}

Substituting ansatz (\ref{eq10C2.3}) into Eq. (\ref{eq10C2.1}), performing the necessary asymptotic calculations and retaining only terms linear in $\hat{\epsilon}$, one obtains the following form of the Hill function:
\begin{equation}
\label{eq10C2.4}
\bar{h}(\bar{\tau}) \underset{\hat{\epsilon}\to 0} \to 1-(3-\beta/4)\hat{\epsilon}-3\hat{\epsilon}\cos{(4\bar{\tau})}
\end{equation}

Next, the tongue boundary slope, $\beta$, can be found by order-comparison asymptotics, using a (correctly constructed) representative part of the  Hill determinant. In line with the aforementioned remark regarding the assumed period of $\tilde{y}(\bar{\tau})$, one can now construct the Hill determinant, in which, following the notation in Eqs. (\ref{eq9.4})-(\ref{eq9.5}), only $\hat{s}_2$  remains -- in addition to $\hat{r}$ -- (of all the coefficients, unlike in the Mathieu equation, where only $\hat{s}_1$ does not vanish).

This leads to a matrix, in which in addition to the main diagonal, there are also nonzero coefficients in the diagonals that are $next$-$to$-$adjacent$ (rather than $adjacent$, as in the Mathieu case) to the principle diagonal.

Now, since what can make the determinant of this matrix vanish in the zero amplitude limit are two terms in the diagonal, corresponding to $m=0,-1$ (see Eq. (\ref{eq10C2.2})), it would be reasonable to take a "chunk" of the matrix centered around the two rows corresponding to these values of $m$. Such typical a "chunk" of the Hill determinant with $\bar{r}$ and $\bar{s}_2$, takes the following form:
\begin{equation}
\label{eq10C2.5}
\text{det} \begin{pmatrix}
       {\bar{r}-9} & 0 & \bar{s}_2 & 0          \\
       0 & {\bar{r}-1} & 0 &\bar{s}_2         \\
       \bar{s}_2 & 0 & {\bar{r}-1} & 0    \\
       0 & \bar{s}_2 & 0  & {\bar{r}-9}
     \end{pmatrix}=0
\end{equation}

The solution of the minimum-representation determinant vanishing problem given in Eq. (\ref{eq10C2.5}) can be easily obtained by expanding in minors, rearranging to quadratic form and extracting the double root, which results in:
\begin{equation}
\label{eq10C2.6}
(\bar{r}-9)(\bar{r}-1)=\bar{s}_2^2
\end{equation}

Extracting $\bar{r}$ and $\bar{s}_2$ from Eq. (\ref{eq10C2.4}), substituting into Eq. (\ref{eq10C2.6}) and rearranging, one gets the following quadratic equation for $\beta$:
\begin{equation}
\label{eq10C2.7}
\hat{\epsilon}\beta^2-\left(8+6\hat{\epsilon}\right)\beta+96=0
\end{equation}
which has the zero-amplitude limit solution of $\beta=12$, leading to the following approximation for the boundary of the second instability tongue:
\begin{equation}
\label{eq10C2.8}
\hat{\eta}_{\pm}^{(2)} \underset{\hat{\epsilon}\to 0} \to 3-12 \hat{\epsilon}
\end{equation}

This approximation is plotted along with the numerical solution in Fig. \ref{Figure5}(b) in Section \ref{sect10}, showing good correspondence for dimensionless amplitudes up to about 0.2.

A few notes should be made. First, a second tongue was obtained but it was not seen that it is a non-collapsed one from the asymptotic analysis. Indeed, the numerical analysis in Section \ref{sect10} verifies the existence of a second tongue, which is a degenerate, or a collapsed one. This means that the Hill determinant not only has a root along the tongue, but that this root is also a local extremum (in this view it was indeed helpful that the secant method was employed to find the root, rather than say, Newton's method). Therefore, both the canonic Mathieu equation analysis, the presented asymptotic treatment, and numerical search showed the same thing: there are no additional non-degenerate instability tongues.

The importance of identifying tongues besides the first, strongest one, is revealed in Section \ref{sect11}. It can be argued that, as aforementioned, a stripe of nonlinear instability surrounds a degenerate linear instability tongue. This nonlinear instability tongue can manifest itself say, in altering the phase space topology (such as changing the torus genus, or creating KAM islands). Indeed, the second tongue is special in that sense, since apart from being identified asymptotically, it is also detected numerically by the full Hill determinants method, and is characterized by the fuller dynamics on Poincar\'e surface sections, as the center of emanation of topological changes (KAM islands). Thus it may well be that a degenerate linear instability tongue corresponds to a non-degenerate nonlinear instability tongue, which is related to topological bifurcations in the associated general motion phase-space dynamics.

\section*{References}

\bibliography{NPOVG_JSV_P_Ref}

\end{document}